\documentclass[preprint,8pt,3p,authoryear]{elsarticle}

\pdfoutput=1

\usepackage[hyperref]{xcolor}
\usepackage[colorlinks = false]{hyperref}
\usepackage{bm}
\usepackage{amsmath}
\usepackage{mathrsfs}
\usepackage{siunitx}
\usepackage{soul}
\usepackage{subfig}
\usepackage{booktabs}
\usepackage[export]{adjustbox}
\usepackage{array}
\usepackage{booktabs}
\usepackage[nameinlink,capitalise]{cleveref}
\usepackage{multirow}
\usepackage{physics}
\usepackage{amssymb}
\usepackage{amsthm}

\DeclareMathOperator*{\argmin}{arg\,min}

\newdefinition{rmk}{Remark}

\usepackage{lineno}
\usepackage{graphicx}

\biboptions{comma, semicolon, sort}

\journal{Journal of the Mechanics and Physics of Solids}

\begin{document}

\begin{frontmatter}

\title{A statistical mechanics framework for polymer chain scission, based on the concepts of distorted bond potential and asymptotic matching} 

\author[address1]{Jason Mulderrig}
\author[address2]{Brandon Talamini}
\author[address1]{Nikolaos Bouklas\corref{corr}}
\address[address1]{Sibley School of Mechanical and Aerospace Engineering, Cornell University, NY 14853, USA}
\address[address2]{Lawrence Livermore National Laboratory, Livermore, CA 94550, USA}
\cortext[corr]{Corresponding author}
\ead{nb589@cornell.edu}

\begin{abstract}
	 To design increasingly tough, resilient, and fatigue-resistant elastomers and hydrogels, the relationship between controllable network parameters at the molecular level (bond type, non-uniform chain length, entanglement density, etc.) to macroscopic quantities that govern damage and failure must be established. Many of the most successful constitutive models for elastomers have been rooted in statistical mechanical treatments of polymer chains. Typically, such constitutive models have used variants of the freely jointed chain model with rigid links. However, since the free energy state of a polymer chain is dominated by enthalpic bond distortion effects as the chain approaches its rupture point, bond extensibility ought to be accounted for if the model is intended to capture chain rupture. To that end, a new bond potential is supplemented to the freely jointed chain model (as derived in the $u$FJC framework of \citet{buche2021chain} and \citet{buche2022freely}), which we have extended to yield a tractable, closed-form model that is amenable to constitutive model development. Inspired by the asymptotically matched $u$FJC model response in both the low/intermediate chain force and high chain force regimes, a simple, quasi-polynomial bond potential energy function is derived. This bond potential exhibits harmonic behavior near the equilibrium state and anharmonic behavior for large bond stretches tending to a characteristic energy plateau (akin to the Lennard-Jones and Morse bond potentials). Using this bond potential, approximate yet highly-accurate analytical functions for bond stretch and chain force dependent upon chain stretch are established. Then, using this polymer chain model, a stochastic thermal fluctuation-driven chain rupture framework is developed. This framework is based upon a force-modified tilted bond potential that accounts for distortional bond potential energy, allowing for the derivation and subsequent calculation of the dissipated chain scission energy. The cases of rate-dependent and rate-independent scission are accounted for throughout the rupture framework. The impact of Kuhn segment number on chain rupture behavior is also investigated. The model is fit to single-chain mechanical response data collected from atomic force microscopy tensile tests for validation and to glean deeper insight into the molecular physics taking place. Due to their analytical nature, this polymer chain model and the associated rupture framework can be straightforwardly implemented in finite element models accounting for fracture and fatigue in polydisperse elastomer networks. 
\end{abstract}


\end{frontmatter}

\section{Introduction} \label{introduction}

Elastomers are materials composed of flexible entropic polymer chains oriented randomly, cross-linked together, and arranged in a network structure. Due to their resilience and ability to undergo large and recoverable deformations, elastomers have been used as components in traditional engineering applications, such as tires, belts, and sealers \citep{gent2012engineering}. More recently, elastomers have emerged as ideal candidate materials for next-generation soft robotics components and biomedical devices \citep{zhalmuratova2020reinforced}. As elastomers with covalent cross-links become increasingly deformed, polymer chains begin to elongate, and bonds composing the backbone of these chains become stretched. Eventually, these bonds or the bonds of the cross-links rupture. These discrete rupture events collectively build up in the elastomer network and ultimately lead to macroscale failure of the bulk specimen. Additionally, these chain rupture events serve to dissipate energy imparted to the network, and enhance network toughness. This work will focus on rupture events impacting the chain backbone directly (and not the cross-links).

Understanding the fundamentals of energy dissipation, network toughness, and fracture mechanics in elastomers and gels has been an active area of research in recent years \citep{long2021fracture, zhao2021soft, bai2019fatigue, creton201750th, creton2016fracture, long2016fracture, zhao2014multi}. One overarching takeaway from this thrust of research is that the fracture energy of a soft elastomer network, $\Gamma$, namely, the energy required to form new unit crack surface area in the bulk material, can be considered as the sum result of two contributions \citep{long2021fracture, zhao2021soft, long2016fracture, zhao2014multi, tanaka2007local}
\begin{equation}
    \Gamma = \Gamma_0 + \Gamma_{diss},
\end{equation}
where $\Gamma_0$ is the intrinsic fracture energy and $\Gamma_{diss}$ is the mechanical dissipation that takes place in a process zone surrounding the crack tip. Due to the random topology of the chain network, and the naturally occurring polydispersity of chain lengths that arises due to polymerization statistics, which can be thought of as network imperfections, chain rupture can occur in a delocalized manner in the elastomer \citep{itskov2016rubber, yang2019polyacrylamide}. $\Gamma_{diss}$ is generated from any dissipative phenomena that occurs during crack propagation, such as delocalized chain scission, molecular friction due to entanglements and chain pullout, viscoelasticity, poroelasticity, strain-induced crystallization, and polymer-filler interactions. The size of this process zone that is responsible for $\Gamma_{diss}$ can range in length as different dissipation mechanisms become prominent. Various elastomer network characteristics govern how large $\Gamma_{diss}$ is compared to $\Gamma_0$.\footnote{Different dissipation mechanisms contributing to $\Gamma_{diss}$ can be further separated from one another and assumed to operate in different process zones. For instance, \citet{lin2022extreme} proposed a decoupling of highly entangled chain pullout and delocalized chain scission effects taking place in a ``near-crack zone'' from other bulk hysteretic dissipation effects which occur in a much larger process zone. The near-crack dissipation acting in the near-crack zone is $\Gamma_{diss}^{tip}$, while the bulk hysteretic dissipation associated with the general process zone is $\Gamma_{diss}^{bulk}$. $\Gamma_{diss}$ is the sum result of these two dissipation contributions: $\Gamma_{diss} = \Gamma_{diss}^{tip} + \Gamma_{diss}^{bulk}$. \citet{slootman2022molecular} has also proposed a similar mechanism governing bond scission in viscoelastic interpenetrating network elastomers.} In ideal elastomer networks, $\Gamma_{diss} \approxeq 0$ and $\Gamma/\Gamma_0 \approxeq 1$, even with a small density of topological defects incorporated in the network \citep{lin2021fracture}. $\Gamma_{diss} \approxeq 0$ and $\Gamma/\Gamma_0 \approxeq 1$ also holds true for nearly unentangled networks, even with structural heterogeneities and topological defects present \citep{zheng2022fracture}. On the contrary, in entangled elastomer networks, $\Gamma_{diss} > 0$ and $\Gamma/\Gamma_0 > 1$, where the fracture toughness enhancement is postulated to originate from chain pullout and delocalized chain damage in a near-crack process zone \citep{zheng2022fracture, kim2021fracture}. In elastomer networks containing notable amounts of network imperfections from structural heterogeneities, topological defects, and even pre-existing cracks in the bulk material, $\Gamma_{diss} > 0$ and $\Gamma/\Gamma_0 > 1$ is also found to be the case \citep{yang2019polyacrylamide, liu2019polyacrylamide}. Here, fracture toughness enhancement and elastic dissipation are believed to stem from delocalized chain scission \citep{yang2019polyacrylamide, liu2019polyacrylamide}.

To reveal and better understand the complex molecular behavior of polymer chains in fractured, and more generally, damaged elastomers, mechanophores have been incorporated into elastomer networks to probe force-activated events originating at the chain level \citep{chen2021mechanochemical, stratigaki2020methods, simon2017mechanochemistry}. Mechanophore motifs that alter their optical properties with respect to applied load -- mechanochromophores and/or mechanofluorophores (hereafter referred to as luminescent mechanophores) -- have emerged as the preferred tool to visualize such force-sensitive chain level behavior \citep{chen2021mechanochemical, gostl2017optical}. Embedded luminescent mechanophores in elastomer networks have confirmed that chains located far from the crack surface (compared to the chain length) become ruptured during crack advance \citep{ducrot2014toughening, slootman2020quantifying, matsuda2021revisiting, boots2022quantifying, matsuda2020crack}, supporting the postulated mechanism of dissipation and toughness enhancement. Unsurprisingly, viscoelastic effects and microscopic dynamics have been found to influence and interplay with the extent of delocalized chain scission measured via luminescent mechanophores \citep{slootman2020quantifying, slootman2022molecular}. Embedded luminescent mechanophores have also been used to visualize delocalized chain scission in elastomer networks undergoing cavitation \citep{morelle20213d, kim2020extreme} and fatigue \citep{sanoja2021mechanical}.

To develop a model that accounts for the collective impact delocalized chain rupture events play on the bulk material response, and more specifically on the fracture energy, several key building blocks are needed. Incorporating a chain length distribution to reflect the structural heterogeneity of non-uniform chain length in an elastomer network is a vital starting point that has already been proven to impact elastomer mechanics and fracture \citep{falender1979effect, mark2003elastomers, dargazany2009network, wang2015mechanics, itskov2016rubber, diani2019fully, lavoie2019modeling, li2020variational, lu2020pseudo, mulderrig2021affine, guo2021micromechanics, xiao2021modeling}. Incorporation of chain length polydispersity into the network requires defining the chain-level load sharing behavior. The equal strain assumption is commonly employed, where all chains, independent of initial length, are assumed to be deformed to the same stretch \citep{itskov2016rubber, tehrani2017effect, diani2019fully, mulderrig2021affine}. The equal force assumption may also be implemented, where all chains are assumed to bear the same force (via a virtual series arrangement of chains with respect to the loading mechanism) \citep{verron2017equal, li2020variational, mulderrig2021affine}. Connecting the chain-level deformation with the continuum-level deformation has been possible through homogenization assumptions, including the affine three-chain model \citep{wang1952statistical}, non-affine four-chain model \citep{flory1943statistical}, non-affine Arruda-Boyce eight-chain model \citep{arruda1993three}, affine full-network microsphere model \citep{treloar1979non, wu1992improved, wu1993improved}, and non-affine full-network microsphere models \citep{miehe2004micro, tkachuk2012maximal, diani2019fully, ghaderi2020physics, mulderrig2021affine, rastak2018non, arunachala2021energy, guo2021micromechanics}. The interplay between chain-level load sharing with the macro-to-micro deformation relationship has been shown to exert a significant role in local strain-stiffening and delocalized chain rupture \citep{tauber2021sharing, tauber2022stretchy, basu2011nonaffine, black2011molecular, chen2021mechanochemistry, chen2020force, mulderrig2021affine}.

A molecular description of chain rupture leading to macroscopic damage and failure is the final building block to address. Such a description of chain rupture requires acccounting for bond extensibility in single chain elasticity and relating the state of the chain to its state of rupture. Bond extensibility is an inherent necessity in a chain rupture model because the rupture of chains along the fracture plane in elastomer networks is enthalpically dominated (not entropically dominated), as recognized by \citet{lake1967strength}. \citet{smith1996overstretching} proposed a phenomenological modification of the Langevin-statistics based freely-jointed chain (FJC) model of \citet{kuhn1942beziehungen} to account for the influence of bond extensibility. Inspired by this work, \citet{mao2017rupture} took the Langevin-statistics based FJC Helmholtz free energy function, permitted bonds to vary in length, added an internal potential energy of bond stretching to the free energy, and heuristically forced the bond stretch to be the minimizer of this modified Helmholtz free energy. Bond extensibility can also be incorporated in a statistical mechanics-consistent extensible FJC model, as achieved recently by \citet{buche2021chain} and \citet{buche2022freely} utilizing asymptotic matching. Notably, the $u$FJC model of \citet{buche2021chain} and \citet{buche2022freely} accounts for arbitrary bond extensibility, i.e., the influence of extensible bonds are theoretically incorporated in the $u$FJC model prior to any particularization of a governing bond potential energy function. Once the entropic and enthalpic contributions are defined, the statistics of rupture for a given population of chains can be studied. One way to resolve the statistics of chain rupture is to consider fully intact chains as automatically ruptured when their free energy or internal energy exceeds some rupture energy (or when their chain force exceeds its maximum value) \citep{mao2017rupture, arunachala2021energy, lamont2021rate, xiao2021modeling, zhao2021multiscale, dal2009micro, buche2021chain}. Another approach is to define a thermodynamically-consistent damage law (often dependent on the bond stretch) that accounts for network softening. This can also be employed in a phase field fracture setting \citep{mao2018theory, talamini2018progressive, li2020variational, mulderrig2021affine}. Alternatively, consistent with statistical thermodynamics, chain scission can be treated as a stochastic process driven by thermal oscillations \citep{arora2020fracture, arora2021coarse, lu2020pseudo, yang2020multiscale, lei2022multiscale, guo2021micromechanics}. Strikingly, the extensible chain model used within the last two (more descriptive) treatments of chain rupture is, to date, solely the \citet{mao2017rupture} phenomenologically modified FJC model. In other words, an extensible FJC model derived thoroughly upon statistical mechanics principles has yet to be embedded within either a damage law-based treatment or a stochastic treatment of chain rupture. 

In this manuscript, we develop a framework to study polymer chain scission, where its novelty lies upon two key pillars: (i) an arbitrarily-extensible FJC model derived entirely through statistical mechanics that also respects the principles of statistical thermodynamics and asymptotic matching, and (ii) a probabilistic understanding of chain scission, starting from a consideration of thermal oscillations and rupture at the segment level, consistent with the principles of mechanochemistry. To address the first pillar, we extend the $u$FJC model from \citet{buche2021chain} and \citet{buche2022freely} to yield an analytical form for the chain force and the Helmholtz free energy function (also seeking to connect to the prevalent phenomenological functional form proposed in \citet{mao2017rupture}). Using the principles of asymptotic matching (as employed in the derivation of the $u$FJC model), a simple, anharmonic, quasi-polynomial potential energy function for a segment (or a bond) in a chain is derived. From this potential energy function, a highly-accurate approximate analytical form for segment stretch as a function of chain stretch is obtained. To address the second pillar, rate-dependent and rate-independent segment scission is treated as a stochastic, energy-activated process as captured via the force-modified tilted potential energy. Principles of scission physics as described in \citet{wang2019quantitative} are evoked to derive the governing equations for rate-dependent and rate-independent dissipated chain scission energy. A functional form for the reference end-to-end chain distance is then derived that fully accounts for segment extensibility and chain length polydispersity. This proposed polymer chain rupture framework exhibits many beneficial properties; it is clearly based upon statistical mechanics principles accounting for bond extensibility, satisfies principles of statistical thermodynamics and asymptotic matching, accounts for scission originating from the molecular level of the bond, and provides a clear multiscale connection between the physics at the bond-level, segment-level, and chain-level. Ideally, the proposed chain rupture framework can be incorporated into future polymer damage and fracture models to gain insight into the complex nature of polymer network fracture toughness.

This manuscript is organized as follows: In \cref{sec:ufjc-model-review}, the fundamentals of the $u$FJC model from \citet{buche2021chain} and \citet{buche2022freely} are reviewed. The $u$FJC model is then extended in \cref{sec:ufjc-model-extension} to ensure that an upscaled continuum model is defined in terms of tractable closed-form solutions. The analytical form for the chain force is determined in \cref{subsection:chain-force-analytical-form}. The asymptotically-matched segment potential energy function is derived in \cref{subsubsec:segment-potential-function} and \cref{app:modulation-parameter-definition}, leading to the highly-accurate approximate analytical form for segment stretch as a function of chain stretch as derived in \cref{subsubsec:segment-stretch-function} and \cref{app:analytical-form-segment-stretch-function}. With the chain model complete, the chain rupture framework is developed in \cref{sec:single-chain-scission}. Starting at the segment level in \cref{subsec:segment-scission}, the probability of rate-dependent and rate-independent segment scission is adopted from the principles of mechanochemistry, which consider segment scission as a stochastic process driven by thermal fluctuations and dependent upon the applied load to the segment. Governing equations for rate-dependent and rate-independent dissipated segment scission energy are then formulated. The segment scission framework is pushed up to the chain level via probabilistic considerations in \cref{subsec:chain-scission}. In \cref{subsec:chain-scission-informed-equil-prob-dist}, the functional form for the reference end-to-end chain distance dependent upon segment extensibility and segment number is derived through statistical mechanics considerations from \citet{buche2021chain}. Verification and validation take place in \cref{sec:single-chain-model-behavior-results}. \cref{subsec:single-chain-mechanical-response} presents the single chain model mechanical response. Implications of the chain scission framework are discussed in \cref{subsec:chain-scission-framework-implications}, and chain rupture behavior for short, intermediately-long, and long chains is investigated in \cref{subsec:single-chain-scission-results}. In \cref{subsec:experimental-fits}, single chain mechanical response data generated from atomic force microscopy (AFM) tensile tests are used to validate the chain model. The chain rupture framework is called upon to uncover the level of dissipated energy and chain scission probability for the chains involved in the AFM tensile tests. Concluding remarks, improvements for future work, and implications for future research in elastomer fracture and fatigue modeling are highlighted in \cref{sec:conclusion}.

\section{$u$FJC model review} \label{sec:ufjc-model-review}

Before defining a framework for the statistics of chain scission, the constitutive behavior of a single polymer chain must be specified. Since polymer chain rupture is an enthalpically dominated process \citep{lake1967strength}, the chain model must account for bond extensibility. The $u$FJC model has recently emerged as the first polymer chain model to intrinsically account for arbitrary segment extensibility within a statistical mechanics framework. In the following, a brief review of the $u$FJC model framework is provided as a means of establishing the proper context for the remainder of this work. This review (which will occupy the entirety of \cref{sec:ufjc-model-review}) is a summary of the most relevant results from \citet{buche2021chain} and \citet{buche2022freely} (unless otherwise noted).

\subsection{Statistical thermodynamics foundation} \label{subsec:ufjc-stat-thermo}

The $u$FJC is a freely jointed chain of $\nu$ massless, flexible, and stretchable links, or Kuhn segments, connecting $\nu+1$ point masses with mass $m$ and momentum $p$. The segment stretch $\lambda_{\nu}$ is taken as the ratio of the segment length $l_{\nu}$ with the equilibrium segment length $l_{\nu}^{eq}$, $\lambda_{\nu} = l_{\nu}/l_{\nu}^{eq}$. The energy state of each segment is described by the (arbitrary and non-particularized) segment potential $U_{\nu}$, which inherently exhibits some characteristic segment potential energy scale $E_{\nu}^{char}$ and segment stiffness $k_{\nu}$ defined as
\begin{equation}
	k_{\nu} \equiv U_{\nu}^{\prime\prime}(l_{\nu}^{eq}) = \pdv[2]{U_{\nu}(l_{\nu})}{l_{\nu}}\bigg|_{l_{\nu} = l_{\nu}^{eq}},
\end{equation}
where primes imply derivatives. Assuming fixed absolute temperature $T$, a fixed inverse energy scale can be defined as $\beta = 1/[k_B T]$ where $k_B$ is the Boltzmann constant. Then, the nondimensional characteristic segment potential energy scale and nondimensional segment stiffness are respectively defined as $\zeta_{\nu}^{char} \equiv \beta E_{\nu}^{char}$ and $\kappa_{\nu} \equiv \beta [l_{\nu}^{eq}]^2 k_{\nu}$. The single chain Hamiltonian of the $u$FJC model is
\begin{equation}
    H(\mathscr{P}) = \sum_{i=1}^{\nu+1}\frac{p_i^2}{2m} + \sum_{i=1}^{\nu}U_{\nu}(l_{\nu i}) = \sum_{i=1}^{\nu+1}\frac{p_i^2}{2m} + \sum_{i=1}^{\nu}U_{\nu i},
\end{equation}
where $\mathscr{P}$ is the phase space state of the chain. The single-segment isotensional configuration partition function is \citep{buche2020statistical}
\begin{equation} \label{eq:single-segment-isotension-config-partition-function}
    \mathcal{Z}_{\nu}(\xi_c) = \int_{0}^{\pi}\int_{0}^{2\pi}\int_{0}^{\infty} e^{\beta f_c l_{\nu} \cos{\theta}}e^{-\beta U_{\nu}(l_{\nu})} dl_{\nu} d\phi d\theta,
\end{equation}
where $f_c$ is the chain force, $\xi_c = \beta f_c l_{\nu}^{eq}$ is its nondimensional counterpart, and $\theta$ is the angle between the segment and the chain force. The chain isotensional configuration partition function is $\mathcal{Z}(\xi_c) = [\mathcal{Z}_{\nu}(\xi_c)]^{\nu}$ \citep{fiasconaro2019analytical}. The mechanical response of the chain \textit{vis-\`a-vis} the equilibrium chain stretch $\lambda_c^{eq}$ is defined and provided \citep{buche2020statistical} as
\begin{equation}
    \lambda_c^{eq}(\xi_c) = \lambda_c(\xi_c)*\mathcal{A}_{\nu} = \frac{1}{\nu}\pdv{\ln(\mathcal{Z}(\xi_c))}{\xi_c} = \pdv{\ln(\mathcal{Z}_{\nu}(\xi_c))}{\xi_c} = \frac{r_{\nu}(\xi_c)}{\nu l_{\nu}^{eq}}.
\end{equation}
Here, $r_{\nu}$ is the end-to-end chain distance, $\lambda_c = r_{\nu}/r_{\nu}^{ref}$ is the chain stretch, $r_{\nu}^{ref}$ is the reference end-to-end chain distance, and $\mathcal{A}_{\nu} = r_{\nu}^{ref}/[\nu l_{\nu}^{eq}]$ is the reference equilibrium chain stretch.

Obtaining an analytical form for $\lambda_c^{eq}$ valid over all $\xi_c$ regimes and intrinsically accounting for arbitrary segment extensibility (as governed by $U_{\nu}(l_{\nu})$) is seemingly an impossible task. However, an analytical form is needed here for computational simplicity. To resolve this conundrum, asymptotic approximations for $\lambda_c^{eq}$ are derived in the low-to-intermediate chain force regime and in the high chain force regime, both of which are valid only under steep segment potentials $U_{\nu}(l_{\nu})$. The segment potential is considered to be steep if it is deep and narrow, which is true when $\zeta_{\nu}^{char}$ and $\kappa_{\nu}$ are large (i.e., $\zeta_{\nu}^{char},\kappa_{\nu} \gg 1$). The asymptotic approximations for $\lambda_c^{eq}$ are then combined via Prandtl's method of asymptotic matching into a composite function valid for all $\xi_c$. In the limit of sufficiently steep segment potentials, this asymptotically-matched equilibrium chain stretch function is further reduced into a particularly useful analytical form.

\subsection{Low-to-intermediate chain force regime} \label{subsec:low-intmed-chain-force-regime}

For chain forces that are both low (i.e., $\xi_c < 1$) and intermediate (i.e., $1 < \xi_c \ll \zeta_{\nu}^{char},\kappa_{\nu}$) and considering steep segment potentials ($\zeta_{\nu}^{char},\kappa_{\nu} \gg 1$), Laplace's method and various asymptotic considerations are employed to evaluate \cref{eq:single-segment-isotension-config-partition-function} to the following asymptotic relation:
\begin{equation}
    \mathcal{Z}_{\nu}(\xi_c) \approx 4\pi [l_{\nu}^{eq}]^3\sqrt{\frac{2\pi}{\kappa_{\nu}}}\frac{\sinh(\xi_c)}{\xi_c}e^{\frac{\xi_c^2}{2\kappa_{\nu}}}\left[1 + \frac{1}{\tilde{c}_{\nu}}\frac{\xi_c}{\kappa_{\nu}}\coth(\xi_c)\right],
\end{equation}
where $1/\tilde{c}_{\nu} \equiv 1 - u_{\nu}^{\prime\prime\prime}(1)/(2u_{\nu}^{\prime\prime}(1))$,  $u_{\nu}(x) \equiv \beta U_{\nu}(x l_{\nu}^{eq})$ is the nondimensional segment potential, and $x$ is a dummy variable (for now. Later on, $x$ will be identified as the segment stretch $\lambda_{\nu}$). The corresponding asymptotic relation for the equilibrium chain stretch is
\begin{equation}
    \lambda_c^{eq}(\xi_c) \approx \mathcal{L}(\xi_c) + \frac{\xi_c}{\kappa_{\nu}}\left[1+\frac{1-\mathcal{L}(\xi_c)\coth(\xi_c)}{\tilde{c}_{\nu}+\frac{\xi_c}{\kappa_{\nu}}\coth(\xi_c)}\right],
\end{equation}
where $\mathcal{L}(y) = \coth{y} - \frac{1}{y}$ is the Langevin function.

\subsection{High chain force regime} \label{subsec:high-chain-force-regime}

For high chain forces (i.e.,  $\xi_c = \text{ord}(\zeta_{\nu}^{char})$) and with steep segment potentials ($\zeta_{\nu}^{char},\kappa_{\nu} \gg 1$), asymptotic considerations are employed to simplify \cref{eq:single-segment-isotension-config-partition-function} to the following asymptotic relation:
\begin{equation}
    \mathcal{Z}_{\nu}(\xi_c) \approx \frac{2\pi[l_{\nu}^{eq}]^3}{\xi_c}\int_{0}^{\infty} e^{-u_{\nu}^{tot}(x)}x dx,
\end{equation}
where $u_{\nu}^{tot}$ is the nondimensional total segment potential, defined as
\begin{equation} \label{eq:nondim-total-segment-potential-buche}
    u_{\nu}^{tot}(x) \equiv u_{\nu}(x) - \xi_c x.
\end{equation}
The corresponding asymptotic relation for $\lambda_c^{eq}$ is
\begin{equation} \label{eq:high-force-equil-chain-stretch-not-simplified}
    \lambda_c^{eq}(\xi_c) \approx \frac{\int_{0}^{\infty} e^{-u_{\nu}^{tot}(x)}x^2 dx}{\int_{0}^{\infty} e^{-u_{\nu}^{tot}(x)}x dx} - \frac{1}{\xi_c}.
\end{equation}
At this point, the implicit definition of the segment stretch $\lambda_{\nu}$ as the minimizer of $u_{\nu}^{tot}$ is introduced:
\begin{equation} \label{eq:segment-stretch-definition}
    \pdv{}{x}\left(u_{\nu}^{tot}(x)\right)\bigg|_{x = \lambda_{\nu}} = 0 \iff \xi_c = \xi_{\nu} \equiv \pdv{u_{\nu}(x)}{x}\bigg|_{x = \lambda_{\nu}} = \pdv{u_{\nu}}{\lambda_{\nu}}.
\end{equation}
Here, $\xi_{\nu} = \beta f_{\nu} l_{\nu}^{eq}$ is the nondimensional segment force, and $f_{\nu}$ is its dimensional counterpart. Imposing the definition of $\lambda_{\nu}$ and utilizing similar considerations from before, \cref{eq:high-force-equil-chain-stretch-not-simplified} simplifies to
\begin{equation} \label{eq:high-force-equil-chain-stretch-simplified}
    \lambda_c^{eq}(\xi_c) \approx \lambda_{\nu}(\xi_c) - \frac{1}{\xi_c}.
\end{equation}

\subsection{Asymptotic matching for all forces} \label{subsec:asymptotic-matching-all-forces}

Utilizing Prandtl's method of asymptotic matching \citep{powers2015mathematical}, a composite asymptotic relation for the equilibrium chain stretch is found that is applicable for all chain forces with steep segment potential ($\zeta_{\nu}^{char},\kappa_{\nu} \gg 1$):
\begin{equation} \label{eq:asymptotic-matched-equil-chain-stretch-not-simplified}
    \lambda_c^{eq}(\xi_c) \approx \mathcal{L}(\xi_c) + \frac{\xi_c}{\kappa_{\nu}}\left[\frac{1-\mathcal{L}(\xi_c)\coth(\xi_c)}{\tilde{c}_{\nu}+\frac{\xi_c}{\kappa_{\nu}}\coth(\xi_c)}\right] + \lambda_{\nu}(\xi_c) - 1.
\end{equation}
When $\kappa_{\nu}$ is sufficiently large, causing the second term to become negligible, \cref{eq:asymptotic-matched-equil-chain-stretch-not-simplified} can be simplified to a useful form
\begin{equation} \label{eq:asymptotic-matched-equil-chain-stretch-reduced}
    \lambda_c^{eq}(\xi_c) = \mathcal{L}(\xi_c) + \lambda_{\nu}(\xi_c) - 1.
\end{equation}
With $\lambda_c^{eq}$ in hand, the nondimensional Helmholtz free energy per segment, $\psi_{c\nu}$, is desired. Employing the Legendre transform, which is asymptotically valid in the thermodynamic limit of sufficiently long chains and appreciable chain forces \citep{buche2020statistical}, the Helmholtz free energy $\Psi_c$ is given as
\begin{equation} \label{eq:legendre-transform}
    \Psi_c \approx r_{\nu}f_c(r_{\nu}) - \int r_{\nu}(f_c) df_c.
\end{equation}
Substituting \cref{eq:asymptotic-matched-equil-chain-stretch-reduced} in a nondimensional form of \cref{eq:legendre-transform} and performing integration by parts leads to
\begin{equation} \label{eq:nondim-helmholtz-free-energy}
    \psi_{c\nu}(\lambda_c^{eq}) = \xi_c(\lambda_c^{eq})\mathcal{L}(\xi_c(\lambda_c^{eq})) + \ln(\frac{\xi_c(\lambda_c^{eq})}{\sinh(\xi_c(\lambda_c^{eq}))}) + u_{\nu}(\lambda_{\nu}(\xi_c(\lambda_c^{eq}))),
\end{equation}
where $\psi_{c\nu} \equiv \beta\Psi_c/\nu$. 

At this point, \citet{buche2021chain} and \citet{buche2022freely} use numerics to determine $\xi_c$ from  \cref{eq:asymptotic-matched-equil-chain-stretch-reduced}. However, this does not lend itself to a tractable closed-form model when the single chain model is upscaled to a continuum model. Ultimately, this limits the impact of the careful statisical mechanics analysis on the design of materials and resilient structures.

\section{Extension of the $u$FJC model} \label{sec:ufjc-model-extension}

To yield a tractable, closed-form continuum level model based upon the $u$FJC model, we extend the $u$FJC model by seeking an analytical form for $\xi_c(\lambda_c^{eq})$ and $\lambda_{\nu}(\lambda_c^{eq})$. Along the way, the principles of asymptotic matching and the use of highly accurate function approximations will be employed.

\subsection{Analytical form for the chain force} \label{subsection:chain-force-analytical-form}

The analytical form for $\xi_c(\lambda_c^{eq})$ is found by inverting \cref{eq:asymptotic-matched-equil-chain-stretch-reduced}
\begin{align}
    \lambda_c^{eq} & = \mathcal{L}(\xi_c(\lambda_c^{eq})) + \lambda_{\nu}(\lambda_c^{eq}) - 1 = \mathcal{L}(\xi_c(\lambda_c^{eq})) + \lambda_{\nu} - 1, \nonumber \\
    \xi_c(\lambda_c^{eq}) & = \mathcal{L}^{-1}(\lambda_c^{eq} - \lambda_{\nu} + 1), \label{eq:nondim-chain-force}
\end{align}
where $\lambda_{\nu}$ is written with the understanding that $\lambda_{\nu} = \lambda_{\nu}(\lambda_c^{eq})$. Substituting \cref{eq:nondim-chain-force} into \cref{eq:nondim-helmholtz-free-energy} leads to a useful analytical form for $\psi_{c\nu}$
\begin{align} \label{eq:nondim-helmholtz-free-energy-simplified}
    & \psi_{c\nu}(\lambda_{\nu}, \lambda_c^{eq}) = s_{c\nu}(\lambda_{\nu}, \lambda_c^{eq}) + u_{\nu}(\lambda_{\nu}), \\
    & s_{c\nu}(\lambda_{\nu}, \lambda_c^{eq}) = [\lambda_c^{eq} - \lambda_{\nu} + 1]\mathcal{L}^{-1}(\lambda_c^{eq} - \lambda_{\nu} + 1) + \ln(\frac{\mathcal{L}^{-1}(\lambda_c^{eq} - \lambda_{\nu} + 1)}{\sinh(\mathcal{L}^{-1}(\lambda_c^{eq} - \lambda_{\nu} + 1))}), \\
    & \psi_{c\nu}(\lambda_{\nu}, \lambda_c^{eq}) = [\lambda_c^{eq} - \lambda_{\nu} + 1]\mathcal{L}^{-1}(\lambda_c^{eq} - \lambda_{\nu} + 1) + \ln(\frac{\mathcal{L}^{-1}(\lambda_c^{eq} - \lambda_{\nu} + 1)}{\sinh(\mathcal{L}^{-1}(\lambda_c^{eq} - \lambda_{\nu} + 1))}) + u_{\nu}(\lambda_{\nu}). \label{eq:nondim-helmholtz-free-energy-simplified}
\end{align}
where $s_{c\nu}$ is the nondimensional chain-level entropic contributions per segment, and $u_{\nu}$ is the nondimensional segment-level enthalpic contributions. Now, recall that the chain force $f_c$ can also be calculated with respect to the Helmholtz free energy $\Psi_c$ as
\begin{equation} \label{eq:chain-force-function}
    f_c = \pdv{\Psi_c}{r_{\nu}}.
\end{equation}
Pushing \cref{eq:nondim-helmholtz-free-energy-simplified} through a nondimensional form of \cref{eq:chain-force-function} returns \cref{eq:nondim-chain-force}, thereby verifying the analytical form for $\xi_c(\lambda_c^{eq})$.

\begin{figure}[t]
	\centering
	\includegraphics[width=0.75\textwidth]{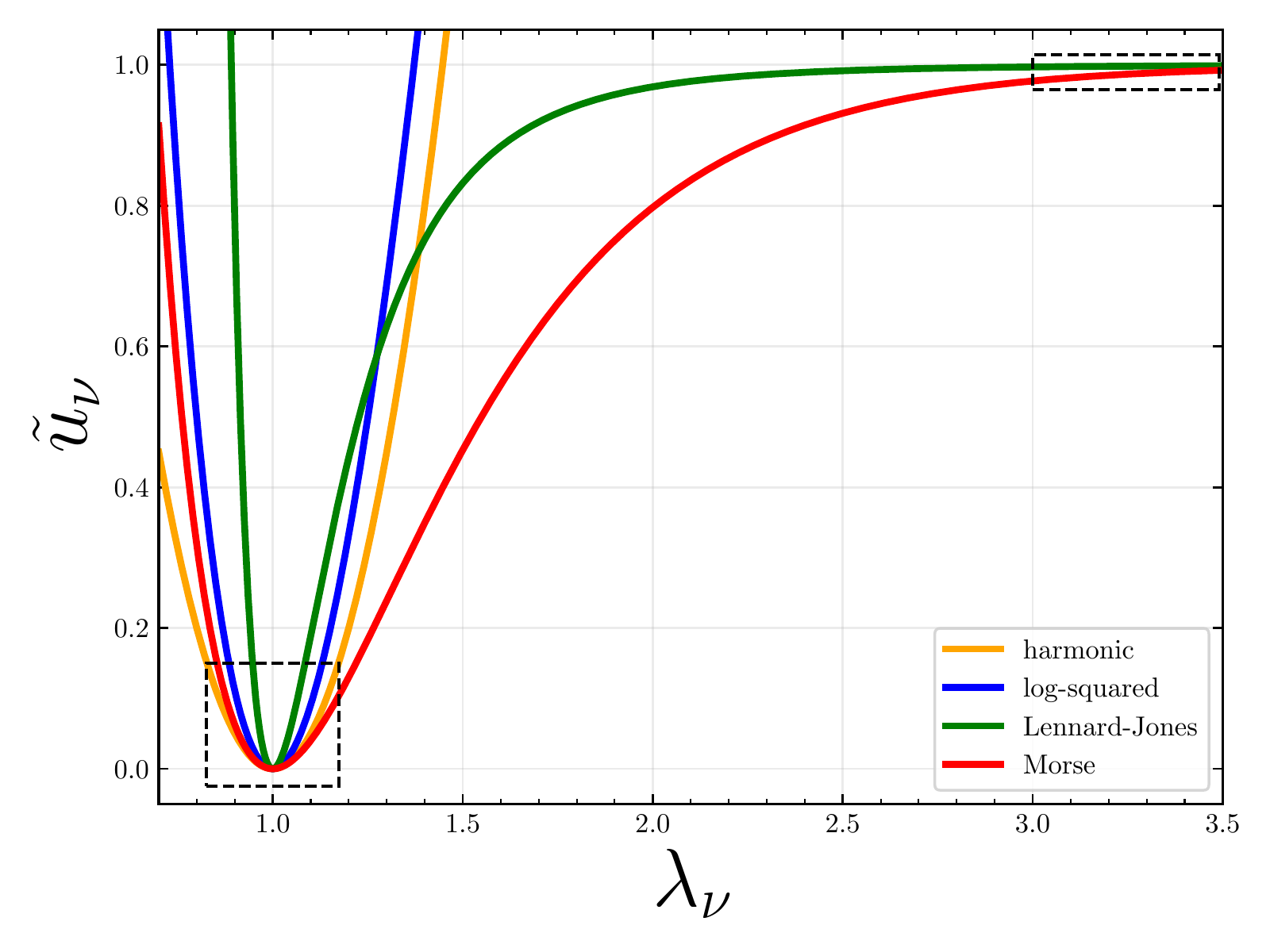}
	\caption{Shifted nondimensional scaled harmonic, log-squared, Lennard-Jones, and Morse segment potentials $\tilde{u}_{\nu}$ as a function of segment stretch $\lambda_{\nu}$. $\zeta_{\nu}^{char} = 100$ for each potential, and $\kappa_{\nu} = 1000$ for each potential (excluding the Lennard-Jones potential, where, by definition, $\kappa_{\nu} = 72\zeta_{\nu}^{char}$).}
	\label{fig:tilde_u_nu-vs-lmbda_nu}
\end{figure}

\subsection{Asymptotically matched segment behavior} \label{subsec:asymptotically-matched-segment}

\subsubsection{Segment potential function} \label{subsubsec:segment-potential-function}

In order for the nondimensional chain force function $\xi_c(\lambda_c^{eq})$ as presented in \cref{eq:nondim-chain-force} to truly be an analytical function of equilibrium chain stretch $\lambda_c^{eq}$, it is required that the segment stretch $\lambda_{\nu}$ also be an analytical function of $\lambda_c^{eq}$. The functional form of $\lambda_{\nu}$, defined as per \cref{eq:segment-stretch-definition}, is dependent upon the complexity of the functional form of the segment potential $U_{\nu}(l_{\nu})$ and its first derivative. Several functional forms of $U_{\nu}(l_{\nu})$ have been proposed and utilized thus far in the literature, including the harmonic potential $U_{\nu}^{har}$ \citep{mao2017large, mao2018theory, talamini2018progressive, li2020variational, mulderrig2021affine, arunachala2021energy, lamont2021rate}, log-squared potential $U_{\nu}^{\ln^2}$ \citep{mao2017rupture, arora2020fracture, arora2021coarse, xiao2021modeling, lu2020pseudo}, Lennard-Jones potential $U_{\nu}^{lj}$ \citep{jones1924determination, yang2020multiscale, zhao2021multiscale, feng2022rigorous, lei2022multiscale}, and Morse potential $U_{\nu}^{morse}$ \citep{morse1929diatomic, dal2009micro, buche2021chain, lavoie2019modeling, guo2021micromechanics}:
\begin{align}
    & U_{\nu}^{har}(l_{\nu}) = E_{\nu}^{char}\left[\frac{1}{2}\frac{k_{\nu}}{E_{\nu}^{char}}\left[l_{\nu} - l_{\nu}^{eq}\right]^2 - 1\right],\qquad U_{\nu}^{\ln^2}(l_{\nu}) = E_{\nu}^{char}\left[\frac{1}{2}\frac{[l_{\nu}^{eq}]^2 k_{\nu}}{E_{\nu}^{char}}\left[\ln(\frac{l_{\nu}}{l_{\nu}^{eq}})\right]^2 - 1\right], \label{eq:harmonic-potential-and-log-squared-potential} \\
    & U_{\nu}^{lj}(l_{\nu}) = E_{\nu}^{char}\left[\left[\frac{l_{\nu}^{eq}}{l_{\nu}}\right]^{12} - 2\left[\frac{l_{\nu}^{eq}}{l_{\nu}}\right]^6\right],\qquad\qquad  U_{\nu}^{morse}(l_{\nu}) = E_{\nu}^{char}\left[\left[1 - e^{-a_{\nu}[l_{\nu} - l_{\nu}^{eq}]}\right]^2 - 1\right],\label{eq:lj-potential-and-morse-potential}
\end{align}
where $a_{\nu}$ is the Morse parameter and is related to $E_{\nu}^{char}$ and $k_{\nu}$ via $k_{\nu} = 2 a_{\nu}^2E_{\nu}^{char}$. The nondimensional Morse parameter is defined as $\alpha_{\nu} \equiv l_{\nu}^{eq} a_{\nu}$. The nondimensional scaled segment potential $\overline{u}_{\nu}$, and its (non-negative) shifted counterpart $\tilde{u}_{\nu}$ are respectively defined as $\overline{u}_{\nu} \equiv u_{\nu}/\zeta_{\nu}^{char}$ and $\tilde{u}_{\nu} \equiv \overline{u}_{\nu} + 1$. The proposed segment potentials from \cref{eq:harmonic-potential-and-log-squared-potential} and \cref{eq:lj-potential-and-morse-potential} are graphically presented in \cref{fig:tilde_u_nu-vs-lmbda_nu} (in $\tilde{u}_{\nu}$ form).

The harmonic potential is the simplest and most commonly used potential to account for segment extensibility, and it approximately captures the general behavior of the other potentials in a neighborhood of small $\lambda_{\nu}$ about the equilibrium state. However, for large $\lambda_{\nu}$, segment potentials are expected to escape the harmonic potential energy well and exhibit anharmonic behavior. By definition, only the log-squared, Lennard-Jones, and Morse potentials capture this behavior. Additionally, for large $\lambda_{\nu}$, segment potentials are expected to ultimately escape to an energy plateau equal to $E_{\nu}^{char}$. Only the Lennard-Jones and Morse potentials capture this behavior. To retain all of the characteristics of the ideal segment potential, we desire to use either the Lennard-Jones or Morse potential in the model framework to yield an analytical form for $\lambda_{\nu}$. 

Unfortunately, due to the squared-exponential term in the Morse potential and two-term polynomial form of the Lennard-Jones potential, it is mathematically impossible for each of these potentials to lead to an approximate analytical expression for $\lambda_{\nu}$ as a function of $\lambda_c^{eq}$ as per \cref{eq:segment-stretch-definition} and \cref{eq:nondim-chain-force} (where the inverse Langevin function involved with $\xi_c$ is represented by some approximant). In order to overcome this obstacle, we seek to derive a simple, quasi-polynomial segment potential which generally captures the essential aforementioned characteristics exhibited by the Lennard-Jones and Morse potentials. This derivation will necessarily evoke the principle of asymptotic matching, consistent to and allowing for integration within the $u$FJC framework. Using this derived segment potential, an expression for $\lambda_{\nu}$ as a function of $\lambda_c^{eq}$ will be reached. 

To begin, consider the behavior of $\lambda_{\nu}$ in the low-to-intermediate chain force state as compared to that in the high chain force state. For low and intermediate chain forces, $\lambda_{\nu}$ resides in a neighborhood about the equilibrium state ($\lambda_{\nu} \approx 1$). The Lennard-Jones and Morse segment potentials in this neighborhood can be considered to be approximated via the harmonic potential (as highlighted by the black dashed box in the lower left of \cref{fig:tilde_u_nu-vs-lmbda_nu})
\begin{equation} \label{eq:tilde-u_nu-low-intermediate-chain-force}
    \tilde{u}_{\nu}(\lambda_{\nu}) \approx \tilde{u}_{\nu}^{har}(\lambda_{\nu}) = \frac{1}{2}\frac{\kappa_{\nu}}{\zeta_{\nu}^{char}}\left[\lambda_{\nu} - 1\right]^2.
\end{equation}
For high chain forces, $\lambda_{\nu}$ is large ($\lambda_{\nu} \gg 1$), and the Lennard-Jones and Morse segment potentials have reached their energy plateau equal to $E_{\nu}^{char}$ (as highlighted by the black dashed box in the upper right of \cref{fig:tilde_u_nu-vs-lmbda_nu})
\begin{equation} \label{eq:tilde-u_nu-high-chain-force}
    \tilde{u}_{\nu}(\lambda_{\nu}) \approx 1.
\end{equation}
When na\"{i}vely applying Prandtl's method of asymptotic matching to the above low-to-intermediate chain force and high chain force segment potentials \citep{powers2015mathematical}, it is self-evident that an asymptotically-matched composite potential is prohibited. To go about creating a simple, quasi-polynomial, and asymptotically-matched composite potential using the fundamental building blocks available to us in \cref{eq:tilde-u_nu-low-intermediate-chain-force} and \cref{eq:tilde-u_nu-high-chain-force}, we define a modulation parameter $\mu_{\nu} = \mu_{\nu}(\lambda_{\nu})$. $\mu_{\nu}$ is strictly a function of $\lambda_{\nu}$ with range $\mu_{\nu}\in[0,1]$. $\mu_{\nu}(\lambda_{\nu})$ is taken to be a monotonically-increasing function of $\lambda_{\nu}$. When $\mu_{\nu}=0$, the composite potential directly returns the harmonic potential ($\tilde{u}_{\nu}(\lambda_{\nu}) = \frac{1}{2}\frac{\kappa_{\nu}}{\zeta_{\nu}^{char}}\left[\lambda_{\nu} - 1\right]^2$), and when $\mu_{\nu}=1$, the composite potential directly returns the energy plateau $E_{\nu}^{char}$ ($\tilde{u}_{\nu}(\lambda_{\nu}) = 1$). Furthermore, $\mu_{\nu}\to 1$ faster than $\lambda_{\nu}\to \infty$, i.e., $\lim_{\lambda_{\nu}\to \infty}([1-\mu_{\nu}][\lambda_{\nu}-1]) = 0$. Finally, the overall composite potential $\tilde{u}_{\nu}(\mu_{\nu}, \lambda_{\nu})$ is a monotonically-increasing function of $\lambda_{\nu}$ for $\lambda_{\nu}\geq 1$, as per \cref{fig:tilde_u_nu-vs-lmbda_nu}. With all this taken into account, the low-to-intermediate chain force segment potential is modulated via $\mu_{\nu}$ as
\begin{equation} \label{eq:low-intermed-chain-force-contribution}
     \tilde{u}_{\nu}(\mu_{\nu}, \lambda_{\nu}) \approx [1-\mu_{\nu}]^f\tilde{u}_{\nu}^{har}(\lambda_{\nu}) = [1-\mu_{\nu}]^f\frac{1}{2}\frac{\kappa_{\nu}}{\zeta_{\nu}^{char}}\left[\lambda_{\nu} - 1\right]^2,
\end{equation}
and the high chain force segment potential is modulated via $\mu_{\nu}$ as
\begin{equation} \label{eq:high-chain-force-contribution}
    \tilde{u}_{\nu}(\mu_{\nu}, \lambda_{\nu}) \approx \mu_{\nu}^g,
\end{equation}
where $f$ and $g$ are integers. Now, we apply Prandtl's method of asymptotic matching to the modulated low-to-intermediate chain force and high chain force segment potentials \citep{powers2015mathematical}, which involves satisfying
\begin{equation}
    \lim_{\lambda_{\nu}\to \infty}[1-\mu_{\nu}]^f\frac{1}{2}\frac{\kappa_{\nu}}{\zeta_{\nu}^{char}}\left[\lambda_{\nu} - 1\right]^2 = \lim_{\lambda_{\nu}\to 0}\mu_{\nu}^g = 0.
\end{equation}
From the above, $f=2$ and $g\neq 0$ must hold true. In order to achieve the simplest composite function, we set $g=1$. The composite potential is consequentially written as
\begin{equation}
    \tilde{u}_{\nu}(\mu_{\nu}, \lambda_{\nu}) = [1-\mu_{\nu}]^2 \frac{1}{2}\frac{\kappa_{\nu}}{\zeta_{\nu}^{char}}[\lambda_{\nu} - 1]^2 + \mu_{\nu}.
\end{equation}
All that remains to be satisfied is the condition that the composite potential is a monotonically-increasing function of $\lambda_{\nu}$ for $\lambda_{\nu}\geq 1$:
\begin{equation}
    \pdv{}{\lambda_{\nu}}[\tilde{u}_{\nu}(\mu_{\nu}, \lambda_{\nu})] = \pdv{\tilde{u}_{\nu}}{\mu_{\nu}}\pdv{\mu_{\nu}}{\lambda_{\nu}} + \pdv{\tilde{u}_{\nu}}{\lambda_{\nu}} \geq 0~\text{for}~\lambda_{\nu}\geq 1.
\end{equation}
Since it is assumed that $\mu_{\nu}(\lambda_{\nu})$ is a monotonically-increasing function of $\lambda_{\nu}$, we seek to strongly satisfy the above by forcing the following equations to individually hold true
\begin{align}
    & \pdv{\tilde{u}_{\nu}}{\lambda_{\nu}} = [1-\mu_{\nu}]^2 \frac{\kappa_{\nu}}{\zeta_{\nu}^{char}}[\lambda_{\nu} - 1] \geq 0~\text{for}~\lambda_{\nu}\geq 1, \label{eq:pdv-u_nu-lambda_nu} \\ 
    & \pdv{\tilde{u}_{\nu}}{\mu_{\nu}} = -[1-\mu_{\nu}]\frac{\kappa_{\nu}}{\zeta_{\nu}^{char}}[\lambda_{\nu} - 1]^2 + 1 \geq 0~\text{for}~\lambda_{\nu}\geq 1. \label{eq:pdv-u_nu-mu_nu}
\end{align}
Using the properties of $\mu_{\nu}(\lambda_{\nu})$, \cref{eq:pdv-u_nu-lambda_nu} automatically holds true. The satisfaction of \cref{eq:pdv-u_nu-mu_nu}, as detailed in \cref{app:modulation-parameter-definition}, yields the functional form for $\mu_{\nu} = \mu_{\nu}(\lambda_{\nu})$:
\begin{equation}
\mu_{\nu} = \begin{cases}
0,& \text{if~}\lambda_{\nu} < \lambda_{\nu}^{crit} \\
1-\frac{\zeta_{\nu}^{char}}{\kappa_{\nu}\left[\lambda_{\nu}-1\right]^2},& \text{if~}\lambda_{\nu} \geq \lambda_{\nu}^{crit} \\
\end{cases}, \label{eq:mu_nu-conditional}
\end{equation}
where $\lambda_{\nu}^{crit} \equiv 1+\sqrt{\frac{\zeta_{\nu}^{char}}{\kappa_{\nu}}}$ is called the critical segment stretch (and $(\lambda_c^{eq})^{crit}$ is the corresponding critical equilibrium chain stretch). 
\begin{figure}[ht]
	\centering
	\includegraphics[width=0.75\textwidth]{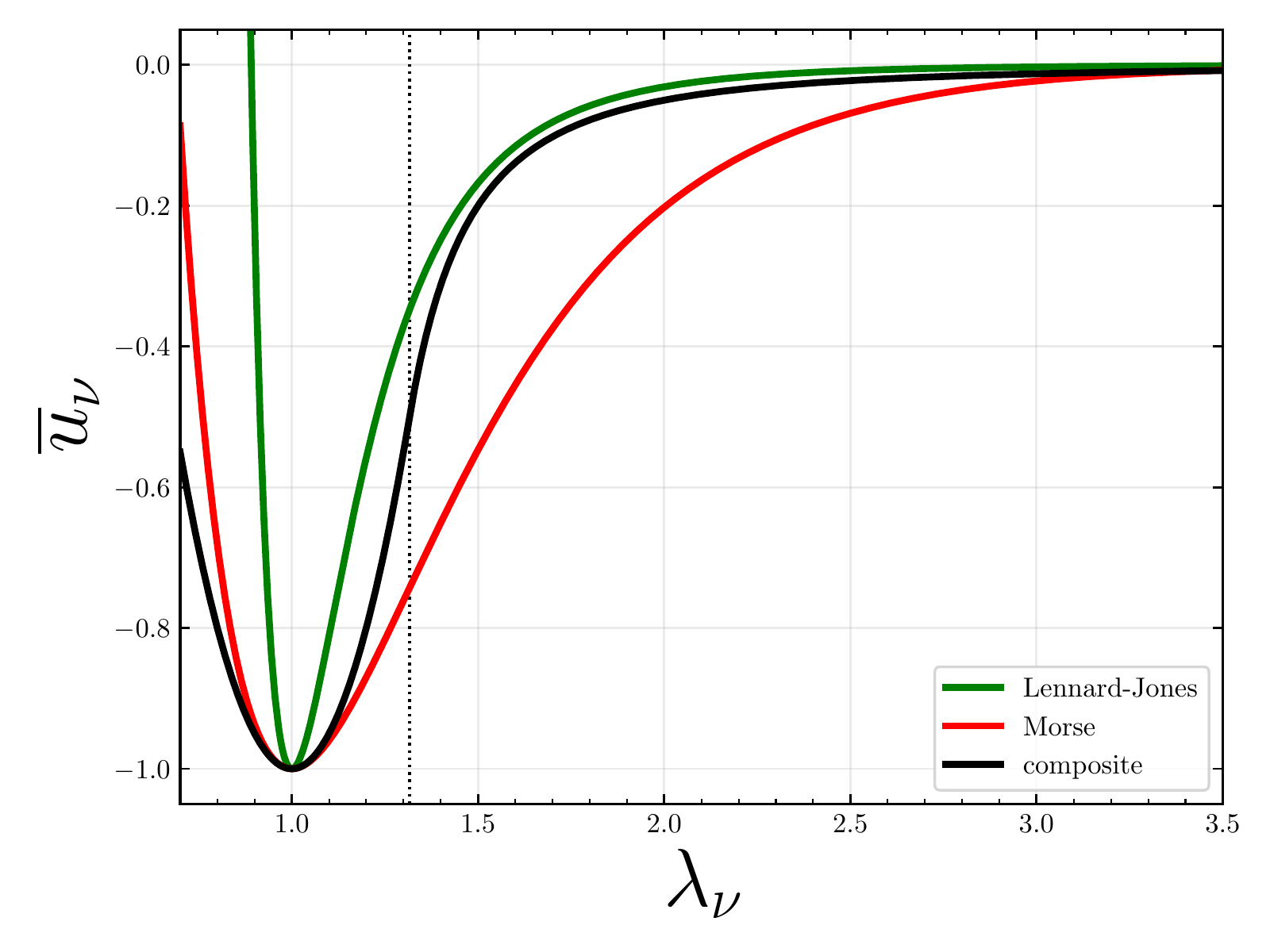}
	\caption{Lennard-Jones, Morse, and composite nondimensional scaled segment potentials $\overline{u}_{\nu}$ versus $\lambda_{\nu}$. The same $\zeta_{\nu}^{char}$ and $\kappa_{\nu}$ values are used here as in \cref{fig:tilde_u_nu-vs-lmbda_nu}. The black dotted line denotes the critical segment stretch value $\lambda_{\nu}^{crit}$ for the composite segment potential.}
	\label{fig:composite-u_nu-figs}
\end{figure}
With this, the composite nondimensional scaled segment potential $\overline{u}_{\nu}$ can be written as a function of $\lambda_{\nu}$
\begin{equation} \label{eq:nondim-scaled-u_nu-conditional}
    \overline{u}_{\nu}(\lambda_{\nu}) = \begin{cases}
	\frac{1}{2}\frac{\kappa_{\nu}}{\zeta_{\nu}^{char}}[\lambda_{\nu} - 1]^2 - 1,& \text{if~}\lambda_{\nu} < \lambda_{\nu}^{crit}\\
	-\frac{\zeta_{\nu}^{char}}{2\kappa_{\nu}\left[\lambda_{\nu}-1\right]^2},& \text{if~}\lambda_{\nu} \geq \lambda_{\nu}^{crit} \\
	\end{cases}.
\end{equation}
The composite nondimensional segment potential $u_{\nu}$ is simply \cref{eq:nondim-scaled-u_nu-conditional} multiplied by $\zeta_{\nu}^{char}$. Using the functional forms for the modulation parameter in \cref{eq:mu_nu-conditional} and the segment potential (e.g., \cref{eq:nondim-scaled-u_nu-conditional}), each presuppositional property of the modulation parameter $\mu_{\nu}$ and the composite segment potential $u_{\nu}$ can be trivially verified. Furthermore, it can also be verified that $u_{\nu}$ is continuous up to the first derivative at $\lambda_{\nu} = \lambda_{\nu}^{crit}$, i.e.,
\begin{equation}
    u_{\nu}(\lambda_{\nu}^{crit}) = -\frac{\zeta_{\nu}^{char}}{2}\text{~and~}\pdv{u_{\nu}}{\lambda_{\nu}}\bigg|_{\lambda_{\nu} = \lambda_{\nu}^{crit}} = \sqrt{\kappa_{\nu}\zeta_{\nu}^{char}}.
\end{equation}
The composite $\overline{u}_{\nu}$ is plotted alongside its Lennard-Jones and Morse potential counterparts in \cref{fig:composite-u_nu-figs}. From this figure, it is clear that the composite potential exhibits the desired characteristics of the Lennard-Jones and Morse potentials: harmonic behavior near the equilibrium state and anharmonic behavior for large $\lambda_{\nu}$ tending to an energy plateau of $E_{\nu}^{char}$. These desired characteristics are expressed purely as a consequence of the derivation undertaken here, without any consideration of the specific functional form of the Lennard-Jones and Morse potentials. Conveniently, the derivation results in a composite potential with a simple functional form, as per \cref{eq:nondim-scaled-u_nu-conditional}.

\begin{figure}[t]
	\centering
	\includegraphics[width=0.75\textwidth]{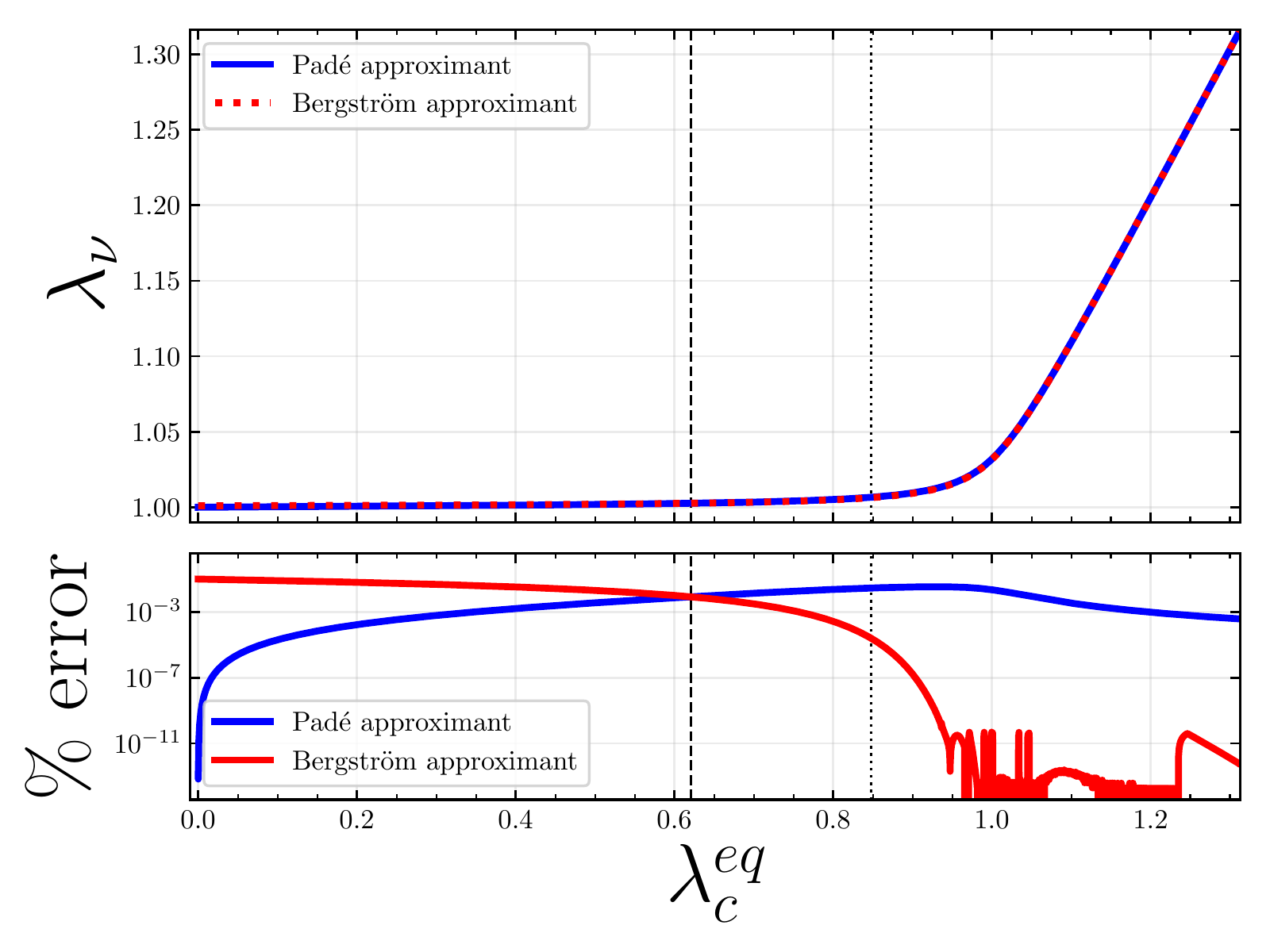}
	\caption{Segment stretch evolution prior to the critical chain state. Here, $\kappa_{\nu} = 1000$, as per the composite segment potential displayed in \cref{fig:composite-u_nu-figs}. The black dashed line denotes the crossover equilibrium chain stretch $(\lambda_c^{eq})^{P2B}$. Additionally, the black dotted line denotes the equilibrium chain stretch $\lambda_c^{eq}$ value at which $\lambda_c^{eq} - \lambda_{\nu} + 1 = 0.84136$. For $\lambda_c^{eq}$ values to the left of the black dotted line, $\lambda_c^{eq} - \lambda_{\nu} + 1 < 0.84136$. (top) Segment stretch $\lambda_{\nu}$ as a function of $\lambda_c^{eq}$ in the domain $0\leq \lambda_c^{eq} < (\lambda_c^{eq})^{crit}$ using the Pad\'e approximant from \cref{eq:lmbda_nu-func-pade-subcrit} and the Bergstr\"{o}m approximant from \cref{eq:lmbda_nu-func-bergstrom-subcrit}. (bottom) Percent error of $\lambda_{\nu}$ in \cref{eq:lmbda_nu-func-pade-subcrit} and \cref{eq:lmbda_nu-func-bergstrom-subcrit} relative to $\lambda_{\nu}$ calculated using a highly accurate numerical solution for the inverse Langevin function. (Beyond $\lambda_c^{eq} \approx 1$ the error in the Bergstr\"{o}m approximant is sufficiently small such that its evaluation is dominated by floating-point error.)}
	\label{fig:lmbda_nu-vs-lmbda_c_eq-approximation-comparison}
\end{figure}

\begin{figure}[t]
	\centering
	\includegraphics[width=0.75\textwidth]{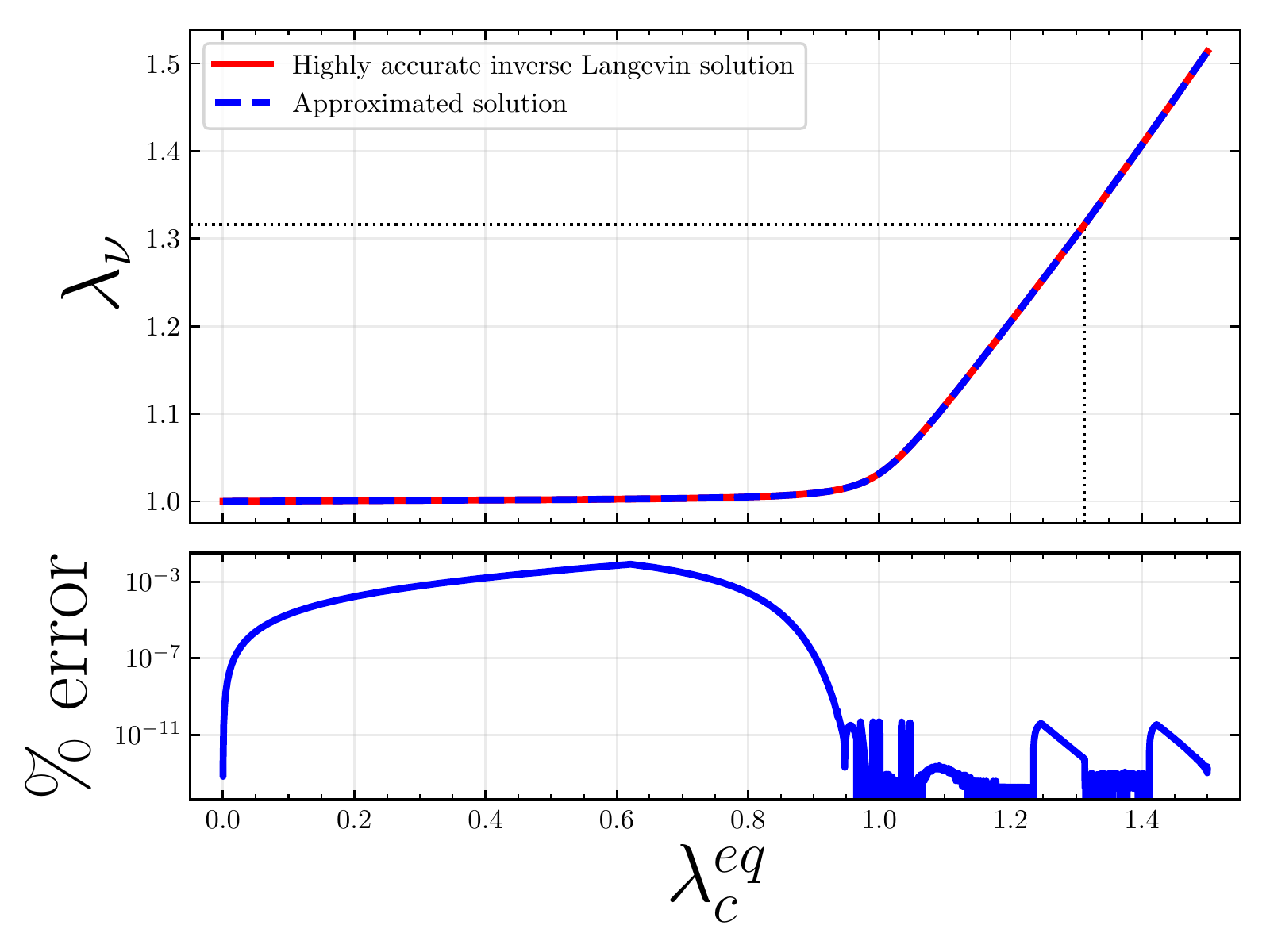}
	\caption{Segment stretch evolution. Here, $\zeta_{\nu}^{char} = 100$ and $\kappa_{\nu} = 1000$, as per the composite segment potential displayed in \cref{fig:composite-u_nu-figs}. The black dotted lines denote the critical equilibrium chain stretch $(\lambda_c^{eq})^{crit}$ and the critical segment stretch $\lambda_{\nu}^{crit}$. (top) Segment stretch $\lambda_{\nu}$ as a function of equilibrium chain stretch $\lambda_c^{eq}$ as per the approximated analytical $\lambda_{\nu}$ function from \cref{eq:segment-stretch-function} along with $\lambda_{\nu}$ calculated using a highly accurate numerical solution for the inverse Langevin function. (bottom) Percent error of the approximated analytical $\lambda_{\nu}$ function in \cref{eq:segment-stretch-function} relative to $\lambda_{\nu}$ calculated using a highly accurate numerical solution for the inverse Langevin function. (Beyond $\lambda_c^{eq} \approx 1$ the error in the approximated solution is sufficiently small such that its evaluation is dominated by floating-point error.)}
	\label{fig:lmbda_nu-vs-lmbda_c_eq-exact-and-approximated-solutions-percent-error}
\end{figure}

\subsubsection{Segment stretch function} \label{subsubsec:segment-stretch-function}

With the simple, quasi-polynomial composite segment potential in hand, a highly accurate approximated analytical relationship between segment stretch and equilibrium chain stretch can now be derived. Substituting the composite $u_{\nu}$ (\cref{eq:nondim-scaled-u_nu-conditional} multiplied by $\zeta_{\nu}^{char}$) into the definition of $\lambda_{\nu}$ in \cref{eq:segment-stretch-definition} and simplifying leads to
\begin{equation} \label{eq:segment-stretch-composite-u_nu}
\xi_c = \mathcal{L}^{-1}(\lambda_c^{eq} - \lambda_{\nu} + 1) = \pdv{u_{\nu}}{\lambda_{\nu}} = \begin{cases}
\kappa_{\nu}[\lambda_{\nu} - 1],& \text{if~}\lambda_c^{eq} < (\lambda_c^{eq})^{crit} \\
\frac{[\zeta_{\nu}^{char}]^2}{\kappa_{\nu}\left[\lambda_{\nu}-1\right]^3},& \text{if~}\lambda_c^{eq} \geq (\lambda_c^{eq})^{crit} \\
\end{cases}.
\end{equation}
Note that $\lambda_c^{eq} < (\lambda_c^{eq})^{crit}$ corresponds one-to-one to the case when $\lambda_{\nu} < \lambda_{\nu}^{crit}$, and $\lambda_c^{eq} \geq (\lambda_c^{eq})^{crit}$ corresponds one-to-one to the case when $\lambda_{\nu} \geq \lambda_{\nu}^{crit}$. We now seek to employ an approximation for the inverse Langevin function in order to yield an approximate analytical solution. Two candidates stand out for the task: the Pad\'e approximant \citep{cohen1991pade}
\begin{equation} \label{eq:pade-approx}
\mathcal{L}^{-1}(y) \approxeq y\frac{3-y^2}{1-y^2},
\end{equation}
and the Bergstr\"{o}m approximant \citep{bergstrom2000large}
\begin{equation} \label{eq:bergstrom-approx}
\mathcal{L}^{-1}(y) \approxeq \frac{1}{\text{sgn}(y)-y},~\text{for~}0.84136<\abs{y}<1,
\end{equation}
where $\text{sgn}(y)$ is the sign function:
\begin{equation} 
\text{sgn}(y) = \begin{cases}
-1,& \text{if~}y<0 \\
0,& \text{if~}y=0 \\
1,& \text{if~}y>0 \\
\end{cases}.
\end{equation}
Using the Pad\'e approximant for the $\lambda_c^{eq} < (\lambda_c^{eq})^{crit}$ case and performing an appropriate cubic root analysis \citep{zwillinger2002crc} leads to
\begin{equation} \label{eq:lmbda_nu-func-pade-subcrit}
    \lambda_{\nu} = \lambda_{\nu}^{PSB}(\kappa_{\nu}; \lambda_c^{eq}),
\end{equation}
where the analytical form of $\lambda_{\nu}^{PSB}$ is provided in \cref{app:analytical-form-segment-stretch-function}. Likewise, using the Bergstr\"{o}m approximant for the $\lambda_c^{eq} < (\lambda_c^{eq})^{crit}$ case and performing an appropriate quadratic root analysis leads to
\begin{equation} \label{eq:lmbda_nu-func-bergstrom-subcrit}
    \lambda_{\nu} = \lambda_{\nu}^{BSB}(\kappa_{\nu}; \lambda_c^{eq}),
\end{equation}
where the analytical form of $\lambda_{\nu}^{BSB}$ is provided in \cref{app:analytical-form-segment-stretch-function}.

The top panel in \cref{fig:lmbda_nu-vs-lmbda_c_eq-approximation-comparison} displays $\lambda_{\nu}^{PSB}$ and $\lambda_{\nu}^{BSB}$ in the domain $0\leq \lambda_c^{eq} < (\lambda_c^{eq})^{crit}$, and the bottom panel displays the error of each approximation compared to $\lambda_{\nu}$ calculated using a highly accurate numerical solution for the inverse Langevin function. The bottom panel in \cref{fig:lmbda_nu-vs-lmbda_c_eq-approximation-comparison} clearly shows that $\lambda_{\nu}^{PSB}$ is more accurate than $\lambda_{\nu}^{BSB}$ to the left of the black dashed line. On the contrary, $\lambda_{\nu}^{BSB}$ is more accurate than $\lambda_{\nu}^{PSB}$ to the right of the black dashed line. This black dashed line denotes $(\lambda_c^{eq})^{P2B}$, the equilibrium chain stretch value at which this crossover in numerical accuracy takes place (see Remarks 1 and 2 for more details).

Using the Bergstr\"{o}m approximant for the $\lambda_c^{eq} \geq (\lambda_c^{eq})^{crit}$ case and performing an appropriate cubic root analysis \citep{zwillinger2002crc} leads to
\begin{equation} \label{eq:lmbda_nu-func-bergstrom-supercrit}
    \lambda_{\nu} = \lambda_{\nu}^{BSP}(\zeta_{\nu}^{char}, \kappa_{\nu}; \lambda_c^{eq}),
\end{equation}
where the analytical form of $\lambda_{\nu}^{BSP}$ is provided in \cref{app:analytical-form-segment-stretch-function}. Unfortunately, using the Pad\'e approximant for the $\lambda_c^{eq} \geq (\lambda_c^{eq})^{crit}$ case results in a sixth-order polynomial in $\lambda_{\nu}$, which does not possess a general form for an analytical solution. 

Considering all of this, the approximated analytical form of $\lambda_{\nu}$ as a function of $\lambda_c^{eq}$ is 
\begin{align}
& \lambda_{\nu} = \begin{cases}
\lambda_{\nu}^{PSB}(\kappa_{\nu}; \lambda_c^{eq}),& \text{if~}\lambda_c^{eq} < (\lambda_c^{eq})^{P2B} \\
\lambda_{\nu}^{BSB}(\kappa_{\nu}; \lambda_c^{eq}),& \text{if~}(\lambda_c^{eq})^{P2B} \leq \lambda_c^{eq} < (\lambda_c^{eq})^{crit} \\
\lambda_{\nu}^{BSP}(\zeta_{\nu}^{char}, \kappa_{\nu}; \lambda_c^{eq}),& \text{if~}\lambda_c^{eq} \geq (\lambda_c^{eq})^{crit}
\end{cases}. \label{eq:segment-stretch-function}
\end{align}

\noindent \cref{fig:lmbda_nu-vs-lmbda_c_eq-exact-and-approximated-solutions-percent-error} displays the approximated $\lambda_{\nu}$ function as per \cref{eq:segment-stretch-function}, $\lambda_{\nu}$ calculated using a highly accurate numerical solution for the inverse Langevin function, and the percent error of the approximation. This figure convincingly verifies that the approximated analytical $\lambda_{\nu}$ function is highly accurate with respect to the highly accurate numerical solution in the domain of physically-sensible $\lambda_c^{eq}$. 

\begin{figure*}[t]
	\centering
	\subfloat[]{
		\includegraphics[width=0.495\textwidth]{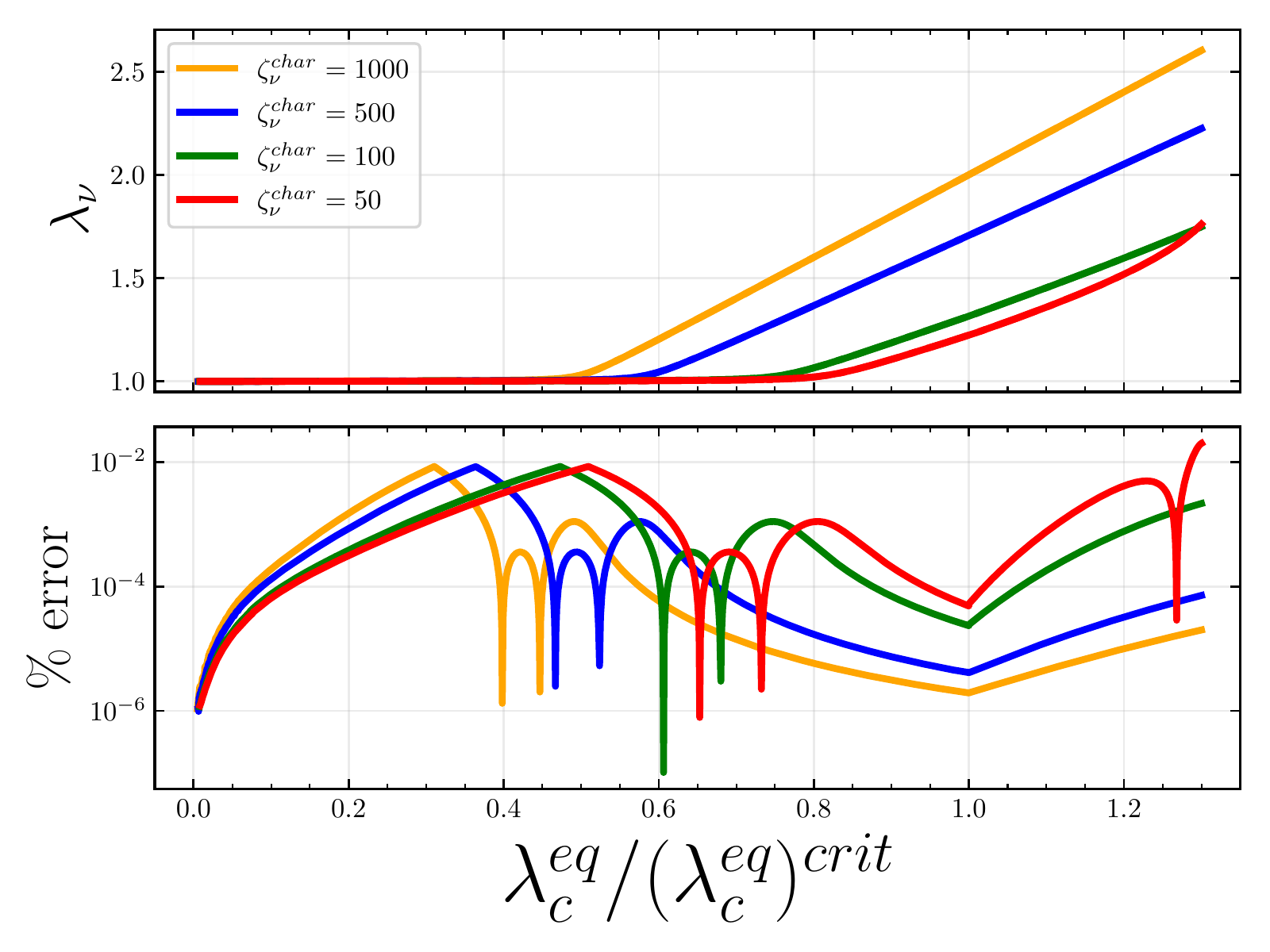}
		\label{fig:zeta_nu_char_lmbda_nu-vs-lmbda_c_eq__lmbda_c_eq_crit-method-comparison}}
	\subfloat[]{
		\includegraphics[width=0.495\textwidth]{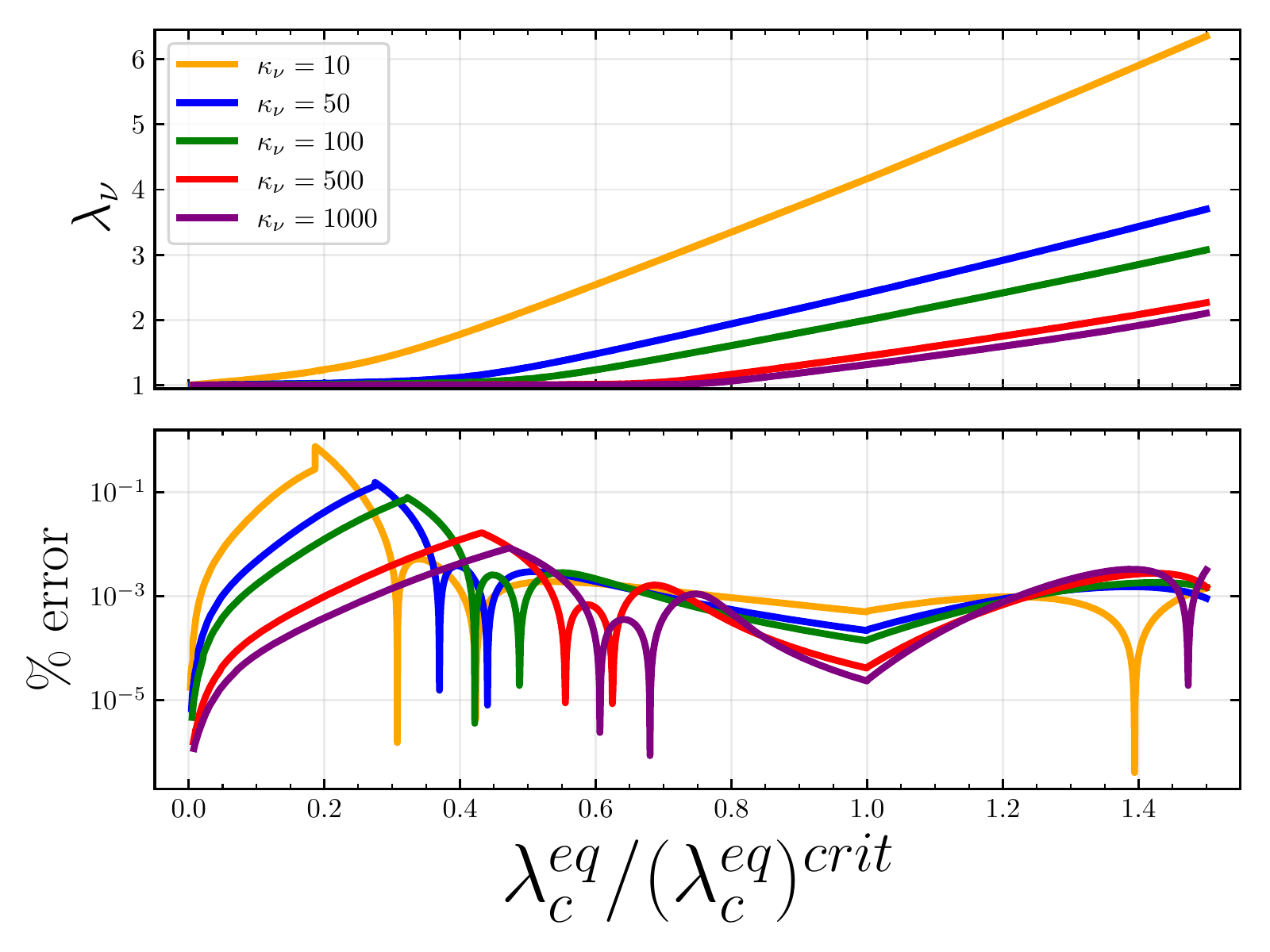}
		\label{fig:kappa_nu_lmbda_nu-vs-lmbda_c_eq__lmbda_c_eq_crit-method-comparison}}
	\caption{Comparison of segment stretches calculated via the $u$FJC chain force satisfaction method and the chain free energy minimization method. (a, top) Segment stretch $\lambda_{\nu}$ calculated via the $u$FJC chain force satisfaction method as a function of equilibrium chain stretch normalized by the critical equilibrium chain stretch, $\lambda_c^{eq}/(\lambda_c^{eq})^{crit}$, for $\kappa_{\nu} = 1000$ and varying $\zeta_{\nu}^{char}$. (b, top) $\lambda_{\nu}$ calculated via the $u$FJC chain force satisfaction method as a function of $\lambda_c^{eq}/(\lambda_c^{eq})^{crit}$, for $\zeta_{\nu}^{char} = 100$ and varying $\kappa_{\nu}$. (bottom) Percent error of $\lambda_{\nu}$ calculated via the chain free energy minimization method (with respect to $\lambda_{\nu}$ calculated via the $u$FJC chain force satisfaction method) as a function of $\lambda_c^{eq}/(\lambda_c^{eq})^{crit}$. The sharp dips in these percent error curves take place where the $\lambda_{\nu}$ values calculated by the chain free energy minimization method cross over or under the $\lambda_{\nu}$ values calculated by the $u$FJC chain force satisfaction method.}
	\label{fig:segment-stretch-method-comparison}
\end{figure*}

In accordance with the $u$FJC model framework as introduced in \citet{buche2021chain} and \citet{buche2022freely} (the reader is referred to the arguments detailed therein), $\lambda_{\nu}$ is implicitly defined to satisfy the equality between the chain force and the segment force, as per \cref{eq:segment-stretch-definition}. This $u$FJC chain force satisfaction method differs from the condition that $\lambda_{\nu}$ be the minimizer of the chain Helmholtz free energy $\Psi_c$, as proposed by \citet{mao2017rupture} and widely adopted in the literature since:
\begin{equation} \label{eq:chain-free-energy-minimization-method}
    \lambda_{\nu} = \argmin_{\lambda_{\nu} \geq 1} \hat{\Psi}_c(\lambda_{\nu}),\text{~where~}\hat{\Psi}_c(\lambda_{\nu}) = \Psi_c(\lambda_{\nu}, \lambda_c=\hat{\lambda}_c),
\end{equation}
where $\lambda_c=\hat{\lambda}_c$ is a given imposed chain stretch.

In an effort to compare the numerical results of the chain free energy minimization method to the $u$FJC chain force satisfaction method, $\lambda_{\nu}$ is calculated via both methods for increasingly steep segments, with the latter method presented in the top panels in \cref{fig:segment-stretch-method-comparison}. The percent error of the $\lambda_{\nu}$ calculation from the chain free energy minimization method (with respect to the $u$FJC chain force satisfaction method) is provided in the bottom panels. In order to ensure an equal comparison between both $\lambda_{\nu}$ calculation methods, $\Psi_c$ involved in the chain free energy minimization method obeys \cref{eq:nondim-helmholtz-free-energy-simplified} multiplied by $\nu/\beta$. In addition, $\lambda_{\nu}$ is presented as a function of $\lambda_c^{eq}$, which normalizes the impact of chain segment number on the $\lambda_{\nu}$ calculations. $\lambda_c^{eq}$ is then further divided by $(\lambda_c^{eq})^{crit}$ as a means of normalizing the impact of differing $(\lambda_c^{eq})^{crit}$ values for segments with varying $\zeta_{\nu}^{char}$ and $\kappa_{\nu}$. As indicated by the bottom panels in \cref{fig:segment-stretch-method-comparison}, the chain free energy minimization method numerically complies quite well with the $u$FJC chain force satisfaction method for $\lambda_c^{eq}$ less than, equal to, and slightly greater than $(\lambda_c^{eq})^{crit}$, i.e., for low and intermediate chain forces $\xi_c$. For extremely large $\lambda_c^{eq}$, i.e., for extremely high chain forces $\xi_c$, the chain free energy minimization method will numerically diverge from the $u$FJC chain force satisfaction method. This is implied by the trend in the percent error curves for increasing $\lambda_c^{eq}/(\lambda_c^{eq})^{crit} \gg 1$.

\rmk The black dashed line in \cref{fig:lmbda_nu-vs-lmbda_c_eq-approximation-comparison} denotes $(\lambda_c^{eq})^{P2B}$, the equilibrium chain stretch value at which the Pad\'e approximant $\lambda_{\nu}$ solution becomes less numerically accurate than the Bergstr\"{o}m approximant $\lambda_{\nu}$ solution (hence $P2B$). Given \cref{eq:lmbda_nu-func-pade-subcrit} and \cref{eq:lmbda_nu-func-bergstrom-subcrit}, $(\lambda_c^{eq})^{P2B}$ solely depends on $\kappa_{\nu}$. A curve fit analysis was used to determine the relationship between $(\lambda_c^{eq})^{P2B}$ and $\kappa_{\nu}$:
\begin{equation}
    (\lambda_c^{eq})^{P2B} = \frac{1}{\kappa_{\nu}^n} + b,\text{~where~}n = 0.8187\text{~and~}b = 0.6176.
\end{equation}
 
\rmk According to \cref{eq:bergstrom-approx}, the Bergstr\"{o}m approximant, as originally proposed, is only valid for $\lambda_c^{eq} - \lambda_{\nu} + 1 \geq 0.84136$ \citep{bergstrom2000large}. $\lambda_c^{eq}$ values that exist at and to the right of the black dotted line in \cref{fig:lmbda_nu-vs-lmbda_c_eq-approximation-comparison} correspond to the situation where $\lambda_c^{eq} - \lambda_{\nu} + 1 \geq 0.84136$. However, for $\lambda_c^{eq}$ values that exist to the left of the black dotted line (where $\lambda_c^{eq} - \lambda_{\nu} + 1 < 0.84136$) and to the right of the black dashed line (where $\lambda_c^{eq} > (\lambda_c^{eq})^{P2B}$), $\lambda_{\nu}^{BSB}$ is highly accurate in this $\lambda_c^{eq}$ regime (even more accurate than $\lambda_{\nu}^{PSB}$ is). This can be visually confirmed in the bottom panel of \cref{fig:lmbda_nu-vs-lmbda_c_eq-approximation-comparison}. As a result, $\lambda_{\nu}^{BSB}$ is validated to be used in this $\lambda_c^{eq}$ regime (where $\lambda_c^{eq} > (\lambda_c^{eq})^{P2B}$ and $\lambda_c^{eq} - \lambda_{\nu} + 1 < 0.84136$ are true).

\rmk Use of the approximated (and highly accurate) analytical form of $\lambda_{\nu}$ as a function of $\lambda_c^{eq}$, as obtained in \cref{eq:segment-stretch-function}, allows one to avoid using numerics to computationally calculate the inverse relationship between $\lambda_c^{eq}$ and $\lambda_{\nu}$. Furthermore, substituting \cref{eq:segment-stretch-function} into \cref{eq:nondim-chain-force} returns an approximated analytical form of $\xi_c$ as a function of $\lambda_c^{eq}$ (where the inverse Langevin function present in \cref{eq:nondim-chain-force} is represented by some approximant). This also allows one to avoid using the tools of numerics to computationally calculate the inverse relationship between $\lambda_c^{eq}$ and $\xi_c$. Considering all of this, a tractable closed-form model emerges when the single chain model is upscaled to a continuum-level model. In this way, costly numerical inverse operations and/or minimization operations are replaced with efficient closed-form solutions in the computational implementation of this continuum model.

\section{Single chain scission energetics and probabilistic considerations} \label{sec:single-chain-scission}

With the single chain model established, a probabilistic description of single chain rupture can now be built upon the principles of mechanochemistry \citep{bell1978models, zhurkov1965kinetic, beyer2005mechanochemistry, ribas2012covalent}. To ultimately provide insight to fracture toughness at the network level, the energy released upon thermally-activated chain scission must be tracked in this framework \citep{wang2019quantitative}. To close the loop in the statistical mechanics-based single chain framework accounting for rupture, the reference end-to-end chain distance is derived to be functionally dependent upon segment extensibility and segment number.

\subsection{Segment scission energetics and probabilistic considerations} \label{subsec:segment-scission}

According to the principles of mechanochemistry, the segments along the backbone of a polymer chain are physically permitted to undergo thermally activated rupture with a probability dependent upon the externally applied force to the chain \citep{bell1978models, zhurkov1965kinetic, beyer2005mechanochemistry, ribas2012covalent}. Using these principles, a variety of rate-dependent bulk and interfacial polymer damage models have been developed \citep{chaudhury1999rate, freund2014brittle, ghatak2000interfacial, hui2004failure, kothari2018mechanical, lavoie2015rate, lavoie2016rate, guo2021micromechanics, yang2019rate, yang2020multiscale, tehrani2017effect, lu2020pseudo, feng2022rigorous, lei2022multiscale}. Notably, \citet{yang2020multiscale} proposed a rate-dependent interfacial polymer damage model where the probability of bond dissociation is calculated using the force-dependent segment-level Gibbs free energy. In this current work, a segment-level rupture framework will be developed utilizing a force-distorted energy landscape. However, we impose the requirement that the segment explore this energy landscape only via its thermally-activated stretch state under a constantly-held externally applied chain force. Since the segment-level Gibbs free energy landscape can only be traversed via varying the applied chain force, a segment-level potential energy landscape dependent on a fixed applied chain force is instead used. This particular energy landscape choice is further substantiated by the fact that enthalpic segment distortion dominates over entropic contributions as the segments in a chain approach the state of rupture \citep{lake1967strength, wang2019quantitative}. Probabilistic considerations will then be used as per \citet{guo2021micromechanics} to develop a chain-level rupture framework.

In the absence of thermal vibrations, a segment in equilibrium must be supplied an activation energy equal to $E_{\nu}^{char}$ in order to undergo rupture. However, this activation energy can be reduced upon the application of a force to the chain. The mechanochemical basis for this activation energy reduction can be found in the ``tilting'' or ``distortion'' of the segment potential energy landscape due to the externally applied chain force. This ``tilted'' segment potential is captured by the previously discussed nondimensional total segment potential in \cref{eq:nondim-total-segment-potential-buche}, following a Legendre transformation of the segment potential energy
\begin{equation}
    \hat{u}_{\nu}^{tot}\left(\hat{\xi}_c, \lambda_{\nu}\right) = u_{\nu}(\lambda_{\nu}) - \hat{\xi}_c\lambda_{\nu},
\end{equation}
where $\hat{\xi}_c$ is the particular force applied to the chain. Unlike earlier in this work, a hat overtop is used hereafter to signify quantities which are in the force controlled setting. The implicit definition of the segment stretch remains unchanged from before in \cref{eq:segment-stretch-definition}, which implies that the application of chain force $\hat{\xi}_c$ corresponds one-to-one to $\hat{\lambda}_{\nu}$. $\hat{\lambda}_{\nu}$ will be referred to as the applied (or particularized) segment stretch. The applied (or particularlized) equilibrium chain stretch $\hat{\lambda}_c^{eq}$ can be calculated from $\hat{\lambda}_{\nu}$ via inverting the $\lambda_{\nu}$ function in \cref{eq:segment-stretch-function}. The approximated analytical relationship between $\lambda_c^{eq}$ as a function of $\lambda_{\nu}$ is provided in \cref{app:equil-chain-segment-stretch-function}. For the sake of notational clarity (since there is a one-to-one relationship between $\hat{\xi}_c$ and $\hat{\lambda}_{\nu}$), $\hat{u}_{\nu}^{tot}$ is equivalently written as $\hat{u}_{\nu}^{tot}\left(\hat{\lambda}_{\nu}, \lambda_{\nu}\right) = u_{\nu}(\lambda_{\nu}) - \hat{\xi}_c\left(\hat{\lambda}_{\nu}\right)*\lambda_{\nu}$ (where the first argument of $\hat{u}_{\nu}^{tot}$ indicates the level of force-driven tilting imposed to $u_{\nu}$ -- which transforms $u_{\nu}$ to $\hat{u}_{\nu}^{tot}$ -- and the second argument specifies a particular state in the energy landscape of $\hat{u}_{\nu}^{tot}$).

\begin{figure}[t]
	\centering
	\includegraphics[width=0.75\textwidth]{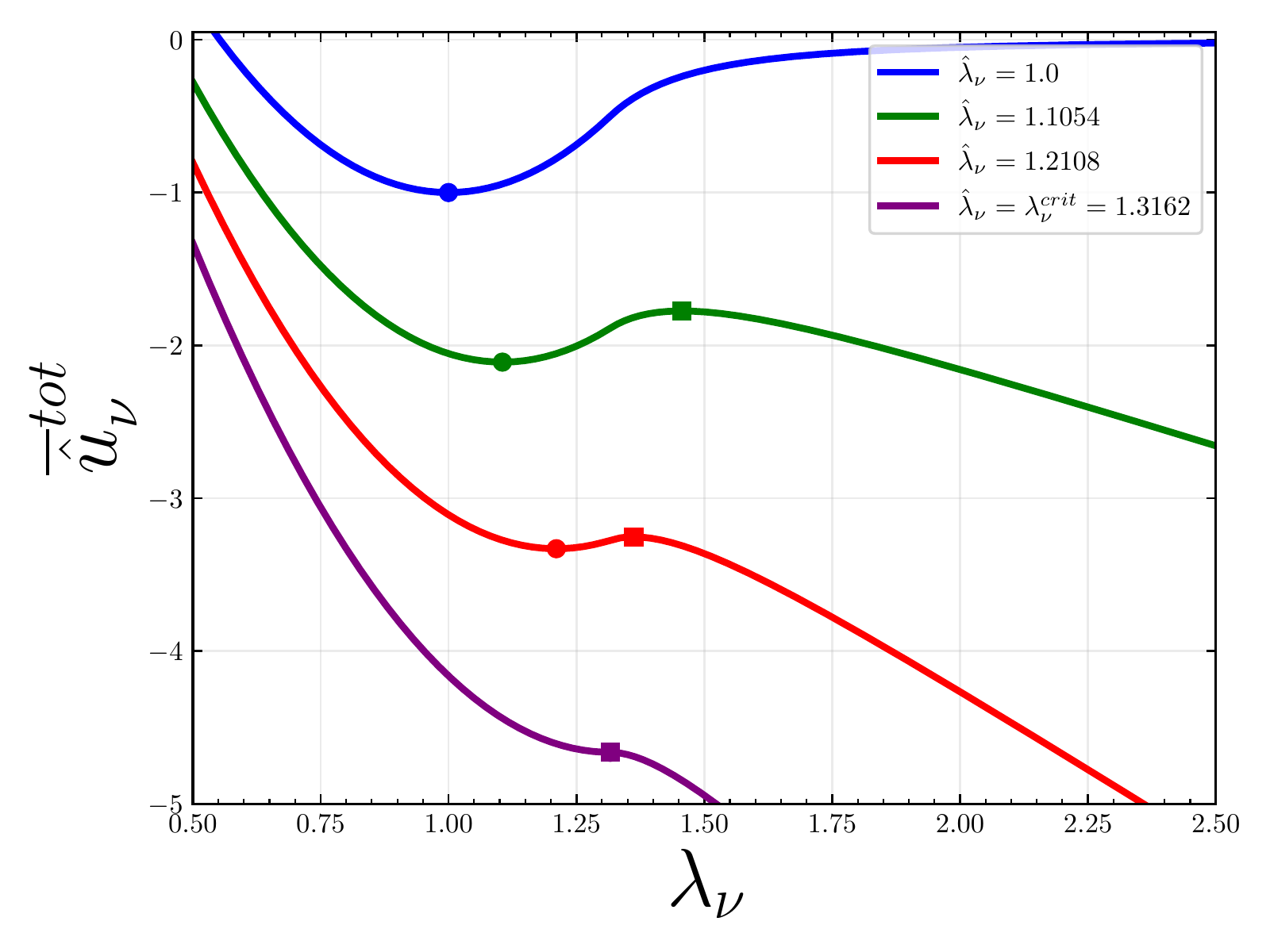}
	\caption{Nondimensional scaled tilted segment potential $\overline{\hat{u}}_{\nu}^{tot}$ as a function of segment stretch $\lambda_{\nu}$ permitted via thermal oscillations. Here, $\zeta_{\nu}^{char} = 100$ and $\kappa_{\nu} = 1000$, as per the composite segment potential displayed in \cref{fig:composite-u_nu-figs}. Several $\overline{\hat{u}}_{\nu}^{tot}$ functions corresponding to various applied segment stretch states $\hat{\lambda}_{\nu}$ are displayed: the blue curve corresponds to a segment under no applied force ($\hat{\xi}_c = 0,~\hat{\lambda}_{\nu} = 1$) and is identical to the nondimensional scaled segment potential $\overline{u}_{\nu}$, the purple curve corresponds to a segment under the critical force $\xi_c^{crit}$ ($\hat{\xi}_c = \xi_c^{crit},~\hat{\lambda}_{\nu} = \lambda_{\nu}^{crit}$), and the green and red curves correspond to a segment under an intermediate force between the force-free state and the critical force state ($0 < \hat{\xi}_c < \xi_c^{crit},~1 < \hat{\lambda}_{\nu} < \lambda_{\nu}^{crit}$). Dots and squares represent the value of $\overline{\hat{u}}_{\nu}^{tot}$ at its local minimum and maximum, respectively, i.e., dots correspond to the point $\left(\hat{\lambda}_{\nu}^{locmin},~\overline{\hat{u}}_{\nu}^{tot}\left(\hat{\lambda}_{\nu}, \hat{\lambda}_{\nu}^{locmin}\right)\right)$ and squares correspond to the point $\left(\hat{\lambda}_{\nu}^{locmax},~\overline{\hat{u}}_{\nu}^{tot}\left(\hat{\lambda}_{\nu}, \hat{\lambda}_{\nu}^{locmax}\right)\right)$. Note that $\hat{\lambda}_{\nu}^{locmax} = \infty$ for the blue curve and $\hat{\lambda}_{\nu}^{locmin} = \hat{\lambda}_{\nu}^{locmax}$ for the purple curve.}
	\label{fig:overline_u_nu_hat_tot-vs-lmbda_nu}
\end{figure}

\cref{fig:overline_u_nu_hat_tot-vs-lmbda_nu} displays the energy landscape for the nondimensional scaled tilted segment potential $\overline{\hat{u}}_{\nu}^{tot} = \hat{u}_{\nu}^{tot}/\zeta_{\nu}^{char}$ for various states of $\hat{\xi}_c$. The blue curve corresponds to $\hat{u}_{\nu}^{tot}$ for a segment in an undisturbed chain in a force-free state, i.e., $u_{\nu}$. Upon the application of $\hat{\xi}_c$, $u_{\nu}$ transforms to $\hat{u}_{\nu}^{tot}$ via tilting the blue curve to a particular non-blue colored curve. Dots and squares correspond to the local minimum energy state and local maximum energy state of $\hat{u}_{\nu}^{tot}$, respectively, and the segment stretch corresponding to these two stationary points express the following form
\begin{equation} \label{eq:hat-lambda_nu-stationary-points}
    \hat{\lambda}_{\nu}^{locmin} = \hat{\lambda}_{\nu} = 1+\frac{\hat{\xi}_c}{\kappa_{\nu}},\qquad\hat{\lambda}_{\nu}^{locmax} = 1+\sqrt[3]{\left[\frac{[\zeta_{\nu}^{char}]^2}{\kappa_{\nu}}\right]\left[\frac{1}{\hat{\xi}_c}\right]}.
\end{equation}
For the segment in an undisturbed, force-free chain, $\hat{\lambda}_{\nu}^{locmin} = 1$ and $\hat{\lambda}_{\nu}^{locmax} = \infty$ (hence why the blue square is not pictured in \cref{fig:overline_u_nu_hat_tot-vs-lmbda_nu}). Also, $\hat{\lambda}_{\nu}^{locmin} = \hat{\lambda}_{\nu}^{locmax} = \lambda_{\nu}^{crit}$ (i.e., $\hat{\lambda}_{\nu} = \lambda_{\nu}^{crit}$) occurs when $\hat{\xi}_c = \sqrt{\kappa_{\nu}\zeta_{\nu}^{char}} = \hat{\xi}_c^{crit}$. 

The activation energy barrier required for a segment to escape from the bottom of the potential energy well and undergo a scission reaction is simply the energy difference between an identically-colored square and dot pair (the difference between the local maximum and local minimum energy states of $\hat{u}_{\nu}^{tot}$). The nondimensional segment scission activation energy barrier thus obeys the following
\begin{equation}
    \hat{e}_{\nu}^{sci} = \hat{u}_{\nu}^{tot}\left(\hat{\lambda}_{\nu}, \hat{\lambda}_{\nu}^{locmax}\right) - \hat{u}_{\nu}^{tot}\left(\hat{\lambda}_{\nu}, \hat{\lambda}_{\nu}^{locmin}\right).
\end{equation}
The energy needed to overcome the activation energy barrier is supplied via segment-level thermal fluctuations. Considering a statistically significant number of segments, the rate-independent probability of segment scission $\hat{p}_{\nu}^{sci}$ and the rate-independent probability of segment survival $\hat{p}_{\nu}^{sur}$ are
\begin{equation}
    \hat{p}_{\nu}^{sci} = \exp{-\hat{e}_{\nu}^{sci}},\qquad\hat{p}_{\nu}^{tot,sur} = 1-\hat{p}_{\nu}^{sci}.
\end{equation}
Clearly, when $\hat{\xi}_c = \hat{\xi}_c^{crit}$, then segment scission occurs automatically. The rate-dependent probability of segment survival $\rho_{\nu}$ follows an Arrhenius law \citep{freund2009characterizing, yang2019rate,yang2020multiscale} and has an associated rate-dependent probability of segment scission $\gamma_{\nu}$
\begin{equation}
    \frac{\dot{\rho}_{\nu}}{\rho_{\nu}} = -\omega_0\hat{p}_{\nu}^{sci},\qquad\gamma_{\nu} = 1-\rho_{\nu} \implies \frac{\dot{\gamma}_{\nu}}{1-\gamma_{\nu}} = \omega_0\hat{p}_{\nu}^{sci},
\end{equation}
where the dot above implies a time derivative (i.e., a time rate-of-change) and $\omega_0$ is a microscopic frequency. This frequency is often interpreted as the natural frequency of atomic oscillation: $\omega_0 = (k_B T)/\hbar \approx 10^{13}~1/sec$, where $\hbar$ is the Planck constant \citep{yang2019rate,yang2020multiscale, hanggi1990reaction}. However, different frequencies for $\omega_0$ have been proposed and investigated \citep{yang2020multiscale,hanggi1990reaction,guo2021micromechanics}. For the rate-dependent scission framework, the segment lifetime decreases as the applied chain force increases.

\begin{figure}[t]
	\centering
	\includegraphics[width=0.75\textwidth]{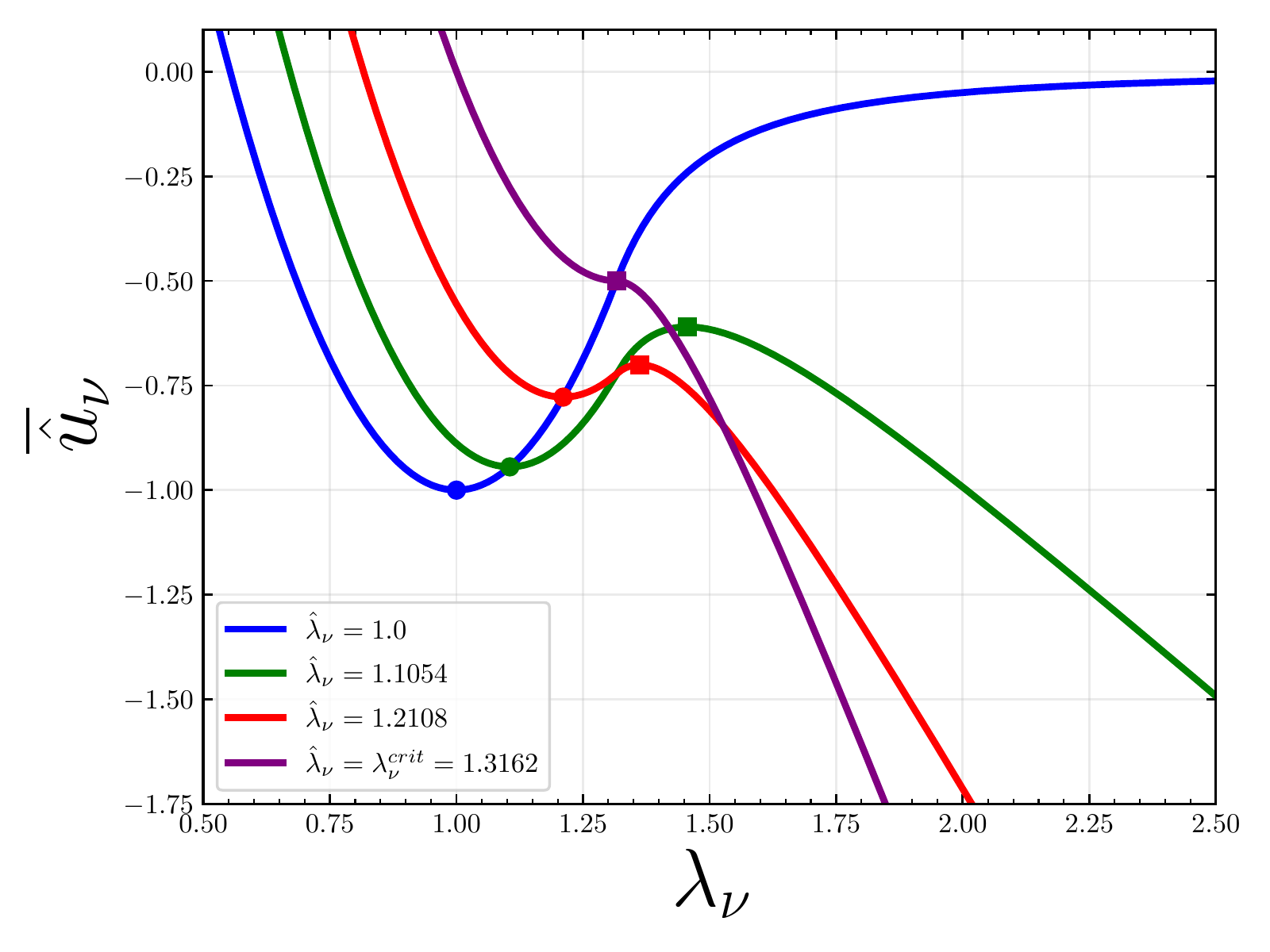}
	\caption{Nondimensional scaled total distorted segment potential $\overline{\hat{u}}_{\nu}$ as a function of segment stretch $\lambda_{\nu}$ permitted via thermal oscillations. Here, $\zeta_{\nu}^{char} = 100$ and $\kappa_{\nu} = 1000$, as per the composite segment potential displayed in \cref{fig:composite-u_nu-figs}. Several $\overline{\hat{u}}_{\nu}$ functions corresponding to various applied segment stretch states $\hat{\lambda}_{\nu}$ are displayed: the blue curve corresponds to a segment under no applied force ($\hat{\xi}_c = 0,~\hat{\lambda}_{\nu} = 1$) and is identical to the nondimensional scaled segment potential $\overline{u}_{\nu}$, the purple curve corresponds to a segment under the critical force $\xi_c^{crit}$ ($\hat{\xi}_c = \xi_c^{crit},~\hat{\lambda}_{\nu} = \lambda_{\nu}^{crit}$), and the green and red curves correspond to a segment under an intermediate force between the force-free state and the critical force state ($0 < \hat{\xi}_c < \xi_c^{crit},~1 < \hat{\lambda}_{\nu} < \lambda_{\nu}^{crit}$). Dots and squares represent the value of $\overline{\hat{u}}_{\nu}$ at its local minimum and maximum, respectively, i.e., dots correspond to the point $\left(\hat{\lambda}_{\nu}^{locmin},~\overline{\hat{u}}_{\nu}\left(\hat{\lambda}_{\nu}, \hat{\lambda}_{\nu}^{locmin}\right)\right)$ and squares correspond to the point $\left(\hat{\lambda}_{\nu}^{locmax},~\overline{\hat{u}}_{\nu}\left(\hat{\lambda}_{\nu}, \hat{\lambda}_{\nu}^{locmax}\right)\right)$. Note that $\hat{\lambda}_{\nu}^{locmax} = \infty$ for the blue curve and $\hat{\lambda}_{\nu}^{locmin} = \hat{\lambda}_{\nu}^{locmax}$ for the purple curve. This figure is effectively analogous to Fig. 4b in \citet{wang2019quantitative}.}
	\label{fig:overline_u_nu_hat-vs-lmbda_nu}
\end{figure}

The final quantity that remains to be ascertained is the amount of potential energy the segment releases upon scission. This energetic quantity is necessary to ultimately calculate the amount of energy dissipated within a polymer network due to chain rupture. Unfortunately, the tilted segment potential does not indicate whatsoever how much energy is released by the segment during a rupture event. To overcome this shortcoming, the conceptual schema provided in the formative work of \citet{wang2019quantitative} is evoked. 

First, \citet{wang2019quantitative} identified the product of the force applied to the segment and the length of the segment as the distorted segment potential energy. More physically, the distorted segment potential energy can be considered as the energy the segment stores due to its distortion under load, i.e., the work done by the applied load to the segment when assuming no thermal fluctuations. This energetic quantity is represented in nondimensional form as $\hat{u}_{\nu}^{distort}\left(\hat{\lambda}_{\nu}\right) = \hat{\xi}_c\left(\hat{\lambda}_{\nu}\right)*\hat{\lambda}_{\nu}$. 

Then, \citet{wang2019quantitative} proposed a reformulated tilted segment potential that captures both the reduction in segment scission activation energy upon applied force and the distorted segment potential energy. This reformulated tilted segment potential, $\hat{u}_{\nu}$, which we call the total distorted segment potential, is simply the sum of the original tilted segment potential and the distorted segment potential, and is provided in nondimensional form
\begin{equation}
    \hat{u}_{\nu}\left(\hat{\lambda}_{\nu}, \lambda_{\nu}\right) = \hat{u}_{\nu}^{tot}\left(\hat{\lambda}_{\nu}, \lambda_{\nu}\right) + \hat{u}_{\nu}^{distort}\left(\hat{\lambda}_{\nu}\right) = u_{\nu}(\lambda_{\nu}) - \hat{\xi}_c\left(\hat{\lambda}_{\nu}\right)\left[\lambda_{\nu} - \hat{\lambda}_{\nu}\right].
\end{equation}
More physically, the total distorted segment potential is the potential energy of the segment in the displacement controlled setting, plus the difference in work done by the load to the segment applied with and without thermal fluctuations. The implicit definition of the segment stretch using $\hat{u}_{\nu}$ also remains unchanged from before in \cref{eq:segment-stretch-definition}.

\cref{fig:overline_u_nu_hat-vs-lmbda_nu} displays the energy landscape for the nondimensional scaled total distorted segment potential $\overline{\hat{u}}_{\nu} = \hat{u}_{\nu}/\zeta_{\nu}^{char}$ for various states of $\hat{\xi}_c$. Identically colored curves in both \cref{fig:overline_u_nu_hat_tot-vs-lmbda_nu} and \cref{fig:overline_u_nu_hat-vs-lmbda_nu} correspond to identical states of $\hat{\xi}_c$. As in \cref{fig:overline_u_nu_hat_tot-vs-lmbda_nu}, dots and squares correspond to the energy state of the two stationarity points of $\hat{u}_{\nu}$, and the segment stretch of the stationarity points of $\hat{u}_{\nu}$ are identical to those for $\hat{u}_{\nu}^{tot}$ in \cref{eq:hat-lambda_nu-stationary-points}.

When comparing the $\hat{u}_{\nu}^{tot}$ curves in \cref{fig:overline_u_nu_hat_tot-vs-lmbda_nu} to the $\hat{u}_{\nu}$ curves in \cref{fig:overline_u_nu_hat-vs-lmbda_nu}, two important takeaways can be made regarding the relationship between $\hat{u}_{\nu}^{tot}$ and $\hat{u}_{\nu}$. First, it is trivial to show that the activation energy barrier of segment scission for $\hat{u}_{\nu}$ is (numerically) identical to that for $\hat{u}_{\nu}^{tot}$. As a consequence, all of the segment scission probabilistics associated with $\hat{u}_{\nu}$ are identical to the segment scission probabilistics calculated from $\hat{u}_{\nu}^{tot}$. Second, $\hat{u}_{\nu}$ is clearly generated by translating $\hat{u}_{\nu}^{tot}$ up in such a way so that its local minimum energy state is its intersection point with $u_{\nu}$, the potential energy curve for a segment in an undisturbed, force-free chain.

\begin{figure}[t]
	\centering
	\includegraphics[width=0.75\textwidth]{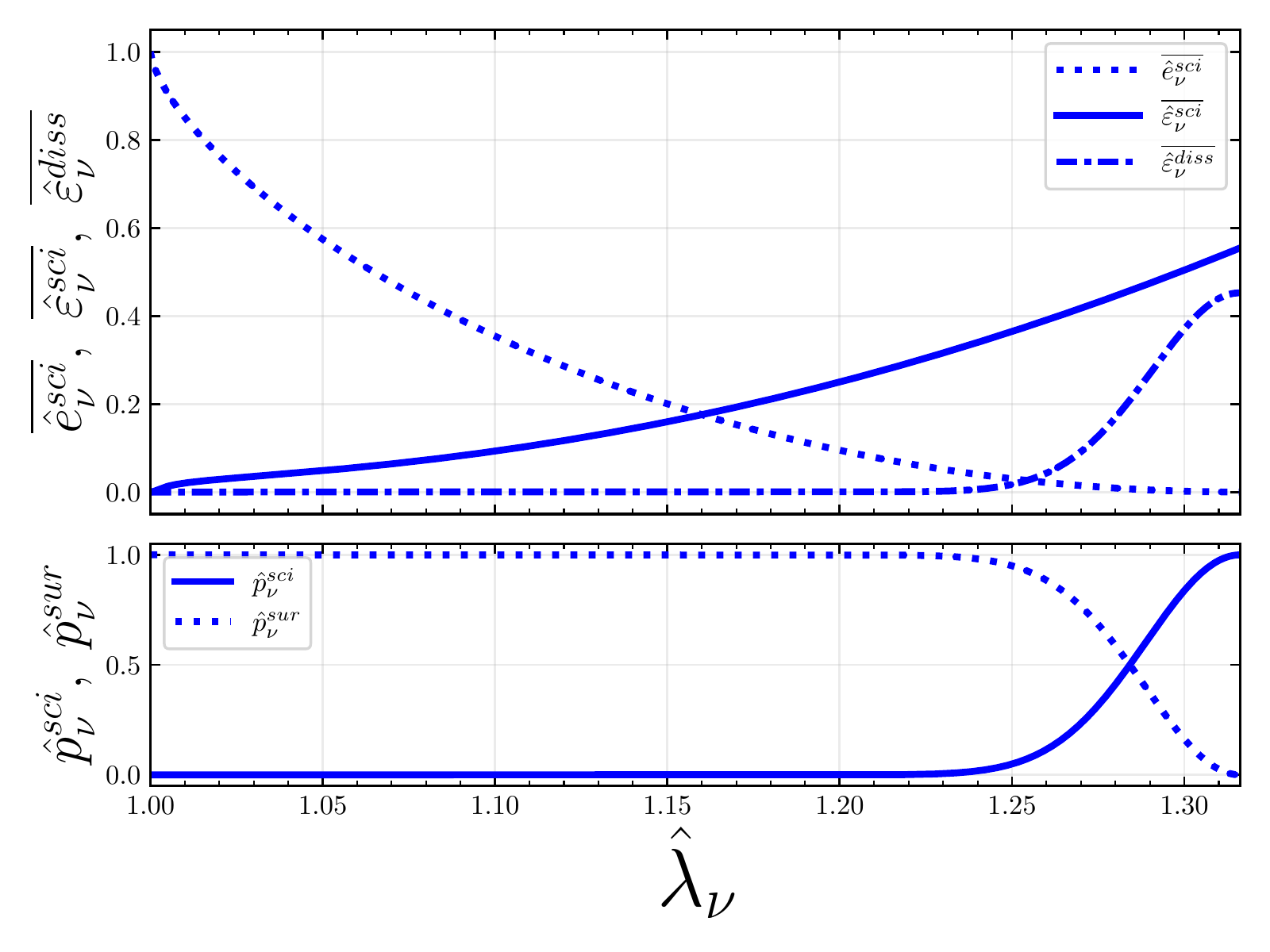}
	\caption{Segment scission energetics and probabilistics. Here, $\zeta_{\nu}^{char} = 100$ and $\kappa_{\nu} = 1000$, as per the composite segment potential displayed in \cref{fig:composite-u_nu-figs}. (top) Nondimensional scaled segment scission activation energy barrier $\overline{\hat{e}_{\nu}^{sci}}$, nondimensional scaled segment scission energy $\overline{\hat{\varepsilon}_{\nu}^{sci}}$, and rate-independent nondimensional scaled dissipated segment scission energy $\overline{\hat{\varepsilon}_{\nu}^{diss}}$ as a function of applied segment stretch $\hat{\lambda}_{\nu}$ in the domain $1 \leq \hat{\lambda}_{\nu} \leq \lambda_{\nu}^{crit}$. (bottom) Rate-independent probability of segment scission $\hat{p}_{\nu}^{sci}$ and rate-independent probability of segment survival $\hat{p}_{\nu}^{sur}$ as a function of $\hat{\lambda}_{\nu}$ in the domain $1 \leq \hat{\lambda}_{\nu} \leq \lambda_{\nu}^{crit}$.}
	\label{fig:segment-scission-indicators-vs-lmbda_nu_hat}
\end{figure}

This last takeaway has serious implications on segment scission energetics, as displayed by \cref{fig:overline_u_nu_hat-vs-lmbda_nu} and as identified previously by \citet{wang2019quantitative}: since a non-blue colored dot (the local minimum energy state of $\hat{u}_{\nu}$) always lies on the blue curve ($u_{\nu}$), then the stored potential energy in a distorted segment is simply the energy difference between a non-blue colored dot with the blue dot. If thermal fluctuations supply a segment the requisite energy to overcome its scission activation energy barrier, then the stored potential energy would be released upon the completion of the scission reaction. \citet{wang2019quantitative} defined this stored segment potential energy as the segment scission potential energy, and is provided in nondimensional form
\begin{equation}
    \hat{u}_{\nu}^{sci} \equiv \hat{u}_{\nu}\left(\hat{\lambda}_{\nu}, \hat{\lambda}_{\nu}^{locmin}\right) - \hat{u}_{\nu}(1,1) = u_{\nu}\left(\hat{\lambda}_{\nu}\right) + \zeta_{\nu}^{char}.
\end{equation}
Clearly, $\hat{u}_{\nu}^{sci}$ is simply $u_{\nu}\left(\hat{\lambda}_{\nu}\right)$ shifted by $\zeta_{\nu}^{char}$. Considering this, the nondimensional segment scission energy is then defined as the nondimensional per segment Helmholtz free energy under $\hat{\xi}_c$ shifted by $\zeta_{\nu}^{char}$
\begin{equation}
    \hat{\varepsilon}_{\nu}^{sci} \equiv s_{c\nu}\left(\hat{\lambda}_{\nu}, \hat{\lambda}_c^{eq}\right) +  u_{\nu}\left(\hat{\lambda}_{\nu}\right) + \zeta_{\nu}^{char} = \psi_{c\nu}\left(\hat{\lambda}_{\nu}, \hat{\lambda}_c^{eq}\right) + \zeta_{\nu}^{char}.
\end{equation}
Note that $\hat{\varepsilon}_{\nu}^{sci}$ here is the energy released by a single segment as it ruptures. However, by considering a statistically significant number of segments, a nondimensional dissipated segment scission energy $\hat{\varepsilon}_{\nu}^{diss}$ can be defined. The rate-dependent $\hat{\varepsilon}_{\nu}^{diss}$ is defined via its time rate-of-change equation
\begin{equation}
    \dot{\hat{\varepsilon}}_{\nu}^{diss} \equiv \dot{\gamma}_{\nu}\hat{\varepsilon}_{\nu}^{sci}.
\end{equation}
In an analogous way, the rate-independent $\hat{\varepsilon}_{\nu}^{diss}$ is defined via its applied segment stretch-based rate-of-change equation
\begin{equation} \label{eq:rate-of-change-rate-independent-segment-scission}
    \left(\hat{\varepsilon}_{\nu}^{diss}\right)^{\prime} \equiv \left(\hat{p}_{\nu}^{sci}\right)^{\prime}\hat{\varepsilon}_{\nu}^{sci},
\end{equation}
where derivatives are taken with respect to $\hat{\lambda}_{\nu}$. The analytical form of $\hat{e}_{\nu}^{sci}$ and $\left(\hat{p}_{\nu}^{sci}\right)^{\prime}$ can be found for $\hat{\lambda}_{\nu} \leq \lambda_{\nu}^{crit}$ (where $\hat{\xi}_c = \kappa_{\nu}[\hat{\lambda}_{\nu}-1]$)
\begin{align}
    & \hat{e}_{\nu}^{sci} = \frac{1}{2}\kappa_{\nu}\left[\hat{\lambda}_{\nu} - 1\right]^2 - \frac{3}{2}\sqrt[3]{\left[\zeta_{\nu}^{char}\right]^2\kappa_{\nu}\left[\hat{\lambda}_{\nu} - 1\right]^2} + \zeta_{\nu}^{char}, \label{eq:activation-energy-barrier} \\
    & \left(\hat{p}_{\nu}^{sci}\right)^{\prime} = \hat{p}_{\nu}^{sci}\left[\sqrt[3]{\frac{\left[\zeta_{\nu}^{char}\right]^2\kappa_{\nu}}{\hat{\lambda}_{\nu} - 1}} - \kappa_{\nu}[\hat{\lambda}_{\nu} - 1]\right].
\end{align}
\cref{eq:rate-of-change-rate-independent-segment-scission} can be rewritten as
\begin{equation}
    \left(\hat{\varepsilon}_{\nu}^{diss}\right)^{\prime} = \hat{p}_{\nu}^{sci}\left[\sqrt[3]{\frac{\left[\zeta_{\nu}^{char}\right]^2\kappa_{\nu}}{\hat{\lambda}_{\nu} - 1}} - \kappa_{\nu}[\hat{\lambda}_{\nu} - 1]\right]\hat{\varepsilon}_{\nu}^{sci},
\end{equation}
where $\hat{\varepsilon}_{\nu}^{diss} = 0$ at $\hat{\lambda}_{\nu} = 1$. If irreversible segment scission is assumed to take place in the network, then $\left(\hat{p}_{\nu}^{sci}\right)^{\prime} \geq 0$ and $\left(\hat{\varepsilon}_{\nu}^{diss}\right)^{\prime} \geq 0$ must hold during deformation.

For segments in a chain between its undisturbed, force-free state and its critical state, \cref{fig:segment-scission-indicators-vs-lmbda_nu_hat} displays the evolution of $\hat{e}_{\nu}^{sci}$, $\hat{\varepsilon}_{\nu}^{sci}$, and rate-independent $\hat{\varepsilon}_{\nu}^{diss}$ in nondimensional scaled form ($\overline{\hat{e}_{\nu}^{sci}} = \hat{e}_{\nu}^{sci}/\zeta_{\nu}^{char}$, $\overline{\hat{\varepsilon}_{\nu}^{sci}} = \hat{\varepsilon}_{\nu}^{sci}/\zeta_{\nu}^{char}$, and $\overline{\hat{\varepsilon}_{\nu}^{diss}} = \hat{\varepsilon}_{\nu}^{diss}/\zeta_{\nu}^{char}$). The associated rate-independent probabilities $\hat{p}_{\nu}^{sci}$ and $\hat{p}_{\nu}^{sur}$ are also displayed. The evolution of $\overline{\hat{e}_{\nu}^{sci}}$ in \cref{fig:segment-scission-indicators-vs-lmbda_nu_hat} confirms the fact that the fundamental principle of the \citet{bell1978models} model holds: an intrinsic activation energy barrier to thermally activated scission in the absence of external forces is reduced by an energetic term dependent upon the applied force. However, in contrast with the \citet{bell1978models} model and in line with the findings in \citet{yang2020multiscale}, the activation energy barrier decreases in a nonlinear fashion upon an increase in applied force. This nonlinear relationship is accounted for in \cref{eq:activation-energy-barrier}, where $\hat{\lambda}_{\nu} = 1+\hat{\xi}_c/\kappa_{\nu}$. 

The evolution of $\overline{\hat{\varepsilon}_{\nu}^{diss}}$ in \cref{fig:segment-scission-indicators-vs-lmbda_nu_hat} confirms that when $\hat{p}_{\nu}^{sci}$ increases from 0 to 1, $\overline{\hat{\varepsilon}_{\nu}^{diss}}$ increases in a manner that simultaneously accounts for increasing $\overline{\hat{\varepsilon}_{\nu}^{sci}}$. At the critical chain state when $\hat{\xi}_c = \hat{\xi}_c^{crit}$, $\overline{\hat{u}_{\nu}^{sci}} = \hat{u}_{\nu}^{sci}/\zeta_{\nu}^{char} = 1/2$, $\overline{\hat{\varepsilon}_{\nu}^{sci}} \gtrapprox \overline{\hat{u}_{\nu}^{sci}}$, and $\overline{\hat{\varepsilon}_{\nu}^{diss}} \lessapprox \overline{\hat{u}_{\nu}^{sci}}$. As a matter of interpretation, recall that $\hat{p}_{\nu}^{sci}$, $\hat{p}_{\nu}^{sur}$, and $\overline{\hat{\varepsilon}_{\nu}^{diss}}$ are intended to describe the probabilitistics and energetics for a statiscially large number of segments.

\subsection{Chain-level scission energetics and probabilistic considerations} \label{subsec:chain-scission}

With the rate-dependent and rate-independent segment-level scission framework established, probabilistic considerations will now be used to establish a chain-level rupture framework. Chain rupture can take place when any one segment along its backbone ruptures. The rate-independent probability of chain survival $\hat{p}_c^{sur}$ and the rate-independent probability of chain scission $\hat{p}_c^{sci}$ are calculated as
\begin{equation} \label{eq:chain-scission-survival-probabilities}
    \hat{p}_c^{sur} = [\hat{p}_{\nu}^{sur}]^{\nu},\qquad\hat{p}_c^{sci} = 1 - \hat{p}_c^{sur}.
\end{equation}
The segment that becomes ruptured contributes an energy dissipation of $\hat{\varepsilon}_{\nu}^{sci}$ (since thermal excitations supply the segment the activation energy $\hat{e}_{\nu}^{sci}$ needed to execute the rupture reaction). Meanwhile, after the segment rupture event takes place, the remaining $\nu-1$ intact segments in the ruptured chain will each traverse the undisturbed segment potential energy landscape back to the minimum energy state at equilibrium, since the ruptured chain is suddenly unable to bear load. Visually, an intact segment will slide down the blue curve from a non-blue colored dot to the blue dot in \cref{fig:overline_u_nu_hat-vs-lmbda_nu} as the chain ruptures at another location. As a result of this, each of the $\nu-1$ intact segments contributes an energy dissipation of $\hat{\varepsilon}_{\nu}^{sci}$ as well. Therefore, the nondimensional chain scission energy is simply the nondimensional chain Helmholtz free energy at the moment of scission shifted by $\nu\zeta_{\nu}^{char}$
\begin{equation}
    \hat{\varepsilon}_c^{sci} \equiv \nu\hat{\varepsilon}_{\nu}^{sci} = \nu\left[\psi_{c\nu}\left(\hat{\lambda}_{\nu}, \hat{\lambda}_c^{eq}\right) + \zeta_{\nu}^{char}\right].
\end{equation}
The nondimensional chain scission energy per segment, $\hat{\varepsilon}_{c\nu}^{sci} \equiv \hat{\varepsilon}_c^{sci}/\nu$, is equivalent to the nondimensional segment scission energy
\begin{equation}
    \hat{\varepsilon}_{c\nu}^{sci} \equiv \frac{\hat{\varepsilon}_c^{sci}}{\nu} = \hat{\varepsilon}_{\nu}^{sci}.
\end{equation}
The rate-dependent probability of chain survival $\rho_c$ is related to the rate-dependent probability of segment survival $\rho_{\nu}$ as $\rho_c = [\rho_{\nu}]^{\nu}$ \citep{yang2020multiscale}. Considering this, the rate-dependent probability of chain survival $\rho_c$ and rate-dependent probability of chain scission $\gamma_c$ equal
\begin{equation}
    \frac{\dot{\rho}_c}{\rho_c} = -\nu\omega_0\hat{p}_{\nu}^{sci},\qquad \gamma_c = 1-\rho_c \implies \frac{\dot{\gamma}_c}{1-\gamma_c} = \nu\omega_0\hat{p}_{\nu}^{sci}.
\end{equation}
Analogous to the segment-level theory, the time rate-of-change equation for the rate-dependent nondimensional dissipated chain scission energy per segment $\hat{\varepsilon}_{c\nu}^{diss}$ is defined as \citep{guo2021micromechanics}
\begin{equation} \label{eq:rate-dependent-Xi_cnu-rate-equation}
    \dot{\hat{\varepsilon}}_{c\nu}^{diss} \equiv \dot{\gamma}_c\hat{\varepsilon}_{c\nu}^{sci}.
\end{equation}
The applied segment stretch-based rate-of-change for the rate-independent $\hat{\varepsilon}_{c\nu}^{diss}$ is defined as
\begin{align}
    & \left(\hat{\varepsilon}_{c\nu}^{diss}\right)^{\prime} \equiv \left(\hat{p}_c^{sci}\right)^{\prime}\hat{\varepsilon}_{c\nu}^{sci},\qquad  \left(\hat{p}_c^{sci}\right)^{\prime}= \nu[1 - \hat{p}_{\nu}^{sci}]^{\nu-1}\left(\hat{p}_{\nu}^{sci}\right)^{\prime}, \\
    &\left(\hat{\varepsilon}_{c\nu}^{diss}\right)^{\prime} =  \nu[1 - \hat{p}_{\nu}^{sci}]^{\nu-1}\hat{p}_{\nu}^{sci}\left[\sqrt[3]{\frac{\left[\zeta_{\nu}^{char}\right]^2\kappa_{\nu}}{\hat{\lambda}_{\nu} - 1}} - \kappa_{\nu}[\hat{\lambda}_{\nu} - 1]\right]\hat{\varepsilon}_{\nu}^{sci},
\end{align}
where $\hat{\varepsilon}_{c\nu}^{diss} = 0$ at $\hat{\lambda}_{\nu} = 1$. If irreversible chain scission is assumed to take place in the polymer network, then $\left(\hat{p}_c^{sci}\right)^{\prime} \geq 0$ and $\left(\hat{\varepsilon}_{c\nu}^{diss}\right)^{\prime} \geq 0$ must hold during deformation.

\rmk Any arbitrary anharmonic segment potential is able to be directly utilized within the segment rupture and chain rupture formulation developed in this work. For instance, consider the Morse segment potential provided in \cref{eq:lj-potential-and-morse-potential}. Using the Morse potential, the local minimum and local maximum for the corresponding $\hat{u}_{\nu}$ take the following functional forms
\begin{equation}
    \hat{\lambda}_{\nu}^{locmin} = \hat{\lambda}_{\nu} = 1+\frac{1}{\alpha_{\nu}}\ln\left(\frac{2}{1 + \sqrt{1 - \frac{\hat{\xi}_c}{\xi_c^{crit}}}}\right),\qquad \hat{\lambda}_{\nu}^{locmax} = 1+\frac{1}{\alpha_{\nu}}\ln\left(\frac{2}{1 - \sqrt{1 - \frac{\hat{\xi}_c}{\xi_c^{crit}}}}\right),
\end{equation}
where $\alpha_{\nu} = \sqrt{\frac{\kappa_{\nu}}{2\zeta_{\nu}^{char}}}$. When $\hat{\xi}_c = 0$, then $\hat{\lambda}_{\nu}^{locmin} = 1$ and $\hat{\lambda}_{\nu}^{locmax} = \infty$, as expected. At the critical point where $\lambda_{\nu}^{crit} = \hat{\lambda}_{\nu}^{locmin} = \hat{\lambda}_{\nu}^{locmax}$, it is found that $\xi_c^{crit} = \sqrt{\frac{\kappa_{\nu}\zeta_{\nu}^{char}}{8}}$, $\lambda_{\nu}^{crit} = 1 + \frac{\ln(2)}{\alpha_{\nu}}$, $\hat{e}_{\nu}^{sci}=0$, and $\hat{u}_{\nu}^{sci} = \zeta_{\nu}^{char}/4$. The analytical form of $\hat{e}_{\nu}^{sci}$ and $\left(\hat{p}_{\nu}^{sci}\right)^{\prime}$ can be found for $\hat{\lambda}_{\nu} \leq \lambda_{\nu}^{crit}$
\begin{align}
    & \hat{e}_{\nu}^{sci} = \zeta_{\nu}^{char}\left[2e^{-\alpha_{\nu}[\hat{\lambda}_{\nu} - 1]} - 1\right] - \xi_c^{crit}\left[1 - \left[2e^{-\alpha_{\nu}[\hat{\lambda}_{\nu} - 1]} - 1\right]^2\right]\left[1 - \frac{1}{\alpha_{\nu}}\ln\left(1 - e^{-\alpha_{\nu}[\hat{\lambda}_{\nu} - 1]}\right) - \hat{\lambda}_{\nu}\right], \\
    & \left(\hat{p}_{\nu}^{sci}\right)^{\prime} = \hat{p}_{\nu}^{sci}\left[-2\alpha_{\nu}\zeta_{\nu}^{char}e^{-\alpha_{\nu}[\hat{\lambda}_{\nu} - 1]} \right. \nonumber \\
    & \left. \qquad\qquad\qquad - \xi_c^{crit}\left[4\alpha_{\nu}e^{-\alpha_{\nu}[\hat{\lambda}_{\nu} - 1]}\right]\left[2e^{-\alpha_{\nu}[\hat{\lambda}_{\nu} - 1]} - 1\right]\left[1 - \frac{1}{\alpha_{\nu}}\ln\left(1 - e^{-\alpha_{\nu}[\hat{\lambda}_{\nu} - 1]}\right) - \hat{\lambda}_{\nu}\right] \right. \nonumber \\
    & \left. \qquad\qquad\qquad - \xi_c^{crit}\left[1 - \left[2e^{-\alpha_{\nu}[\hat{\lambda}_{\nu} - 1]} - 1\right]^2\right]\left[\frac{1}{e^{-\alpha_{\nu}[\hat{\lambda}_{\nu} - 1]} - 1}\right]\right],
\end{align}
where $\hat{e}_{\nu}^{sci} = \zeta_{\nu}^{char}$ at $\hat{\lambda}_{\nu} = 1$. From this point on, the remainder of the segment scission and chain scission formulation can be implemented for the Morse potential, with the caveat that the inverse relationship between $\lambda_c^{eq}$ and $\lambda_{\nu}$ must be calculated computationally. Do take stock of this caveat and its implications: as discussed in Remark 3, using numerics to calculate the inverse relationship between $\lambda_c^{eq}$ and $\lambda_{\nu}$ will ultimately prevent an upscaled continuum damage model to be cast in a tractable closed-form manner. On the contrary, and as implied by Remark 3, the segment rupture and chain rupture formulation developed in this work using the composite segment potential directly leads to a tractable closed-form continuum damage model. In the context of segment rupture and chain rupture, this is truly what sets the composite segment potential apart from other anharmonic segment potentials.

\subsection{Chain scission informed equilibrium probability distribution considerations} \label{subsec:chain-scission-informed-equil-prob-dist}

With a description of the intact and ruptured chain configuration spaces at hand, a formalization of the equilibrium probability distribution for intact chains can now be undertaken.\footnote{In this section, equilibrium refers to time-independent behavior of constituents in a statistical ensemble. This is in contrast to the use of the term equilibrium thus far in this manuscript, which has referred to the minimum potential energy state of a segment in an undisturbed, force-free chain.} Once again, the statistical mechanics theory derived in \citet{buche2021chain} is employed and briefly reviewed.

The total configuration equilibrium probability for a chain to be either intact or ruptured is given by the following conservation law
\begin{equation} \label{eq:tot-config-equil-prob}
    1 = \int\cdots\int \mathcal{P}_{eq}^{intact}(\mathbf{r}_{\nu})d^3\mathbf{r}_{\nu}  + \sum_{j=1}^{\nu}\int\cdots\int \mathcal{P}_{eq}^{rup_j}(\mathbf{r}_{\nu})d^3\mathbf{r}_{\nu},
\end{equation}
where $\mathbf{r}_{\nu}$ is the end-to-end chain vector, $\int\cdots\int d^3\mathbf{r}_{\nu}$ denotes integration about all spatial chain configurations at equilibrium, $\mathcal{P}_{eq}^{intact}(\mathbf{r}_{\nu})$ is the intact chain configuration equilibrium probability distribution, and $\mathcal{P}_{eq}^{rup_j}(\mathbf{r}_{\nu})$ is the chain configuration equilibrium probability distribution for a chain ruptured via the $j$th segment.\footnote{According to \citet{buche2021chain}, the following assumptions underlie the statistical mechanics theory derived up to this point: (i) Polymer chains are noninteracting and considered as members of a classical, canonical statistical mechanics ensemble; (ii) Local equilibrium is independently maintained in the phase space region associated with intact chains and the phase space region associated with ruptured chains, as per transition state theory; (iii) Any arbitrary segment along a polymer chain backbone may become ruptured; (iv) A chain is considered ruptured once only a single segment becomes ruptured; (v) The possibility of two chains ruptured at different segment locations cross-reforming together is neglected.} Multiplying the total configuration equilibrium probability in \cref{eq:tot-config-equil-prob} by the total configuration equilibrium partition function $Z_{eq}^{tot}$ leads to the following relationship
\begin{equation}
    Z_{eq}^{tot} = \int\cdots\int \mathcal{Z}^{intact}(\mathbf{r}_{\nu})d^3\mathbf{r}_{\nu}  + \sum_{j=1}^{\nu}\int\cdots\int \mathcal{Z}^{rup_j}(\mathbf{r}_{\nu})d^3\mathbf{r}_{\nu},
\end{equation}
where $\mathcal{Z}^{intact}(\mathbf{r}_{\nu})$ is the intact chain configuration partition function and $\mathcal{Z}^{rup_j}(\mathbf{r}_{\nu})$ is the chain configuration partition function for a chain that has ruptured via the $j$th segment. Via the principle thermodynamic connection formula $\beta\Psi_c(\mathbf{r}_{\nu}) = -\ln\left(\mathcal{Z}(\mathbf{r}_{\nu})\right)$, the net Helmholtz free energy change is defined solely for the $j$th segment undergoing rupture, provided in nondimensional form
\begin{equation}
    \Delta\psi_{j}^{\nu sci} \equiv -\ln(\frac{\int\cdots\int \mathcal{Z}^{rup_j}(\mathbf{r}_{\nu})d^3\mathbf{r}_{\nu}}{\int\cdots\int \mathcal{Z}^{intact}(\mathbf{r}_{\nu})d^3\mathbf{r}_{\nu}}).
\end{equation}
Using the above, $Z_{eq}^{tot}$ and $\mathcal{P}_{eq}^{intact}(\mathbf{r}_{\nu})$ are respectively calculated as
\begin{align}
    & Z_{eq}^{tot} = \left[1 + \sum_{j=1}^{\nu}e^{-\Delta\psi_{j}^{\nu sci}}\right]\left[\int\cdots\int \mathcal{Z}^{intact}(\mathbf{r}_{\nu})d^3\mathbf{r}_{\nu}\right], \label{eq:tot-config-equil-partition-func} \\
    & \mathcal{P}_{eq}^{intact}(\mathbf{r}_{\nu}) = \frac{\mathcal{Z}^{intact}(\mathbf{r}_{\nu})}{Z_{eq}^{tot}} = \frac{\mathcal{Z}^{intact}(\mathbf{r}_{\nu})}{\left[1 + \sum_{j=1}^{\nu}e^{-\Delta\psi_{j}^{\nu sci}}\right]\left[\int\cdots\int \mathcal{Z}^{intact}(\mathbf{r}_{\nu})d^3\mathbf{r}_{\nu}\right]}. \label{eq:intact-chain-config-equil-prob-dist}
\end{align}
At this point, $\Delta\psi_{j}^{\nu sci}$ is identified as the rate-independent nondimensional dissipated segment scission energy at the critical segment stretch $\hat{\varepsilon}_{\nu}^{diss}(\lambda_{\nu}^{crit})$. The integral of $\mathcal{Z}^{intact}(\mathbf{r}_{\nu})$ is taken over all chain configurations where chain survival is probable. In light of the chain scission framework established in \cref{subsec:segment-scission} and \cref{subsec:chain-scission}, that integral is evaluated as
\begin{align}
    & \int\cdots\int \mathcal{Z}^{intact}(\mathbf{r}_{\nu})d^3\mathbf{r}_{\nu} = 4\pi \int\displaylimits_{0}^{r_{\nu}^{crit}} e^{-\nu\psi_{c\nu}(l_{\nu}, r_{\nu})}r_{\nu}^2 dr_{\nu} \\
    & \qquad = 4\pi e^{\nu \zeta_{\nu}^{char}}[\nu l_{\nu}^{eq}]^3 \int\displaylimits_{0}^{(\lambda_c^{eq})^{crit}} e^{-\nu[\psi_{c\nu}(\lambda_{\nu},\lambda_c^{eq}) + \zeta_{\nu}^{char}]}[\lambda_c^{eq}]^2 d\lambda_c^{eq} = 4\pi e^{\nu \zeta_{\nu}^{char}}[\nu l_{\nu}^{eq}]^3\mathscr{I}(3,\nu), \\
    & \mathscr{I}(n,\nu) \equiv \int\displaylimits_{0}^{(\lambda_c^{eq})^{crit}} e^{-\nu[\psi_{c\nu}(\lambda_{\nu},\lambda_c^{eq}) + \zeta_{\nu}^{char}]} [\lambda_c^{eq}]^{n-1} d\lambda_c^{eq},
\end{align}
where $\lambda_c^{eq}$ is substituted in for $r_{\nu}$. Note that segment extensibility is functionally accounted for in the integral function $\mathscr{I}(n,\nu)$ as per the segment stretch function in \cref{eq:segment-stretch-function}. $Z_{eq}^{tot}$ and $\mathcal{P}_{eq}^{intact}(\mathbf{r}_{\nu})$ can now be simplified
\begin{align}
    & Z_{eq}^{tot} = \left[1 + \nu e^{-\hat{\varepsilon}_{\nu}^{diss}(\lambda_{\nu}^{crit})}\right]4\pi e^{\nu \zeta_{\nu}^{char}}[\nu l_{\nu}^{eq}]^3\mathscr{I}(3,\nu), \label{eq:tot-config-equil-partition-func-simplified} \\
    & \mathcal{P}_{eq}^{intact}(\mathbf{r}_{\nu}) = \frac{\mathcal{Z}^{intact}(\mathbf{r}_{\nu})}{\left[1 + \nu e^{-\hat{\varepsilon}_{\nu}^{diss}(\lambda_{\nu}^{crit})}\right]4\pi e^{\nu \zeta_{\nu}^{char}}[\nu l_{\nu}^{eq}]^3\mathscr{I}(3,\nu)}. \label{eq:intact-chain-config-equil-prob-dist-simplified}
\end{align}
Using $\mathcal{P}_{eq}^{intact}(\mathbf{r}_{\nu})$ and considering the chain configurations corresponding to probable chain survival, the root-mean-square end-to-end chain distance $r_{\nu}^{rms}$ is calculated as
\begin{align}
    & \left(\overline{r_{\nu}^{rms}}\right)^2 = \int\cdots\int r_{\nu}^2 \mathcal{P}_{eq}^{intact}(\mathbf{r}_{\nu})d^3\mathbf{r}_{\nu} = [\nu l_{\nu}^{eq}]^2\left[\frac{1}{1+\nu e^{-\hat{\varepsilon}_{\nu}^{diss}(\lambda_{\nu}^{crit})}}\right]\frac{\mathscr{I}(5,\nu)}{\mathscr{I}(3,\nu)}, \\
    & r_{\nu}^{rms} = \nu l_{\nu}^{eq}\sqrt{\left[\frac{1}{1+\nu e^{-\hat{\varepsilon}_{\nu}^{diss}(\lambda_{\nu}^{crit})}}\right]\frac{\mathscr{I}(5,\nu)}{\mathscr{I}(3,\nu)}}.
\end{align}
\\
\begin{figure}[t]
	\centering
	\includegraphics[width=0.75\textwidth]{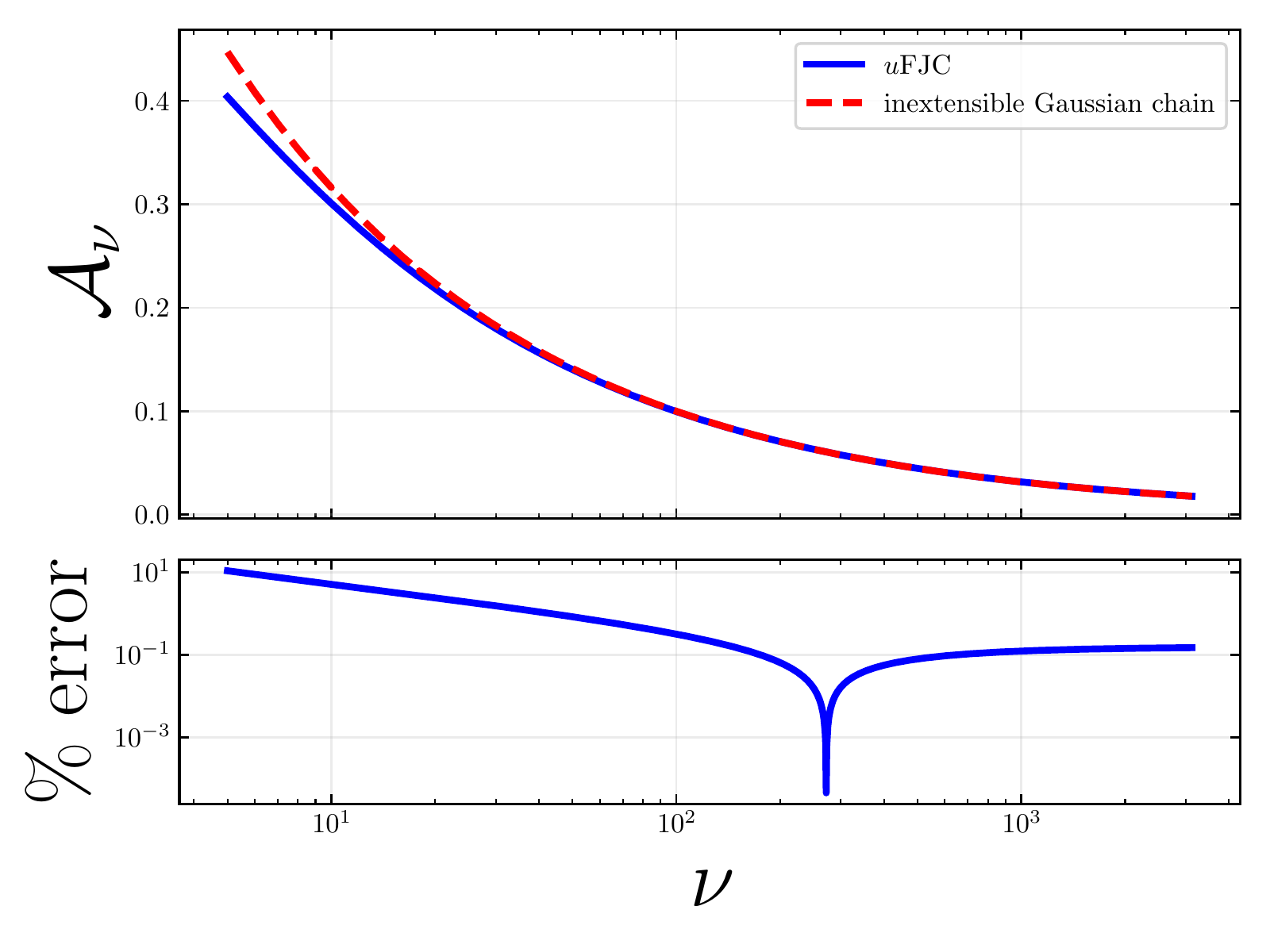}
	\caption{Comparison of reference equilibrium chain stretches calculated via the $u$FJC statistical mechanics framework and the inextensible Gaussian chain assumption. Here, $\zeta_{\nu}^{char} = 100$ and $\kappa_{\nu} = 1000$, as per the composite segment potential displayed in \cref{fig:composite-u_nu-figs}. (top) Reference equilibrium chain stretch $\mathcal{A}_{\nu}$ calculated via the $u$FJC statistical mechanics framework and the inextensible Gaussian chain assumption as a function of segment number $\nu$ in the domain $5\leq\nu\leq 3125$. (bottom) Percent error between $\mathcal{A}_{\nu}$ calculated via the $u$FJC statistical mechanics framework and the inextensible Gaussian chain assumption as a function of $\nu$ in the domain $5\leq\nu\leq 3125$. The sharp dip and rise in the percent error curve takes place where $\mathcal{A}_{\nu}$ calculated via the $u$FJC statistical mechanics framework crosses over $\mathcal{A}_{\nu}$ calculated via the inextensible Gaussian chain assumption.}
	\label{fig:A_nu-gen-ufjc-model-framework-and-inextensible-Gaussian-chain-comparison}
\end{figure}
\\
Taking the root-mean-square end-to-end chain distance $r_{\nu}^{rms}$ as the reference end-to-end chain distance $r_{\nu}^{ref}$, the reference equilibrium chain stretch $\mathcal{A}_{\nu}$ directly falls out from the above
\begin{equation}
    \mathcal{A}_{\nu} = \sqrt{\left[\frac{1}{1+\nu e^{-\hat{\varepsilon}_{\nu}^{diss}(\lambda_{\nu}^{crit})}}\right]\frac{\mathscr{I}(5,\nu)}{\mathscr{I}(3,\nu)}}.
\end{equation}
Segment extensibility is fundamentally incorporated in the statistical mechanics underpinning the $\mathcal{A}_{\nu}$ calculation. In the case of inextensible chains described via Gaussian statistics, then it can be shown that $\mathcal{A}_{\nu} = 1/\sqrt{\nu}$. $\mathcal{A}_{\nu}$ calculated via the $u$FJC-based approach and the inextensible Gaussian chain assumption for a range of segment numbers associated with short, intermediately-long, and long chains are presented in the top panel in \cref{fig:A_nu-gen-ufjc-model-framework-and-inextensible-Gaussian-chain-comparison}. The percent error between the $\mathcal{A}_{\nu}$ calculations of the two approaches is provided in the bottom panel. For short chains, the inextensible Gaussian chain assumption poorly complies with the $u$FJC $\mathcal{A}_{\nu}$, with percent errors up to $\approx 10\%$ for short chains. The percent error only improves slightly for intermediately-long and long chains, converging from above to $\approx 0.1\%$. Since $\mathcal{A}_{\nu}$ is extensively used throughout the model in functions which capture the state of a chain, it is crucial for $\mathcal{A}_{\nu}$ to be precisely calculated.

Finally, as per \citet{guo2021micromechanics}, the reference segment stretch $\Lambda_{\nu}^{ref}$ is taken as the ratio of the reference segment length $l_{\nu}^{ref}$ with the equilibrium segment length $l_{\nu}^{eq}$, $\Lambda_{\nu}^{ref} = l_{\nu}^{ref}/l_{\nu}^{eq}$, and can be straightforwardly calculated as $\Lambda_{\nu}^{ref} = \lambda_{\nu}(\mathcal{A}_{\nu})$. Note that the quantity $\lambda_{\nu}/\Lambda_{\nu}^{ref} = l_{\nu}/l_{\nu}^{ref}$ is the ratio of the segment length $l_{\nu}$ with the reference segment length $l_{\nu}^{ref}$, and may be considered as an alternate segment stretch measure for modeling use. 

\begin{figure*}[t]
	\centering
	\subfloat[]{
		\includegraphics[width=0.495\textwidth]{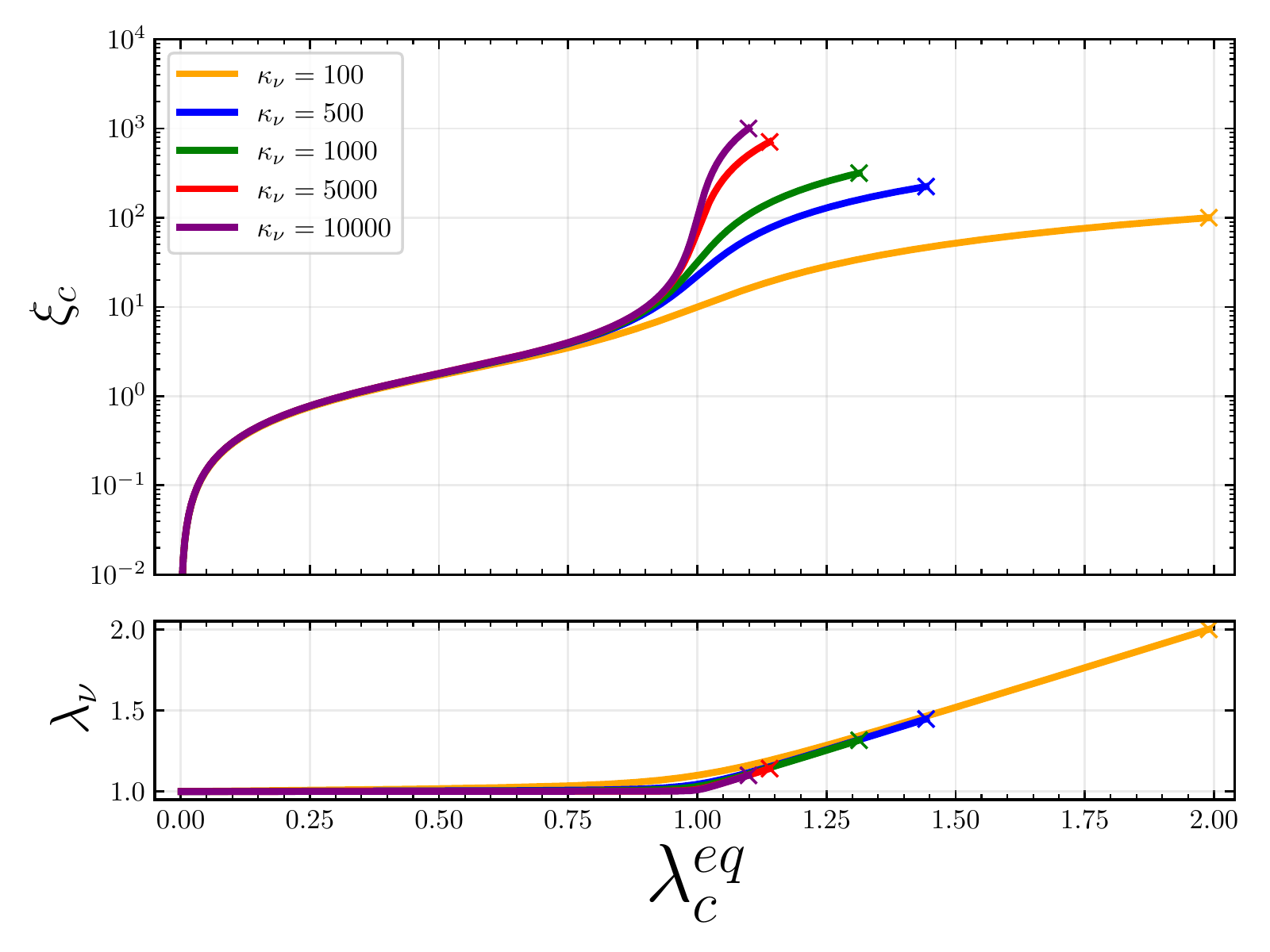}
		\label{fig:kappa_nu-xi_c-lmbda_nu-vs-lmbda_c_eq}}
	\subfloat[]{
		\includegraphics[width=0.495\textwidth]{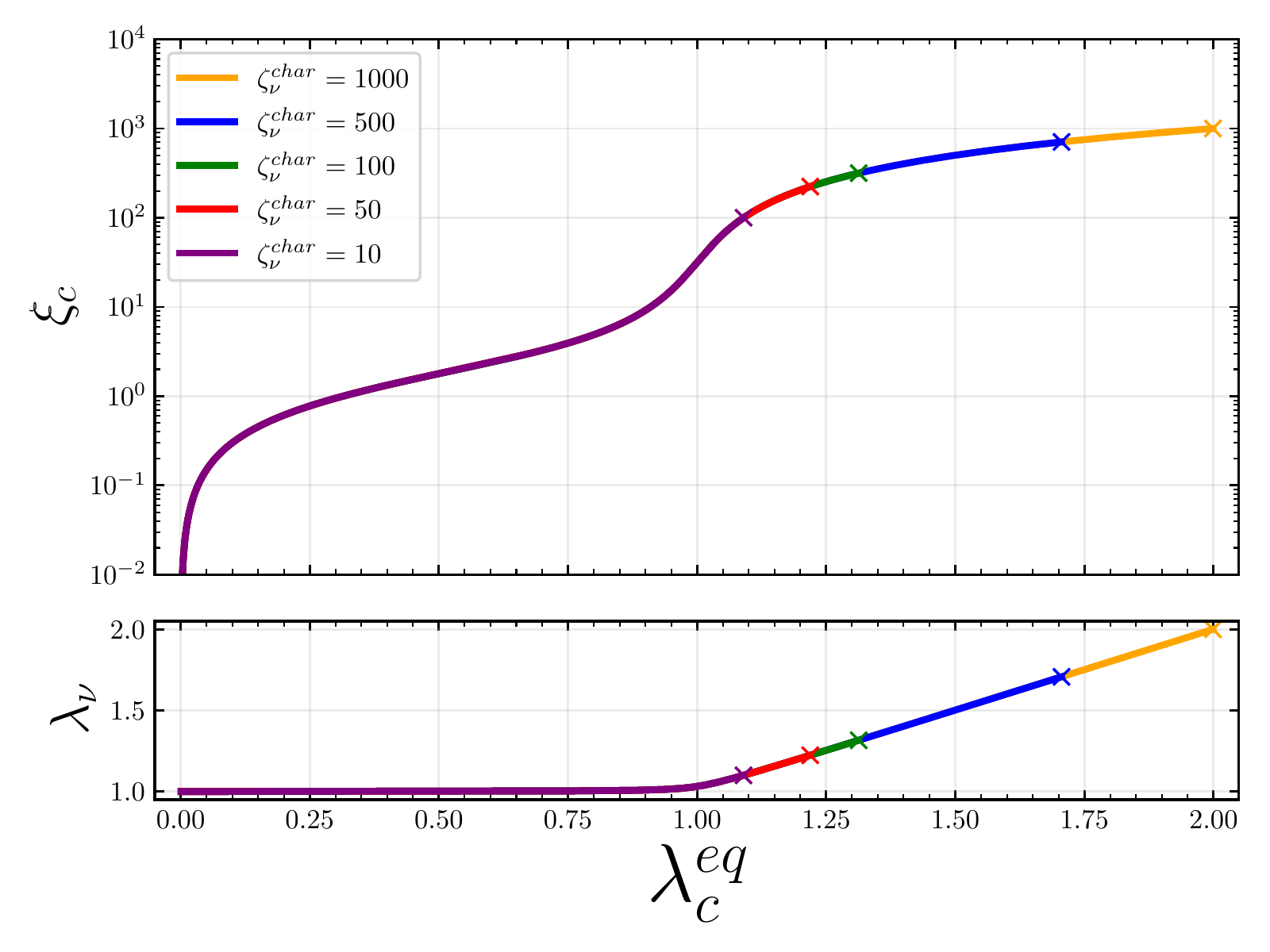}
		\label{fig:zeta_nu_char-xi_c-lmbda_nu-vs-lmbda_c_eq}}
	\caption{Single chain mechanical response for a chain with segments obeying the composite segment potential. (top) Nondimensional chain force $\xi_c$ as a function of equilibrium chain stretch $\lambda_c^{eq}$ in the domain $0\leq \lambda_c^{eq}\leq (\lambda_c^{eq})^{crit}$. The ``x'' markers denotes $\xi_c^{crit}$. (bottom) Segment stretch $\lambda_{\nu}$ as a function of $\lambda_c^{eq}$ in the domain $0\leq \lambda_c^{eq}\leq (\lambda_c^{eq})^{crit}$. The ``x'' markers denotes $\lambda_{\nu}^{crit}$. (a) Single chain mechanical response for $\zeta_{\nu}^{char} = 100$ and varying $\kappa$. (b) Single chain mechanical response for $\kappa = 1000$ and varying $\zeta_{\nu}^{char}$.}
	\label{fig:xi_c-lmbda_nu-vs-lmbda_cA_nu}
\end{figure*}

\begin{figure*}[t]
	\centering
	\subfloat[]{
		\includegraphics[width=0.495\textwidth]{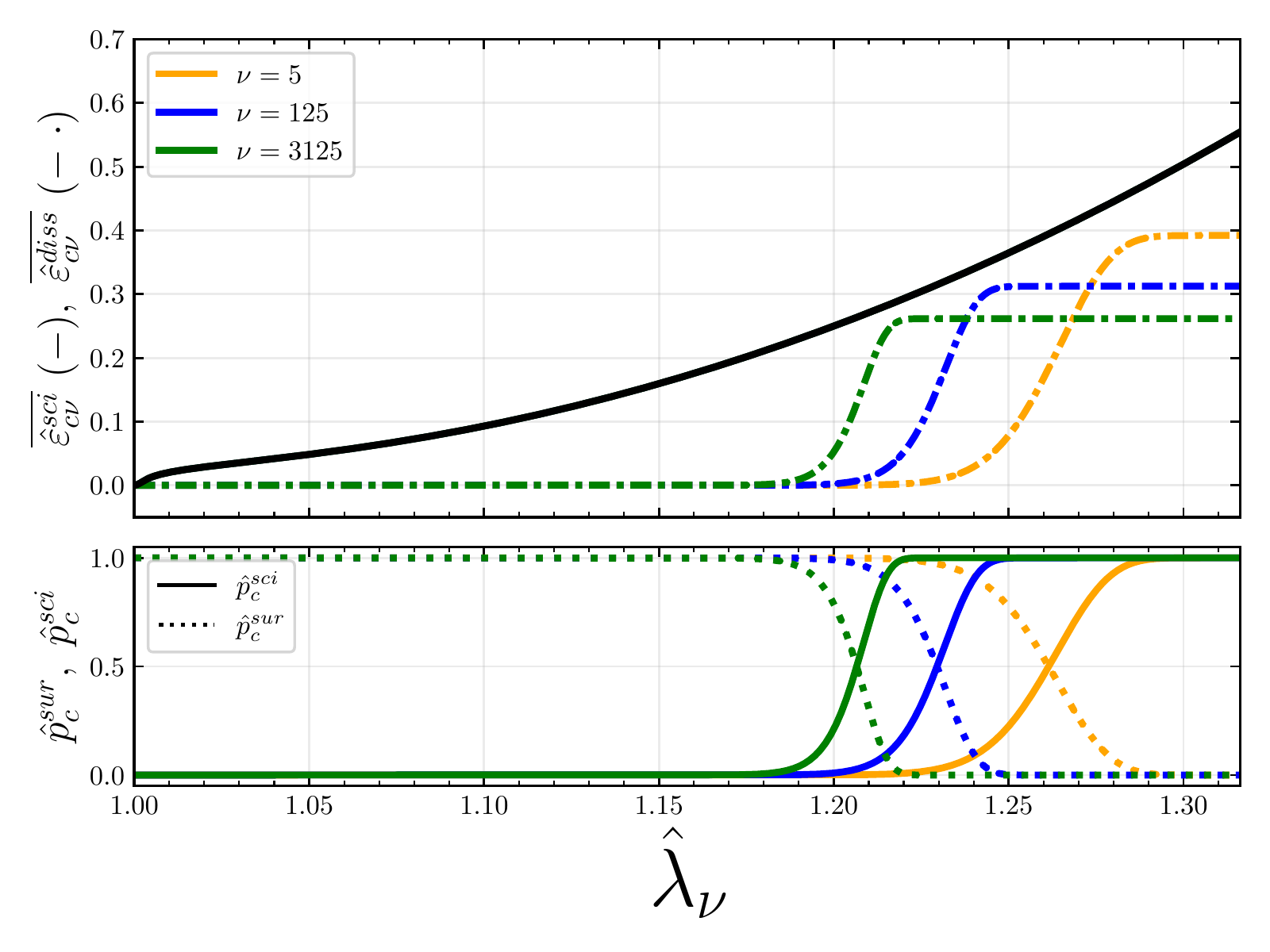}
		\label{fig:chain-scission-indicators-vs-lmbda_nu_hat}}
	\subfloat[]{
		\includegraphics[width=0.495\textwidth]{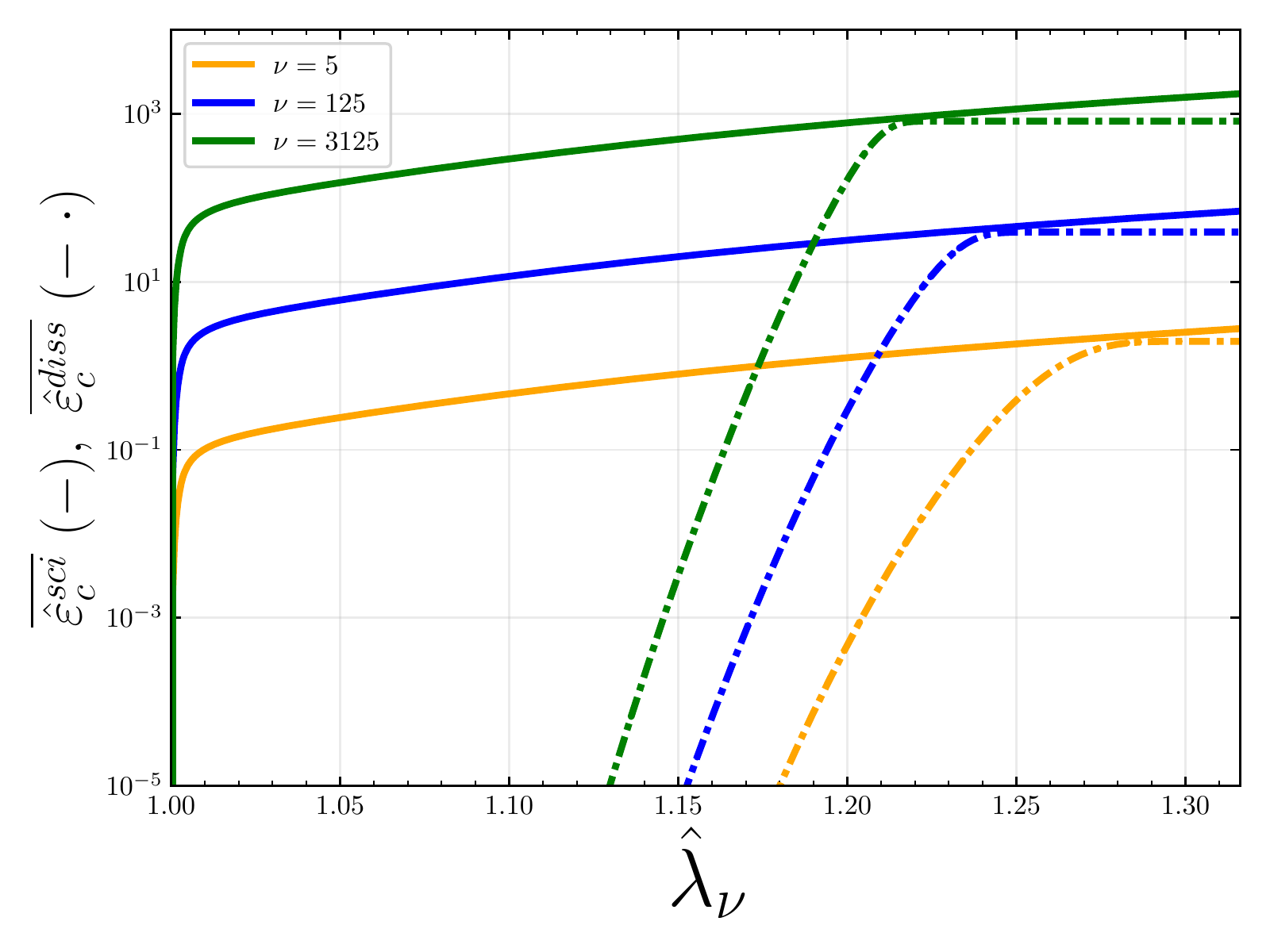}
		\label{fig:chain-scission-energy-vs-lmbda_nu_hat}}
	\caption{Chain scission energetics and probabilistics for short, intermediately-long, and long chains. Here, the short, intermediately-long, and long chains are composed of 5, 125, and 3125 segments, respectively. Also, $\zeta_{\nu}^{char} = 100$ and $\kappa_{\nu} = 1000$, as per the composite segment potential displayed in \cref{fig:composite-u_nu-figs}. (a, top) Nondimensional scaled chain scission energy per segment $\overline{\hat{\varepsilon}_{c\nu}^{sci}}$ and rate-independent nondimensional scaled dissipated chain scission energy per segment $\overline{\hat{\varepsilon}_{c\nu}^{diss}}$ as a function of applied segment stretch $\hat{\lambda}_{\nu}$ in the domain $1 \leq \hat{\lambda}_{\nu} \leq \lambda_{\nu}^{crit}$. $\overline{\hat{\varepsilon}_{c\nu}^{sci}} = \overline{\hat{\varepsilon}_{\nu}^{sci}}$ is represented by the black curve. (a, bottom) Rate-independent probability of chain scission $\hat{p}_c^{sci}$ and rate-independent probability of chain survival $\hat{p}_c^{sur}$ as a function of $\hat{\lambda}_{\nu}$ in the domain $1 \leq \hat{\lambda}_{\nu} \leq \lambda_{\nu}^{crit}$. (b) Nondimensional scaled chain scission energy $\overline{\hat{\varepsilon}_c^{sci}}$ and rate-independent nondimensional scaled dissipated chain scission energy $\overline{\hat{\varepsilon}_c^{diss}}$ as a function of applied segment stretch $\hat{\lambda}_{\nu}$ in the domain $1 \leq \hat{\lambda}_{\nu} \leq \lambda_{\nu}^{crit}$.}
	\label{fig:chain-scission-energetics-probabilistics}
\end{figure*}

\section{Results and discussion} \label{sec:single-chain-model-behavior-results}

At this point, the probabilistic description of single chain rupture is fully established. In light of the rupture framework, the mechanical response of the chain is presented and described. Then, implications of the chain scission framework in the context of polymer chain damage and fracture models are discussed. Chain rupture behavior for short, intermediately-long, and long chains is then investigated. Finally, model validation is achieved by fitting experimental single chain atomic force microscopy tensile test results to the $u$FJC framework with the composite segment potential. Using the fitted model, the dissipated energy and probability associated with chain scission are presented for each chain involved in the AFM tensile tests, providing deeper insight to the experimental results.

\subsection{Single chain mechanical response} \label{subsec:single-chain-mechanical-response}

The mechanical response of a chain with segments obeying the composite potential is presented in \cref{fig:xi_c-lmbda_nu-vs-lmbda_cA_nu}. This figure presents the chain mechanical response only for states where chain survival is probable, i.e., for chain states inclusively between the undisturbed force-free state ($\lambda_c^{eq} = 0$) and the critical state ($\lambda_c^{eq} = (\lambda_c^{eq})^{crit}$). In \cref{fig:kappa_nu-xi_c-lmbda_nu-vs-lmbda_c_eq}, $\zeta_{\nu}^{char}$ is held fixed while $\kappa_{\nu}$ is permitted to vary, and vice versa in \cref{fig:zeta_nu_char-xi_c-lmbda_nu-vs-lmbda_c_eq}. In order to calculate the nondimensional force response, \cref{eq:segment-stretch-function} was used with \cref{eq:nondim-chain-force}.

The $u$FJC composite potential chain mechanical response compares quite well with other extensible chain mechanical response results \citep{mao2017rupture, li2020variational, mulderrig2021affine, buche2021chain}. For low equilibrium chain stretch $\lambda_c^{eq}$ values, the chain mechanical response exhibits compliant, entropically-dominated behavior that is independent of $\zeta_{\nu}^{char}$ and $\kappa_{\nu}$. As the chain is stretched near its inextensible equilibrium contour length, $0.75\leq\lambda_c^{eq}\leq 1$, segments begin to stretch, and enthalpic contributions begin to emerge. For increasing $\kappa_{\nu}$, these enthalpic contributions dominate more and more as $\lambda_c^{eq} \rightarrow 1$, and the chain becomes increasingly stiffer and stiffer. As the chain is stretched past its inextensible equilibrium contour length, $\lambda_c^{eq} > 1$, the chain mechanical response exhibits enthalpically-dominated stiffening behavior as the segments continue to stretch more and more. For increasing $\kappa_{\nu}$, $\xi_c^{crit}$ increases while $\lambda_{\nu}^{crit}$ and $(\lambda_c^{eq})^{crit}$ decreases. Meanwhile, as $\zeta_{\nu}^{char}$ increases, $\lambda_{\nu}^{crit}$, $(\lambda_c^{eq})^{crit}$, and $\xi_c^{crit}$ all increase. However, prior to the critical point, the chain mechanical response remains identical to the case with smaller $\zeta_{\nu}^{char}$.

\subsection{Implications of the chain scission framework} \label{subsec:chain-scission-framework-implications}

In light of polymer chain damage and fracture modeling, two important implications can be highlighted from the chain scission framework developed in \cref{subsec:chain-scission}. First, this framework can account for the damage of polymer networks composed of chains with a polydispersity in segment number. The dependence of chain scission behavior with respect to segment number is further examined in \cref{subsec:single-chain-scission-results}. Second, this chain scission framework can now be directly incorporated in polymer chain damage and fracture models, such as the Lake and Thomas theory of polymer network fracture \citep{lake1967strength}. In these models, the criterion for chain scission and associated energy dissipation fall into one of several categories -- an upper bound, a lower bound, or one of several intermediate limits -- as follows:
\begin{enumerate}
    \item\textbf{Upper bound of chain scission:} All of the segments in the chain are assumed to be identically stretched up to the point of dissociation before one of the segments becomes ruptured. The dissipated scission energy for a chain, $\hat{\varepsilon}_c^{diss}/\beta$, is taken to be approximately equal to $\nu E_{\nu}^{char}$. Clearly, this criterion neglects both thermal fluctuations and the reduction in scission activation energy under an increasing chain force. This criterion was first introduced by \citet{lake1967strength}, and has since been widely used \citep{mao2017rupture, talamini2018progressive, vernerey2018statistical, mao2018theory, li2020variational, mulderrig2021affine, lamont2021rate, arunachala2021energy}.
    \item\textbf{Lower bound of chain scission:} Considering the rapid time scale of thermal fluctuations, it is assumed that the chain becomes ruptured when the energy needed for a single segment to dissociate is imparted to the chain as a whole. $\hat{\varepsilon}_c^{diss}/\beta$ is taken to be approximately equal to $E_{\nu}^{char}$ \citep{lei2020study, lei2021recent}. This criterion neglects both the probabilistic nature of thermal fluctuations and the reduction in scission activation energy under an increasing chain force.
    \item\textbf{Chain scission at maximum segment force:} All of the segments in the chain are assumed to be identically stretched up to the point when the chain force equals the maximum segment force (as determined by the particular anharmonic segment potential in use), at which point one of the segments becomes ruptured \citep{dal2009micro, tehrani2017effect, buche2021chain}. Note that the point when the chain force equals the maximum segment force has been referred to in this work as the critical point, with nondimensional maximum segment force equal to $\xi_c^{crit}$. For segments described by the Morse potential and the composite potential from this work, $\hat{\varepsilon}_c^{diss}/\beta$ is taken to be approximately equal to $\nu E_{\nu}^{char}/4$ and $\nu E_{\nu}^{char}/2$, respectively. This criterion neglects the probabilistic nature thermal fluctuations can play in a scission reaction.
    \item\textbf{Chain scission at a previously measured chain force of rupture:} As proscribed by \citet{wang2019quantitative} and used in \citet{zhang2021relationship}, all of the segments in the chain are assumed to be identically stretched up to the point when the chain force equals a pre-defined force of rupture before one of the segments becomes ruptured. This dissipated energy of chain scission, $\hat{\varepsilon}_c^{diss}/\beta$, is calculated as the work done to extend a polymer chain from a force-free equilibrium state up to this pre-calculated or pre-measured force of rupture $\xi_c^{sci}$ (in nondimensional terms). $\xi_c^{sci}$ can be calculated via density functional theory \citep{grandbois1999strong, beyer2000mechanical} or measured by single-chain atomic force microscopy tensile test results. Since $\xi_c^{sci}$ is typically less than the maximum segment force $\xi_c^{crit}$ from the prior criterion, $\hat{\varepsilon}_c^{diss}/\beta$ for a chain obeying this criterion will be a positive energy value less than $\hat{\varepsilon}_c^{diss}/\beta$ taken at $\xi_c^{crit}$. 
    \item\textbf{Thermally-driven stochastic force-activated chain scission:} Segment rupture is stochastically driven by thermal excitations supplying the energy needed to overcome a force-dependent activation energy barrier of scission. All such rate-dependent bulk and interfacial polymer damage models intrinsically incorporate these features to account for chain rupture. Recently, several models, including this work, have begun to incorporate the effects of segment extensibility in the rate-dependent damage model context \citep{yang2020multiscale, guo2021micromechanics,feng2022rigorous, lei2022multiscale}. Furthermore, only \citet{guo2021micromechanics} and this work (in \cref{eq:rate-dependent-Xi_cnu-rate-equation}) have defined a rate equation for chain scission energy dissipated upon rupture. Notably, this work also proposes a rate-independent polymer damage framework that fully complies with the assumptions underlying this criterion. $\hat{\varepsilon}_c^{diss}/\beta$ for a chain obeying this criterion will be a positive energy value less than $\hat{\varepsilon}_c^{diss}/\beta$ taken at the maximum segment force $\xi_c^{crit}$ from the third criterion.
\end{enumerate}
Excluding the last criterion, the above criteria are often implemented in computational models via calculating a critical segment stretch of rupture $\lambda_{\nu}^{sci}$. The chain is considered ruptured if $\hat{\lambda}_{\nu} \geq \lambda_{\nu}^{sci}$. The nondimensional dissipated chain scission energy is then calculated as $\hat{\varepsilon}_c^{diss} = \nu u_{\nu}(\hat{\lambda}_{\nu} = \lambda_{\nu}^{sci})$ or $\hat{\varepsilon}_c^{diss} = \nu \psi_{c\nu}(\hat{\lambda}_{\nu} = \lambda_{\nu}^{sci})$. An exception to this implementation is found in \citet{dal2009micro}, which considered a segment as ruptured when a numerical discontinuity arose in the $\lambda_{\nu}$ calculation. Note that this numerical discontinuity arose only when $\hat{\xi}_c = \xi_c^{crit}$.

\subsection{Single chain scission} \label{subsec:single-chain-scission-results}

Using the chain scission framework developed in \cref{subsec:chain-scission}, the rupture behavior of short, intermediately-long, and long chains can be probed. \cref{fig:chain-scission-energetics-probabilistics} displays the evolution of $\hat{\varepsilon}_{c\nu}^{sci}$, rate-independent $\hat{\varepsilon}_{c\nu}^{diss}$, $\hat{\varepsilon}_c^{sci}$, and rate-independent $\hat{\varepsilon}_c^{diss}$ in nondimensional scaled form ($\overline{\hat{\varepsilon}_{c\nu}^{sci}} = \hat{\varepsilon}_{c\nu}^{sci}/\zeta_{\nu}^{char}$, $\overline{\hat{\varepsilon}_{c\nu}^{diss}} = \hat{\varepsilon}_{c\nu}^{diss}/\zeta_{\nu}^{char}$, $\overline{\hat{\varepsilon}_c^{sci}} = \hat{\varepsilon}_c^{sci}/\zeta_{\nu}^{char}$, and $\overline{\hat{\varepsilon}_c^{diss}} = \hat{\varepsilon}_c^{diss}/\zeta_{\nu}^{char}$) for a short chain of $\nu=5$ segments, an intermediately-long chain of $\nu=125$ segments, and a long chain of $\nu=3125$ segments. The associated rate-independent probabilities $\hat{p}_c^{sci}$ and $\hat{p}_c^{sur}$ are also displayed. Similarly to \cref{fig:segment-scission-indicators-vs-lmbda_nu_hat}, these measures are displayed for chains between the undisturbed, force free state and the critical state. The impact of chain segment number to chain rupture behavior is strikingly clear: according to the bottom panel in \cref{fig:chain-scission-indicators-vs-lmbda_nu_hat}, chain rupture initiates at smaller applied segment stretches $\hat{\lambda}_{\nu}$ for longer chains. This phenomenon materializes strictly from the probabilistic treatment of chain rupture in \cref{eq:chain-scission-survival-probabilities}, which is built upon the principle that chain rupture takes place when any one segment along its backbone ruptures. However, even though it is favorable for longer chains to rupture at smaller $\hat{\lambda}_{\nu}$, this does not mean that longer chains will undergo rupture before shorter chains within a deformed polymer chain network as $\hat{\lambda}_{\nu}$ is a normalized stretch quantity. Regarding the scission energy dissipated during chain rupture, \cref{fig:chain-scission-energy-vs-lmbda_nu_hat} shows that $\hat{\varepsilon}_c^{diss}$ at the critical state increases with increasing segment number, as expected. However, on a per-segment basis, $\hat{\varepsilon}_{c\nu}^{diss}$ at the critical state actually decreases as the number of segments in a chain increases, as shown in the top panel in \cref{fig:chain-scission-indicators-vs-lmbda_nu_hat}.

\begin{figure*}[t]
	\centering
	\subfloat[]{
		\includegraphics[width=0.495\textwidth]{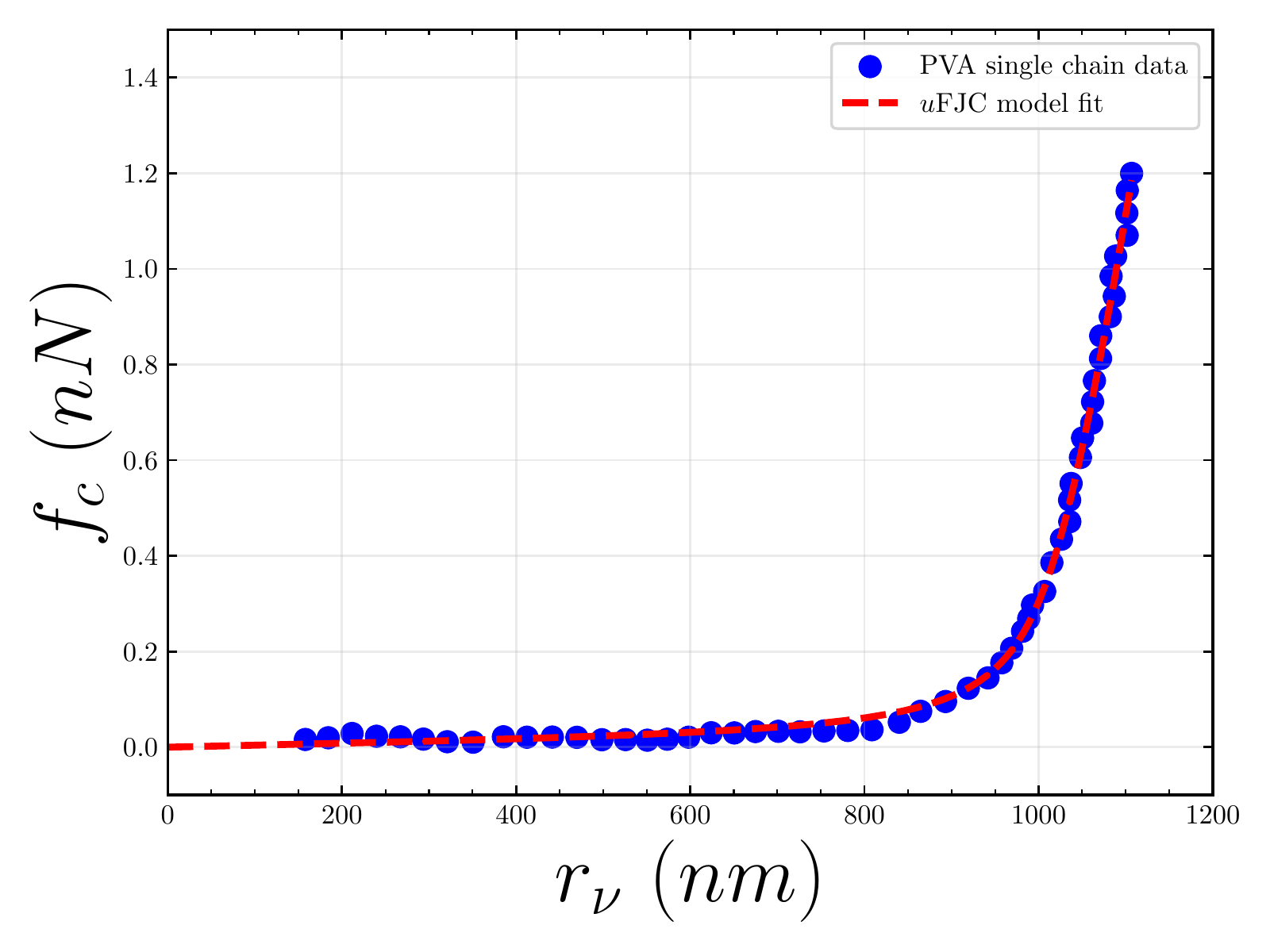}
		\label{fig:hugel-et-al-pva-c-c-chain-a-intgr_nu-f_c-vs-r_nu-gen-uFJC-curve-fit}}
	\subfloat[]{
		\includegraphics[width=0.495\textwidth]{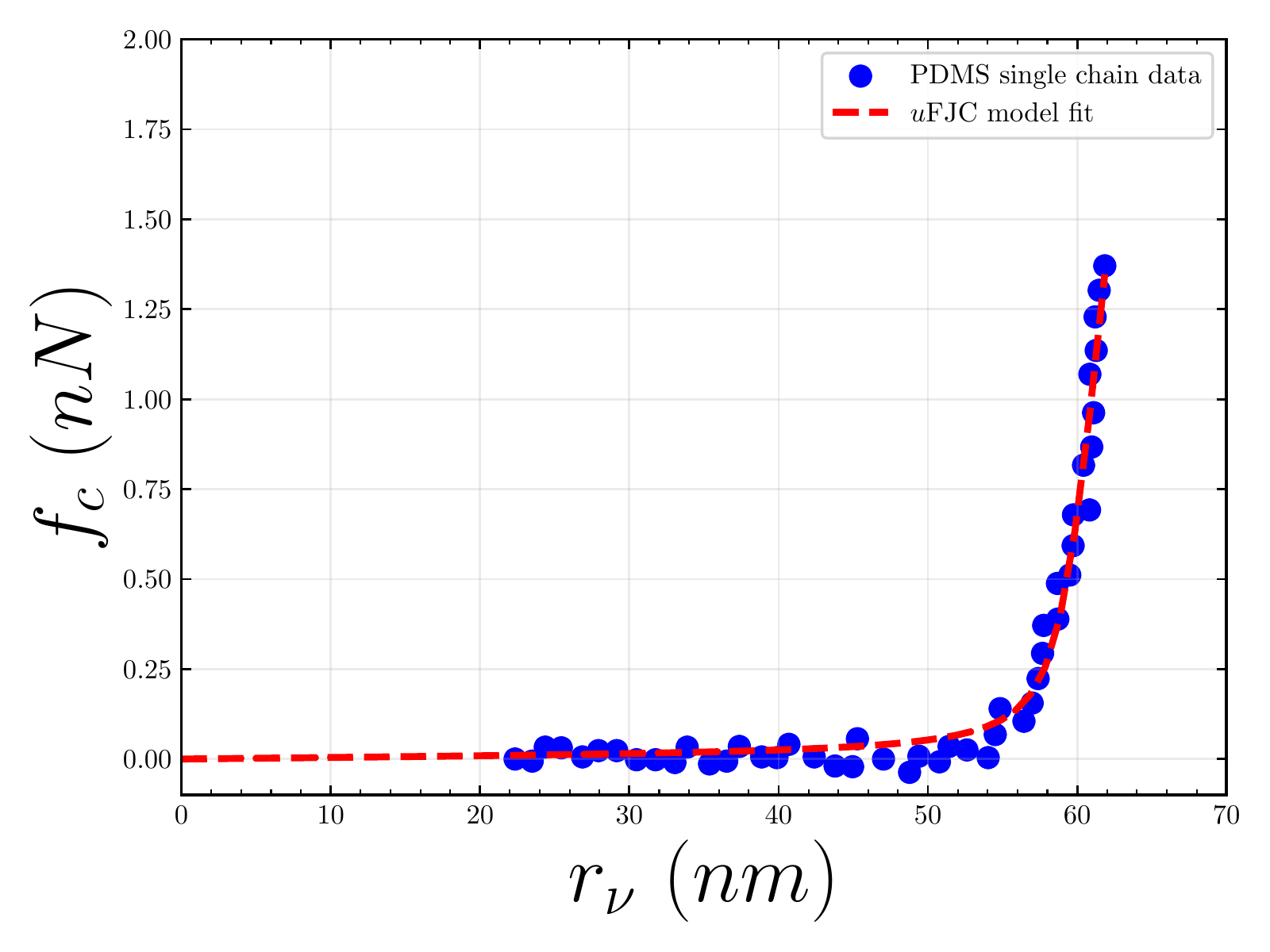}
		\label{fig:al-maawali-et-al-pdms-si-o-chain-a-intgr_nu-f_c-vs-r_nu-gen-uFJC-curve-fit}}
	\caption{Single chain AFM tensile test data -- chain force applied via the AFM $f_c$ as a function of the AFM-measured end-to-end chain distance $r_{\nu}$ -- fit to the $u$FJC model with the composite segment potential. (a) PVA single chain AFM tensile test data, provided in Fig. 5 in \citet{hugel2001elasticity}, alongside the corresponding $u$FJC model fit. (b) PDMS single chain AFM tensile test data, provided in Fig. 3 in \citet{al2001study}, alongside the corresponding $u$FJC model fit.}
	\label{fig:intgr_nu-f_c-vs-r_nu-gen-uFJC-curve-fit}
\end{figure*}

\begin{figure}[t]
	\centering
	\includegraphics[width=0.75\textwidth]{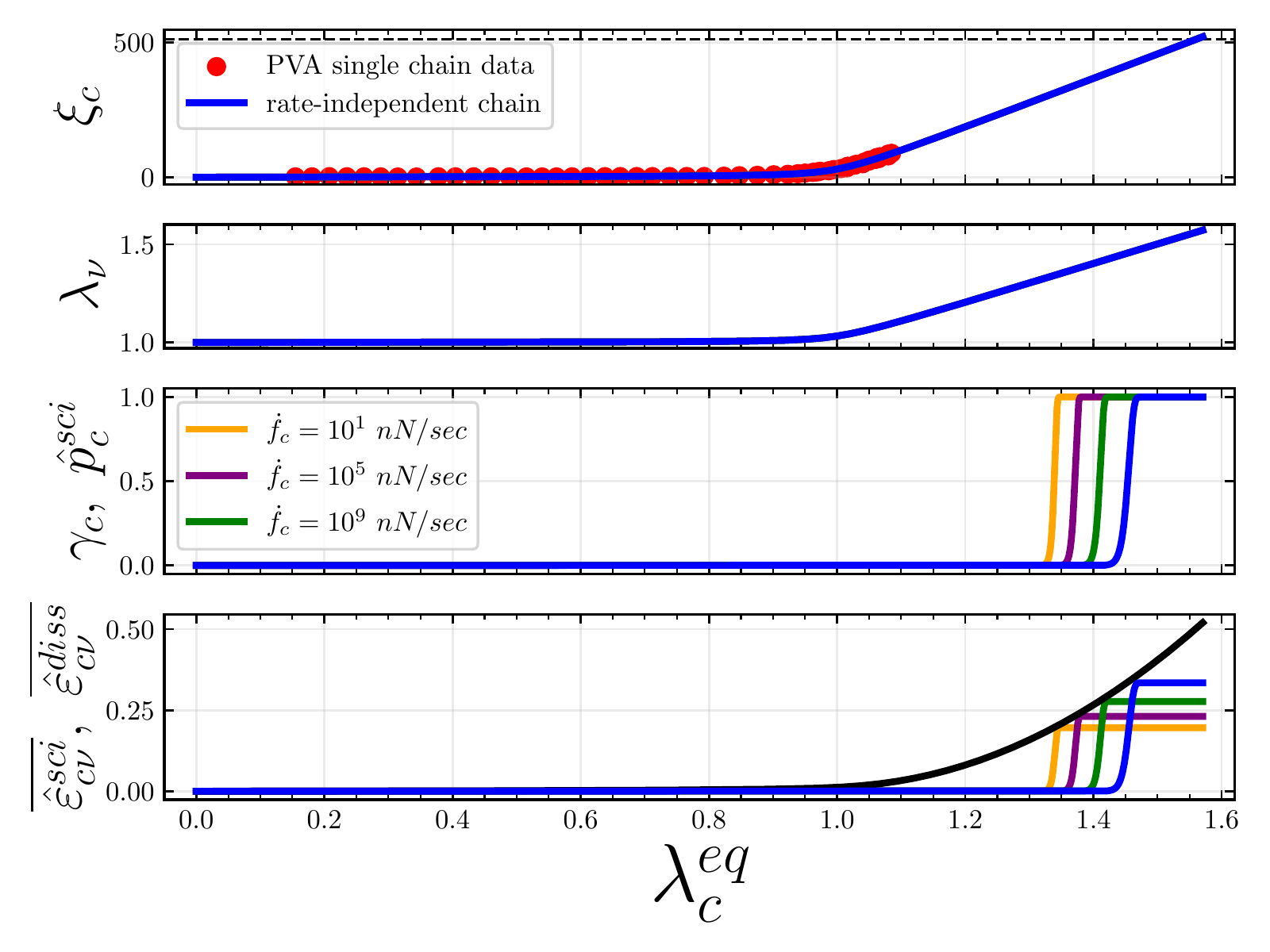}
	\caption{PVA single chain mechanical response, probability of chain scission, and dissipated chain scission energy from rate-dependent and rate-independent single chain AFM tensile test simulations. Each tensile test simulation initially begins in the undisturbed, force-free chain state. Tension is then applied to the chain up to the critical chain state. Several different loading rates are investigated. Fundamental segment and chain parameters are taken from the $u$FJC model fit presented in \cref{fig:hugel-et-al-pva-c-c-chain-a-intgr_nu-f_c-vs-r_nu-gen-uFJC-curve-fit}. (top) Nondimensional chain force $\xi_c$ as a function of equilibrium chain stretch $\lambda_c^{eq}$, which produce identical curves for the rate-dependent and rate-independent cases. The experimental PVA single chain tensile test data from Fig. 5 in \citet{hugel2001elasticity} is presented in nondimensional form here. The horizontal black dashed line is the reported maximum nondimensional chain force a single C-C bond can sustain as calculated from density functional theory calculations in \citet{beyer2000mechanical}. (upper-middle) Segment stretch $\lambda_{\nu}$ as a function of $\lambda_c^{eq}$, which produce identical curves for the rate-dependent and rate-independent cases. (lower-middle) Rate-dependent and rate-independent probability of chain scission, $\gamma_c$ and $\hat{p}_c^{sci}$, respectively, as a function of $\lambda_c^{eq}$. (bottom) Rate-dependent and rate-independent nondimensional scaled dissipated chain scission energy per segment $\overline{\hat{\varepsilon}_{c\nu}^{diss}}$ as a function of $\lambda_c^{eq}$, along with the nondimensional scaled chain scission energy per segment $\overline{\hat{\varepsilon}_{c\nu}^{sci}}$ as a function of $\lambda_c^{eq}$. $\overline{\hat{\varepsilon}_{c\nu}^{sci}} = \overline{\hat{\varepsilon}_{\nu}^{sci}}$ is represented by the black curve. The rate-dependent and rate-independent $\overline{\hat{\varepsilon}_{c\nu}^{diss}}$ are represented by the non-black colored curves.}
	\label{fig:hugel-et-al-pva-c-c-chain-a-rate-independent-and-rate-dependent-chains-vs-lmbda_c_eq}
\end{figure}

\begin{figure}[t]
	\centering
	\includegraphics[width=0.75\textwidth]{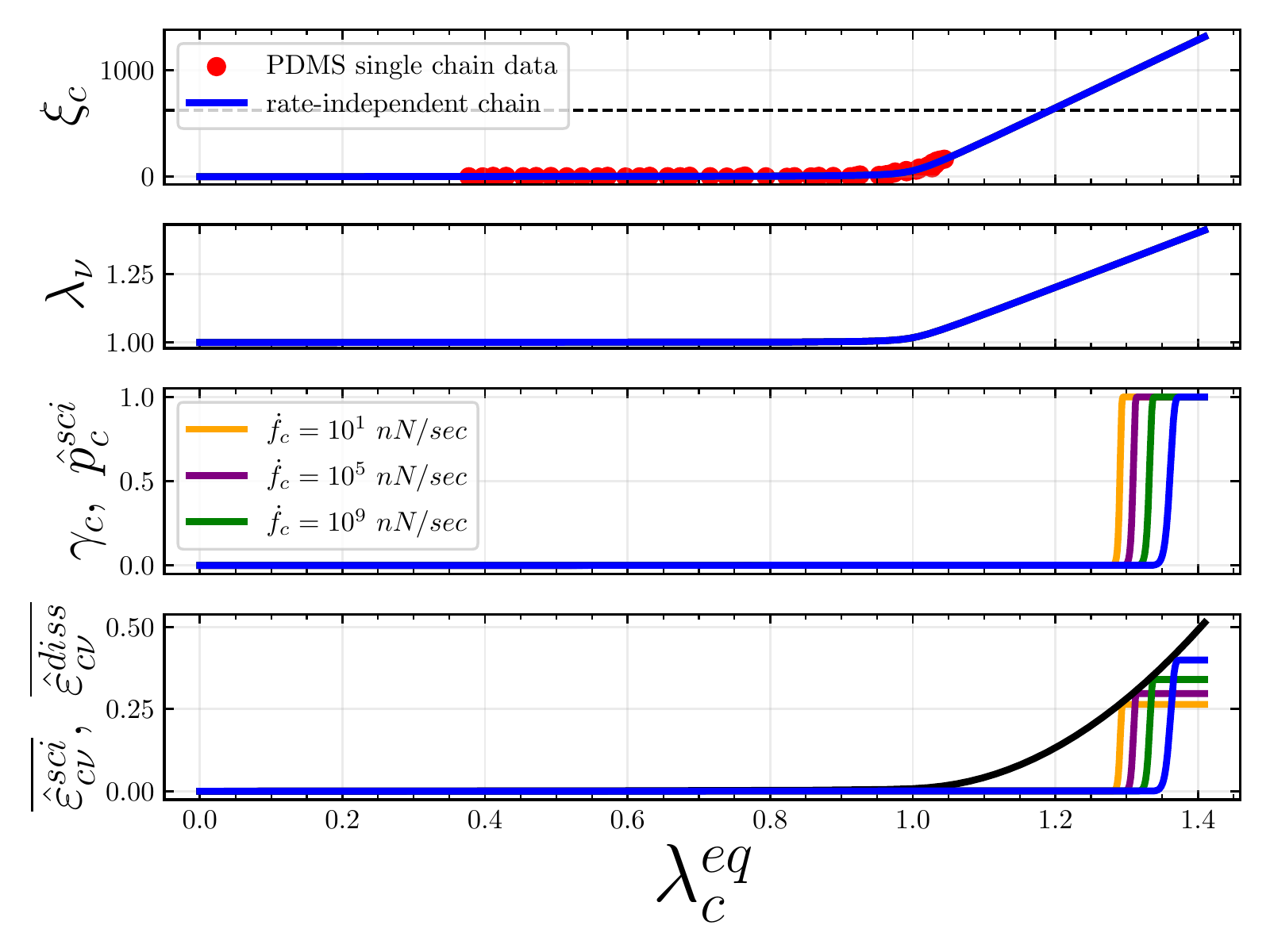}
	\caption{PDMS single chain mechanical response, probability of chain scission, and dissipated chain scission energy from rate-dependent and rate-independent single chain AFM tensile test simulations. Each tensile test simulation initially begins in the undisturbed, force-free chain state. Tension is then applied to the chain up to the critical chain state. Several different loading rates are investigated. Fundamental segment and chain parameters are taken from the $u$FJC model fit presented in \cref{fig:al-maawali-et-al-pdms-si-o-chain-a-intgr_nu-f_c-vs-r_nu-gen-uFJC-curve-fit}. (top) Nondimensional chain force $\xi_c$ as a function of equilibrium chain stretch $\lambda_c^{eq}$, which produce identical curves for the rate-dependent and rate-independent cases. The experimental PDMS single chain tensile test data from Fig. 3 in \citet{al2001study} is presented in nondimensional form here. The horizontal black dashed line is the reported maximum nondimensional chain force a single Si-O bond can sustain as calculated from density functional theory calculations in \citet{beyer2000mechanical}. (upper-middle) Segment stretch $\lambda_{\nu}$ as a function of $\lambda_c^{eq}$, which produce identical curves for the rate-dependent and rate-independent cases. (lower-middle) Rate-dependent and rate-independent probability of chain scission, $\gamma_c$ and $\hat{p}_c^{sci}$, respectively, as a function of $\lambda_c^{eq}$. (bottom) Rate-dependent and rate-independent nondimensional scaled dissipated chain scission energy per segment $\overline{\hat{\varepsilon}_{c\nu}^{diss}}$ as a function of $\lambda_c^{eq}$, along with the nondimensional scaled chain scission energy per segment $\overline{\hat{\varepsilon}_{c\nu}^{sci}}$ as a function of $\lambda_c^{eq}$. $\overline{\hat{\varepsilon}_{c\nu}^{sci}} = \overline{\hat{\varepsilon}_{\nu}^{sci}}$ is represented by the black curve. The rate-dependent and rate-independent $\overline{\hat{\varepsilon}_{c\nu}^{diss}}$ are represented by the non-black colored curves.}
	\label{fig:al-maawali-et-al-pdms-si-o-chain-a-rate-independent-and-rate-dependent-chains-vs-lmbda_c_eq}
\end{figure}

\subsection{Comparison with experiments} \label{subsec:experimental-fits}

Now that the theory underpinning the $u$FJC framework with the composite segment potential is completely established, model validation via fitting single chain mechanical response data from AFM tensile tests is desired. These data also provide a depth of information that we will try to extract with our framework. Polymer chains composed of either single carbon-carbon bonds (C-C) or single silicon-oxygen bonds (Si-O) -- two common backbone bonds found in many elastic polymer chains -- are desired for model fitting. The characteristic bond potential energy scale $E_b^{char}$ and equilibrium bond length $l_b^{eq}$ for the C-C bond are experimentally found to be $370.3~kJ/mol$ \citep{luo2007comprehensive} and $1.524~\si{\angstrom}$ \citep{allen1987tables}, respectively. Likewise, $E_b^{char}$ and $l_b^{eq}$ for the Si-O bond are experimentally found to be $444.0~kJ/mol$ \citep{wiberg2001inorganic} and $1.645~\si{\angstrom}$ \citep{allen1987tables}, respectively. These experimentally-derived bond parameter values are used within the $u$FJC model to fit AFM single chain force versus extension data. Note that the transition from the bond level to the segment level that necessarily takes place in the $u$FJC model fit follows the theory outlined in \cref{sec:bonds-theory}. Through the model fit procedure, the number of bonds in a single segment $\nu_b$, the number of segments in the chain $\nu$, the nondimensional segment stiffness $\kappa_{\nu}$, and the nondimensional characteristic segment potential energy scale $\zeta_{\nu}^{char}$ are obtained. To comply with the physical makeup of a polymer chain, $\nu_b$ and $\nu$ are restricted to integer values. Using single polyvinylamine (PVA) chain mechanical response data from \citet{hugel2001elasticity}, $\kappa_{\nu} = 912.2$ and $\zeta_{\nu}^{char} = 298.9$ are found, along with $\nu_b = 2$ and $\nu = 3347$. Using single polydimethylsiloxane (PDMS) chain mechanical response data from \citet{al2001study}, $\kappa_{\nu} = 3197.5$ and $\zeta_{\nu}^{char} = 537.6$ are found, along with $\nu_b = 3$ and $\nu = 120$. \cref{fig:intgr_nu-f_c-vs-r_nu-gen-uFJC-curve-fit} presents the AFM single chain tensile test force versus extension results along with the respective $u$FJC model fits. In order to yield these well-fit results, AFM tensile tests that forced the chain to its enthalpic-dominated response regime were sought after, which \citet{hugel2001elasticity} and \citet{al2001study} both provide. Do note that the PVA and PDMS chains here were pulled via an AFM tip until a detachment event occurred, where either the chain detached from the AFM tip, the chain detached from a fixed substrate, or a backbone bond ruptured in the chain.

To elucidate deeper insight into the molecular physics taking place in these single chain AFM tensile tests, rate-dependent and rate-independent single chain AFM tensile test calculations were undertaken using each $u$FJC model fit. The rate-dependent and rate-independent single PVA chain mechanical response, probability of chain scission $\gamma_c$ and $\hat{p}_c^{sci}$ (respectively), and nondimensional dissipated chain scission energy per segment $\hat{\varepsilon}_{c\nu}^{diss}$ (provided in scaled form) are presented as a function of $\lambda_c^{eq}$ in \cref{fig:hugel-et-al-pva-c-c-chain-a-rate-independent-and-rate-dependent-chains-vs-lmbda_c_eq}. The PVA chain mechanical response from the AFM tensile test in \citet{hugel2001elasticity} (as presented in \cref{fig:hugel-et-al-pva-c-c-chain-a-intgr_nu-f_c-vs-r_nu-gen-uFJC-curve-fit}) is included in \cref{fig:hugel-et-al-pva-c-c-chain-a-rate-independent-and-rate-dependent-chains-vs-lmbda_c_eq} in nondimensional form. Analogous single PDMS chain results are presented in \cref{fig:al-maawali-et-al-pdms-si-o-chain-a-rate-independent-and-rate-dependent-chains-vs-lmbda_c_eq}. Several chain force loading rates $\dot{f}_c$ are presented, and the chain is pulled up to the critical loading state as per the $u$FJC model fit. $\dot{f}_c = 10~nN/sec$ is a typical AFM force loading rate for these single chain AFM tensile test experiments \citep{beyer2000mechanical, ribas2012covalent}.

Comparison of the AFM tensile test data and the modeled scission probabilistics presented in the top and lower-middle panels of \cref{fig:hugel-et-al-pva-c-c-chain-a-rate-independent-and-rate-dependent-chains-vs-lmbda_c_eq} and \cref{fig:al-maawali-et-al-pdms-si-o-chain-a-rate-independent-and-rate-dependent-chains-vs-lmbda_c_eq} strongly suggests that a backbone bond in the PVA or PDMS chains did not undergo scission at the moment of detachment. This corroborates the intuition of \citet{hugel2001elasticity} and \citet{al2001study}: \citet{hugel2001elasticity} assumed that covalent bonds between PVA chains and an epoxy-functionalized glass surface ruptured during the AFM tensile test, while \citet{al2001study} asserted that PDMS chains detached from the AFM tip. Another chain scission-based comparison to make is with density functional theory: the black dashed line in the top panel in \cref{fig:hugel-et-al-pva-c-c-chain-a-rate-independent-and-rate-dependent-chains-vs-lmbda_c_eq} and \cref{fig:al-maawali-et-al-pdms-si-o-chain-a-rate-independent-and-rate-dependent-chains-vs-lmbda_c_eq} corresponds to the maximum force that a C-C or Si-O bond can respectively withstand as determined via the density functional theory calculations performed in \citet{beyer2000mechanical}. The maximum bond force calculated in \citet{beyer2000mechanical} is analogous to the critical bond force in this work. The critical C-C bond force calculated from the model fit with the PVA chain from \citet{hugel2001elasticity} corresponds quite well with maximum C-C bond force from \citet{beyer2000mechanical}, but the critical Si-O bond force calculated from the model fit with the PDMS chain from \citet{al2001study} is about double the maximum Si-O bond force from \citet{beyer2000mechanical}.

The relationship between rate-dependent chain scission to rate-independent chain scission is clearly implied from the scission probabilities and dissipated chain scission energies presented in the lower-middle and bottom panels in \cref{fig:hugel-et-al-pva-c-c-chain-a-rate-independent-and-rate-dependent-chains-vs-lmbda_c_eq} and \cref{fig:al-maawali-et-al-pdms-si-o-chain-a-rate-independent-and-rate-dependent-chains-vs-lmbda_c_eq}. As $\dot{f}_c$ increases larger and larger, the initiation of non-zero $\gamma_c$ and non-zero $\hat{\varepsilon}_{c\nu}^{diss}$ occurs at larger and larger $\lambda_c^{eq}$, and the critical $\hat{\varepsilon}_{c\nu}^{diss}$ increases in value. Essentially, as $\dot{f}_c$ increases, rate-dependent $\gamma_c$ and $\hat{\varepsilon}_{c\nu}^{diss}$ curves gradually approach their rate-independent counterparts. In effect, this clearly implies that rate-independent chain scission is equivalent to the limiting case of rate-dependent chain scission under an infinitely-large $\dot{f}_c$. Knowing this, rate-independent chain scission can be used to approximate the behavior of chains under massively large loading rates, and it can be used as a means to understand the theoretical limits of rate-dependent $\gamma_c$ and $\hat{\varepsilon}_{c\nu}^{diss}$.

Finally, $\hat{\varepsilon}_{c\nu}^{diss}/\beta$ at the critical chain state for both rate-dependent and rate-independent chain scission is well below the characteristic segment potential energy scale $E_{\nu}^{char}$. According to the bottom panel in \cref{fig:hugel-et-al-pva-c-c-chain-a-rate-independent-and-rate-dependent-chains-vs-lmbda_c_eq}, for a rate-dependent PVA chain with typical AFM force loading rate of $\dot{f}_c = 10~nN/sec$ and for a rate-independent PVA chain, the critical $\hat{\varepsilon}_{c\nu}^{diss}$ equals $0.196\times\zeta_{\nu}^{char}$ and $0.335\times\zeta_{\nu}^{char}$, respectively. The analogous values for a PDMS chain equal $0.264\times\zeta_{\nu}^{char}$ and $0.399\times\zeta_{\nu}^{char}$, respectively (as per the bottom panel in \cref{fig:al-maawali-et-al-pdms-si-o-chain-a-rate-independent-and-rate-dependent-chains-vs-lmbda_c_eq}). This result complies with the findings of \citet{wang2019quantitative}. Do note that $\gamma_c$, $\hat{p}_c^{sci}$, and $\hat{\varepsilon}_{c\nu}^{diss}$ presented in \cref{fig:hugel-et-al-pva-c-c-chain-a-rate-independent-and-rate-dependent-chains-vs-lmbda_c_eq} and \cref{fig:al-maawali-et-al-pdms-si-o-chain-a-rate-independent-and-rate-dependent-chains-vs-lmbda_c_eq} are statistical descriptions of an ensemble, while the experiments are obviously discrete events. If we truly consider an individual single chain in an AFM tensile test and find that the chain ruptures in this test at some $\lambda_c^{eq} = (\lambda_c^{eq})^{sci}$ corresponding to $\lambda_{\nu} = \lambda_{\nu}^{sci}$, then the probability of chain scission will instantly step from 0 to 1 at the moment of rupture. At the same time, $\hat{\varepsilon}_{c\nu}^{diss}$ will instantly step from $\hat{\varepsilon}_{c\nu}^{diss} = 0$ to $\hat{\varepsilon}_{c\nu}^{diss} = \hat{\varepsilon}_{c\nu}^{sci}(\lambda_{\nu}^{sci})$. Graphically speaking, $\overline{\hat{\varepsilon}_{c\nu}^{diss}}$ will jump from the zero line to the black curve at $\lambda_{\nu} = \lambda_{\nu}^{sci}$ in the bottom panel of \cref{fig:hugel-et-al-pva-c-c-chain-a-rate-independent-and-rate-dependent-chains-vs-lmbda_c_eq} and \cref{fig:al-maawali-et-al-pdms-si-o-chain-a-rate-independent-and-rate-dependent-chains-vs-lmbda_c_eq}.

\section{Conclusions} \label{sec:conclusion}

In this manuscript, we developed a chain rupture framework founded upon a freely-jointed chain model derived entirely via statistical mechanics principles accounting for arbitrary bond extensibility. An extended version of the $u$FJC model from \citet{buche2021chain} and \citet{buche2022freely} was first formulated and utilized to establish the statistical mechanics and thermodynamics foundation for the model. Using the principles of asymptotic matching, a simple, anharmonic, quasi-polynomial segment potential energy function was derived and verified. This potential energy function permitted a highly-accurate approximate analytical function to be derived that relates the chain stretch to the segment stretch, which is of critical importance for efficient computations. By considering scission as a stochastic, load-dependent process driven by thermal oscillations, the probability of rate-dependent and rate-independent segment scission was formulated. By considering the physics of segment scission along the lines of \citet{wang2019quantitative}, the dissipated energy of rate-dependent and rate-independent segment scission was also formulated. This segment scission framework was then pushed up to the chain level using probabilistic principles. Implications from this scission framework over statistical mechanics considerations in \citet{buche2021chain} permitted for the reference end-to-end chain distance to be represented as a function of segment extensibility and segment number. The chain model was used to precisely fit single-chain mechanical response data from atomic force microscopy tensile tests, validating the efficacy of the chain model. The scission framework was then called upon to yield valuable insight over the molecular physics taking place in the polymer chains under the AFM tensile test protocol.

Taking this chain modeling framework as a point of departure, a number of research paths are evident. Finite element-based elastomer fracture and fatigue models can seamlessly implement the framework presented in this manuscript since the entire framework is cast in terms of straightforward, computationally-efficient analytical functions. An exciting challenge is to incorporate this rupture framework in phase field fracture models simulating elastomers with a non-uniform distribution of chain length in order to explicitly capture delocalized chain rupture during fracture and fatigue. In addition to fracture and fatigue studies, nucleation and cavitation in elastomer networks can also be investigated. An outstanding challenge for a micromechanically-motivated continuum framework is the network-level representative micro-to-macro transition, which can also be motivated by discrete network calculations, as the bond potential derived here can be easily incorporated in coarse-grained molecular dynamics network calculations.

Clearly, the rupture framework as presented in this manuscript does not account for viscoelastic effects or for chain entanglements, which are increasingly important at the network level. Incorporating these effects to the rupture framework would certainly improve the model by adding a further dimension of relevant molecular physics to the theory. One could potentially go about adding viscoelasticity and chain entanglement effects from the fundamental statistical mechanics level via accounting for rate-dependent molecular friction force contributions to the chain. Alternatively, viscoelastic effects could be incorporated in a more statistical manner via the transient network theory developed by \citet{vernerey2017statistically}. Note that in such a model, reversible bond dissociation must be distinguished from irreversible bond rupture, along the lines of \citet{lamont2021rate}. Meanwhile, chain entanglement effects could be incorporated at the chain level in a more phenomenological manner by modulating the bond force along the chain with respect to the presence of entanglements, relevant to the recent work of \citet{hassan2022polyacrylamide}. \citet{hassan2022polyacrylamide} find that for short-chain elastomer networks, fracture toughness is independent of loading rate, but for long-chain elastomer networks above some critical loading rate, fracture toughness exhibits an inversely proportional relationship to loading rate. \citet{hassan2022polyacrylamide} postulate that for long chains at low loading rates, stresses caused by chains sliding relative to one another are negligible. As a result, the chain sustains tension evenly along its length, and the entire chain dissipates energy upon rupture, leading to high fracture toughness. However, for long chains at high loading rates, entanglements concentrate tension within a short portion of the chain, causing that high-tension region of the chain to rupture and dissipate energy while the rest of the chain remains relatively relaxed, thereby reducing fracture toughness. Incorporating this physical picture to the chain level in some phenomenological way would provide the means to study how chain entanglements impact delocalized chain rupture and fracture toughness in elastomer networks.

\appendix
\gdef\thesection{\Alph{section}}
\makeatletter
\renewcommand\@seccntformat[1]{Appendix \csname the#1\endcsname.\hspace{0.5em}}
\makeatother
\section{Derivation of the modulation parameter $\mu_{\nu}(\lambda_{\nu})$} \label{app:modulation-parameter-definition}

From the derivation of the asymptotically matched segment potential, it is found that \cref{eq:pdv-u_nu-mu_nu} must hold true:
\begin{equation}
    \pdv{\tilde{u}_{\nu}}{\mu_{\nu}} = -[1-\mu_{\nu}]\frac{\kappa_{\nu}}{\zeta_{\nu}^{char}}[\lambda_{\nu} - 1]^2 + 1 \geq 0~\text{for}~\lambda_{\nu}\geq 1. 
\end{equation}
To proceed, consider the satisfaction of the equality and the conditional separately, for $\lambda_{\nu}\geq 1$
\begin{align}
    &\text{(A)}~\pdv{\tilde{u}_{\nu}}{\mu_{\nu}} >0\implies-[1-\mu_{\nu}]\frac{\kappa_{\nu}}{\zeta_{\nu}^{char}}[\lambda_{\nu} - 1]^2 + 1 > 0, \\
    &\text{(B)}~\pdv{\tilde{u}_{\nu}}{\mu_{\nu}} =0\implies-[1-\mu_{\nu}]\frac{\kappa_{\nu}}{\zeta_{\nu}^{char}}[\lambda_{\nu} - 1]^2 + 1 = 0.
\end{align}
Examining the above leads to the following conditions:
\begin{itemize}
    \item[(A)] $\pdv{\tilde{u}_{\nu}}{\mu_{\nu}} >0\implies 1 - [1-\mu_{\nu}][\kappa_{\nu}/\zeta_{\nu}^{char}][\lambda_{\nu} - 1]^2 > 0$. 
    This is satisfied under the following:
    \begin{itemize}
        \item[(A1)] $\mu_{\nu} > 1 - \zeta_{\nu}^{char}/\left[\kappa_{\nu}[\lambda_{\nu} - 1]^2\right]$. However, this does not set any strict regulation on the exact value of $\mu_{\nu}$, and is therefore mathematically unhelpful.
        \item[(A2)] $1-\mu_{\nu}<0$. However, this violates the chain damage condition that $\mu_{\nu}\in\left[0,1\right]$.
        \item[(A3)] $\mu_{\nu} = 1$ or $\lambda_{\nu} = 1$, both of which imply that $1>0$.
        \item[(A4)] $\mu_{\nu} = 0\implies\kappa_{\nu}[\lambda_{\nu} - 1]^2/\zeta_{\nu}^{char}<1$, which potentially holds true depending on the values of $\zeta_{\nu}^{char}$, $\kappa_{\nu}$, and $\lambda_{\nu}\geq 1$.
    \end{itemize}
    \item[(B)] $\pdv{\tilde{u}_{\nu}}{\mu_{\nu}} =0\implies 1 - [1-\mu_{\nu}][\kappa_{\nu}/\zeta_{\nu}^{char}][\lambda_{\nu} - 1]^2 = 0$, i.e., $\pdv{\tilde{u}_{\nu}}{\mu_{\nu}} =0\implies \mu_{\nu} = 1 - \zeta_{\nu}^{char}/\left[\kappa_{\nu}[\lambda_{\nu} - 1]^2\right]$. 
    This is satisfied when (A) does not apply, i.e., when
    \begin{equation*}
        \kappa_{\nu}[\lambda_{\nu} - 1]^2/\zeta_{\nu}^{char}\geq 1.
    \end{equation*}
    However, when $\kappa_{\nu}[\lambda_{\nu} - 1]^2/\zeta_{\nu}^{char} = 1$, then $\mu_{\nu} = 0$.
\end{itemize}
Given this, an unambiguous definition of the modulation parameter $\mu_{\nu}$ is provided
\begin{equation} \label{eq:modulation-parameter-definition}
\mu_{\nu} = \begin{cases}
0,& \text{if~}\lambda_{\nu} < \lambda_{\nu}^{crit} \\
1-\frac{\zeta_{\nu}^{char}}{\kappa_{\nu}\left[\lambda_{\nu}-1\right]^2},& \text{if~}\lambda_{\nu} \geq \lambda_{\nu}^{crit} \\
\end{cases},
\end{equation}
where $\lambda_{\nu}^{crit} \equiv 1+\sqrt{\frac{\zeta_{\nu}^{char}}{\kappa_{\nu}}}$ is called the critical segment stretch. This analysis is exactly analogous to that performed in Appendix E of \citet{mulderrig2021affine}.

\section{Analytical form of the segment stretch function} \label{app:analytical-form-segment-stretch-function}

Substituting the Pad\'e approximant (\cref{eq:pade-approx}) for the $\lambda_c^{eq} < (\lambda_c^{eq})^{crit}$ case in \cref{eq:segment-stretch-composite-u_nu} and performing an appropriate cubic root analysis \citep{zwillinger2002crc} leads to
\begin{equation}
    \lambda_{\nu} = \lambda_{\nu}^{PSB}(\kappa_{\nu}; \lambda_c^{eq}) = 2\sqrt{-\frac{\tilde{\pi}}{3}}\cos(\frac{1}{3}\arccos(\frac{3\tilde{\rho}}{2\tilde{\pi}}\sqrt{-\frac{3}{\tilde{\pi}}}) - \frac{2\pi}{3}) - \frac{\tilde{\beta}}{3\tilde{\alpha}},
\end{equation}
where 
\begin{align*}
& \tilde{\alpha} = 1,~\tilde{\beta} = -\left[\frac{3[\kappa_{\nu} + 1] + \lambda_c^{eq}[2\kappa_{\nu} + 3]}{\kappa_{\nu} + 1}\right],~\tilde{\gamma} = \frac{2\kappa_{\nu} + \lambda_c^{eq}[4\kappa_{\nu} + 6 + \lambda_c^{eq}[\kappa_{\nu} + 3]]}{\kappa_{\nu} + 1}, \\
& \tilde{\delta} = \frac{2 - \lambda_c^{eq}[2\kappa_{\nu} + \lambda_c^{eq}[\kappa_{\nu} + 3 + \lambda_c^{eq}]]}{\kappa_{\nu} + 1},~\tilde{\pi} = \frac{3\tilde{\alpha}\tilde{\gamma} - \tilde{\beta}^2}{3\tilde{\alpha}^2},~\tilde{\rho} = \frac{2\tilde{\beta}^3 - 9\tilde{\alpha}\tilde{\beta}\tilde{\gamma} + 27\tilde{\alpha}^2\tilde{\delta}}{27\tilde{\alpha}^3}.
\end{align*}
The above holds provided that $4\tilde{\pi}^3+27\tilde{\rho}^2<0$, $3\tilde{\alpha}\tilde{\gamma} - \tilde{\beta}^2 < 0$, and
\begin{equation*}
    -1 < \frac{3\tilde{\rho}}{2\tilde{\pi}}\sqrt{-\frac{3}{\tilde{\pi}}} < 1,
\end{equation*}
which are each satisfied for physically-sensible $\lambda_c^{eq}$.

Using the Bergstr\"{o}m approximant (\cref{eq:bergstrom-approx}) for the $\lambda_c^{eq} < (\lambda_c^{eq})^{crit}$ case in \cref{eq:segment-stretch-composite-u_nu} and performing an appropriate quadratic root analysis leads to
\begin{equation}
    \lambda_{\nu} = \lambda_{\nu}^{BSB}(\kappa_{\nu}; \lambda_c^{eq}) = \frac{\lambda_c^{eq} + 1 + \sqrt{[\lambda_c^{eq}]^2 - 2\lambda_c^{eq} + 1 + \frac{4}{\kappa_{\nu}}}}{2}.
\end{equation}
The above holds provided that $[\lambda_c^{eq}]^2 - 2\lambda_c^{eq} + 1 + \frac{4}{\kappa_{\nu}}>0$, which is satisfied for physically-sensible $\lambda_c^{eq}$. 

Using the Bergstr\"{o}m approximant (\cref{eq:bergstrom-approx}) for the $\lambda_c^{eq} \geq (\lambda_c^{eq})^{crit}$ case in \cref{eq:segment-stretch-composite-u_nu} and performing an appropriate cubic root analysis \citep{zwillinger2002crc} leads to
\begin{equation}
    \lambda_{\nu} = \lambda_{\nu}^{BSP}(\zeta_{\nu}^{char}, \kappa_{\nu}; \lambda_c^{eq}) = 2\sqrt{-\frac{\tilde{P}}{3}}\cos(\frac{1}{3}\arccos(\frac{3\tilde{R}}{2\tilde{P}}\sqrt{-\frac{3}{\tilde{P}}}) - \frac{2\pi}{3}) - \frac{\tilde{B}}{3\tilde{A}},
\end{equation}
where
\begin{align*}
& \tilde{A} = 1,~\tilde{B} = -3,~\tilde{C} = 3 - \frac{[\zeta_{\nu}^{char}]^2}{\kappa_{\nu}},~\tilde{D} = \frac{[\zeta_{\nu}^{char}]^2}{\kappa_{\nu}}\lambda_c^{eq} - 1, \\
& \tilde{P} = \frac{3\tilde{A}\tilde{C} - \tilde{B}^2}{3\tilde{A}^2},~\tilde{R} = \frac{2\tilde{B}^3 - 9\tilde{A}\tilde{B}\tilde{C} + 27\tilde{A}^2\tilde{D}}{27\tilde{A}^3}.
\end{align*}
The above holds provided that $4\tilde{P}^3+27\tilde{R}^2<0$, $3\tilde{A}\tilde{C} - \tilde{B}^2 < 0$, and
\begin{equation*}
    -1 < \frac{3\tilde{R}}{2\tilde{P}}\sqrt{-\frac{3}{\tilde{P}}} < 1,
\end{equation*}
which are each satisfied for physically-sensible $\lambda_c^{eq}$.

A note regarding the superscript nomenclature: $P$ and $B$ indicate if the solution for $\lambda_{\nu}$ is derived using either the Pad\'e approximant ($P$) or the Bergstr\"{o}m approximant ($B$). $SB$ indicates if the solution for $\lambda_{\nu}$ is derived for the case where $\lambda_c^{eq}$ is less than its so-called critical value, $(\lambda_c^{eq})^{crit}$, i.e., the segment is in a sub-critical ($SB$) state. $SP$ indicates if the solution for $\lambda_{\nu}$ is derived for the case where $\lambda_c^{eq}$ is greater than or equal to its so-called critical value, i.e., the segment is in a super-critical ($SP$) state.

\section{Analytical form of the equilibrium chain stretch function} \label{app:equil-chain-segment-stretch-function}

\begin{figure}[t]
	\centering
	\includegraphics[width=0.75\textwidth]{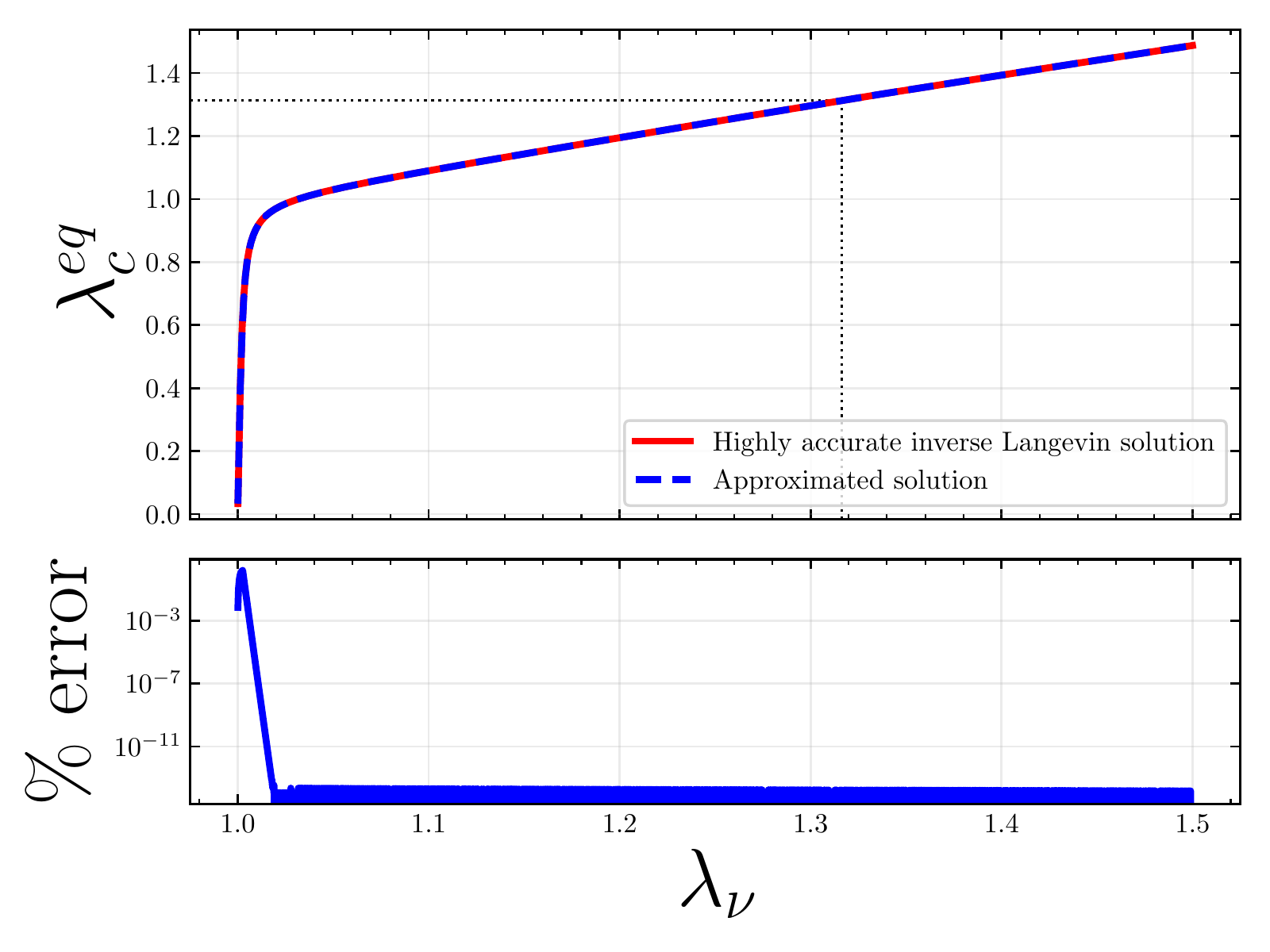}
	\caption{Equilibrium chain stretch evolution. Here, $\zeta_{\nu}^{char} = 100$ and $\kappa_{\nu} = 1000$, as per the composite segment potential displayed in \cref{fig:composite-u_nu-figs}. The black dotted lines denote the critical segment stretch $\lambda_{\nu}^{crit}$ and the critical equilibrium chain stretch $(\lambda_c^{eq})^{crit}$. (top) Equilibrium chain stretch $\lambda_c^{eq}$ as a function of segment stretch $\lambda_{\nu}$ as per the approximated analytical $\lambda_c^{eq}$ function from \cref{eq:equil-chain-stretch-function} along with $\lambda_c^{eq}$ calculated using a highly accurate numerical solution for the inverse Langevin function. (bottom) The percent error of the approximated analytical $\lambda_c^{eq}$ function in \cref{eq:equil-chain-stretch-function} relative to $\lambda_c^{eq}$ calculated using a highly accurate numerical solution for the inverse Langevin function.}
	\label{fig:lmbda_c_eq-vs-lmbda_nu-exact-and-approximated-solutions}
\end{figure}

Recall that substituting the nondimensional composite segment potential into the definition of segment stretch and simplifying leads to
\begin{equation}
\mathcal{L}^{-1}(\lambda_c^{eq} - \lambda_{\nu} + 1) = \pdv{u_{\nu}}{\lambda_{\nu}} = \begin{cases}
\kappa_{\nu}[\lambda_{\nu} - 1],& \text{if~}\lambda_{\nu} < \lambda_{\nu}^{crit} \\
\frac{[\zeta_{\nu}^{char}]^2}{\kappa_{\nu}\left[\lambda_{\nu}-1\right]^3},& \text{if~}\lambda_{\nu} \geq \lambda_{\nu}^{crit} \\
\end{cases},
\end{equation}
where $\lambda_{\nu}^{crit} = 1+\sqrt{\frac{\zeta_{\nu}^{char}}{\kappa_{\nu}}}$. Using the Pad\'e approximant for the $\lambda_{\nu} < \lambda_{\nu}^{P2B}$ case (where $\lambda_{\nu}^{P2B} = \lambda_{\nu}((\lambda_c^{eq})^{P2B})$) and performing an appropriate cubic root analysis leads to
\begin{equation}
    \lambda_c^{eq} = 2\sqrt{-\frac{\tilde{\pi}}{3}}\cos(\frac{1}{3}\arccos(\frac{3\tilde{\rho}}{2\tilde{\pi}}\sqrt{-\frac{3}{\tilde{\pi}}}) - \frac{2\pi}{3}) - \frac{\tilde{\beta}}{3\tilde{\alpha}},
\end{equation}
with 
\begin{align*}
& \tilde{\alpha} = 1,~\tilde{\beta} = [\kappa_{\nu} + 3][1 - \lambda_{\nu}],~\tilde{\gamma} = [2\kappa_{\nu} + 3][\lambda_{\nu}^2 - 2\lambda_{\nu}] + 2\kappa_{\nu}, \\
& \tilde{\delta} = [\kappa_{\nu} + 1][3\lambda_{\nu}^2 - \lambda_{\nu}^3] - 2[\kappa_{\nu}\lambda_{\nu} + 1],~\tilde{\pi} = \frac{3\tilde{\alpha}\tilde{\gamma} - \tilde{\beta}^2}{3\tilde{\alpha}^2},~\tilde{\rho} = \frac{2\tilde{\beta}^3 - 9\tilde{\alpha}\tilde{\beta}\tilde{\gamma} + 27\tilde{\alpha}^2\tilde{\delta}}{27\tilde{\alpha}^3}.
\end{align*}
The above holds provided that $4\tilde{\pi}^3+27\tilde{\rho}^2<0$, $3\tilde{\alpha}\tilde{\gamma} - \tilde{\beta}^2 < 0$, and
\begin{equation*}
    -1 < \frac{3\tilde{\rho}}{2\tilde{\pi}}\sqrt{-\frac{3}{\tilde{\pi}}} < 1,
\end{equation*}
which are each satisfied for physically-sensible $\lambda_{\nu}$. Using the Bergstr\"{o}m approximant for the $\lambda_{\nu}^{P2B} \leq \lambda_{\nu} < \lambda_{\nu}^{crit}$ case and simplifying leads to
\begin{equation}
    \lambda_c^{eq} = \lambda_{\nu} - \left[\frac{1}{\kappa_{\nu}}\right]\left[\frac{1}{\lambda_{\nu} - 1}\right],
\end{equation}
where $\lambda_{\nu} > 1$ holds true here. Using the Bergstr\"{o}m approximant for the $\lambda_{\nu} \geq \lambda_{\nu}^{crit}$ case and simplifying leads to
\begin{equation}
    \lambda_c^{eq} = \lambda_{\nu} - \left[\frac{\kappa_{\nu}}{[\zeta_{\nu}^{char}]^2}\right][\lambda_{\nu} - 1]^3.
\end{equation}
Considering all of this, the approximated analytical form of the equilibrium chain stretch as a function of segment stretch is 
\begin{equation} \label{eq:equil-chain-stretch-function}
\lambda_c^{eq} = \begin{cases}
2\sqrt{-\frac{\tilde{\pi}}{3}}\cos(\frac{1}{3}\arccos(\frac{3\tilde{\rho}}{2\tilde{\pi}}\sqrt{-\frac{3}{\tilde{\pi}}}) - \frac{2\pi}{3}) - \frac{\tilde{\beta}}{3\tilde{\alpha}},& \text{if~}\lambda_{\nu} < \lambda_{\nu}^{P2B} \\
\lambda_{\nu} - \left[\frac{1}{\kappa_{\nu}}\right]\left[\frac{1}{\lambda_{\nu} - 1}\right],& \text{if~}\lambda_{\nu}^{P2B} \leq \lambda_{\nu} < \lambda_{\nu}^{crit} \\
\lambda_{\nu} - \left[\frac{\kappa_{\nu}}{[\zeta_{\nu}^{char}]^2}\right][\lambda_{\nu} - 1]^3,& \text{if~}\lambda_{\nu} \geq \lambda_{\nu}^{crit} \\
\end{cases},
\end{equation}
with
\begin{align*}
& \tilde{\alpha} = 1,~\tilde{\beta} = [\kappa_{\nu} + 3][1 - \lambda_{\nu}],~\tilde{\gamma} = [2\kappa_{\nu} + 3][\lambda_{\nu}^2 - 2\lambda_{\nu}] + 2\kappa_{\nu}, \\
& \tilde{\delta} = [\kappa_{\nu} + 1][3\lambda_{\nu}^2 - \lambda_{\nu}^3] - 2[\kappa_{\nu}\lambda_{\nu} + 1],~\tilde{\pi} = \frac{3\tilde{\alpha}\tilde{\gamma} - \tilde{\beta}^2}{3\tilde{\alpha}^2},~\tilde{\rho} = \frac{2\tilde{\beta}^3 - 9\tilde{\alpha}\tilde{\beta}\tilde{\gamma} + 27\tilde{\alpha}^2\tilde{\delta}}{27\tilde{\alpha}^3}.
\end{align*}
\cref{fig:lmbda_c_eq-vs-lmbda_nu-exact-and-approximated-solutions} displays the approximated equilibrium chain stretch function as per \cref{eq:equil-chain-stretch-function}, the equilibrium chain stretch calculated using a highly accurate numerical solution for the inverse Langevin function, and the percent error between each of these functions. This figure convincingly verifies that the approximated analytical equilibrium chain stretch function is sufficiently accurate with respect to the highly accurate numerical solution in the domain of physically-sensible segment stretches.

As a matter of proper interpretation, \cref{eq:equil-chain-stretch-function} ought to be treated as the inverse segment stretch function given in \cref{eq:segment-stretch-function}.

\section{Bond-level theory} \label{sec:bonds-theory}

To reconcile the transition from the segment level to the bond level, this Appendix section highlights the minor differences that arise for the bond-level theory for the single chain model and the rupture framework.

The transition from the segment level to the bond level is built upon the fact that each individual segment in the chain is identically composed of $\nu_b$ bonds. Given this, then the product $[\nu][\nu_b]$ is the number of bonds composing the entire polymer chain backbone, $n_b$. The bond stretch is the ratio of the bond length $l_b$ with the equilibrium bond length $l_b^{eq}$, $\lambda_b = l_b/l_b^{eq}$. The energy state of each bond is described by the bond potential, $U_b$, which inherently exhibits some characteristic bond potential energy scale $E_b^{char}$ and bond stiffness $k_b$ defined as
\begin{equation}
	k_b \equiv U_b^{\prime\prime}(l_b^{eq}) = \pdv[2]{U_b(l_b)}{l_b}\bigg|_{l_b = l_b^{eq}}.
\end{equation}
The functional form for the bond potential analogously follows that for the segment potential as per \cref{eq:harmonic-potential-and-log-squared-potential} and \cref{eq:lj-potential-and-morse-potential}, with the segment-level parameters $l_{\nu}$, $l_{\nu}^{eq}$, $E_{\nu}^{char}$, and $k_{\nu}$ respectively swapped for the bond-level parameters $l_b$, $l_b^{eq}$, $E_b^{char}$, and $k_b$
\begin{align}
    & U_b^{har}(l_b) = E_b^{char}\left[\frac{1}{2}\frac{k_b}{E_b^{char}}\left[l_b - l_b^{eq}\right]^2 - 1\right],\qquad U_b^{\ln^2}(l_b) = E_b^{char}\left[\frac{1}{2}\frac{[l_b^{eq}]^2 k_b}{E_b^{char}}\left[\ln(\frac{l_b}{l_b^{eq}})\right]^2 - 1\right], \\
    & U_b^{lj}(l_b) = E_b^{char}\left[\left[\frac{l_b^{eq}}{l_b}\right]^{12} - 2\left[\frac{l_b^{eq}}{l_b}\right]^6\right],\qquad\qquad  U_b^{morse}(l_b) = E_b^{char}\left[\left[1 - e^{-a_b[l_b - l_b^{eq}]}\right]^2 - 1\right],
\end{align}
where $a_b$ is the Morse parameter and is related to $E_b^{char}$ and $k_b$ via $k_b = 2 a_b^2E_b^{char}$. The composite bond potential is written as
\begin{equation}
    U_b(l_b) = \begin{cases}
	 E_b^{char}\left[\frac{1}{2}\frac{k_b}{E_b^{char}}\left[l_b - l_b^{eq}\right]^2 - 1\right],& \text{if~}l_b < l_b^{crit} = l_b^{eq}\lambda_b^{crit}\\
	-\frac{E_b^2}{2k_b\left[l_b-l_b^{eq}\right]^2},& \text{if~}l_b \geq l_b^{crit} \\
	\end{cases}.
\end{equation}
The nondimensional bond potential, nondimensional characteristic bond potential energy scale, and nondimensional bond stiffness are respectively defined as $u_b \equiv \beta U_b$, $\zeta_b^{char} \equiv \beta E_b^{char}$, and $\kappa_b \equiv \beta [l_b^{eq}]^2 k_b$. The nondimensional Morse parameter is also defined as $\alpha_b \equiv l_b^{eq} a_b$. The nondimensional scaled bond potential $\overline{u}_b$, and its (non-negative) shifted counterpart $\tilde{u}_b$ are respectively defined as $\overline{u}_b \equiv u_b/\zeta_b^{char}$ and $\tilde{u}_b \equiv \overline{u}_b + 1$. The bond force is given in nondimensional terms as
\begin{equation}
    \xi_b = \pdv{u_b}{\lambda_b}.
\end{equation}
Segment-level length and energy measures multiplicatively scale with their bond-level counterparts by a factor of $\nu_b$. Thus, segment length, equilibrium segment length, and segment potential are respectively defined as $l_{\nu} = \nu_b l_b$,  $l_{\nu}^{eq} = \nu_b l_b^{eq}$, and $U_{\nu} = \nu_b U_b$. As a corollary, the following relations are found to hold true
\begin{align}
    & \lambda_{\nu} = \lambda_b,\qquad E_{\nu}^{char} = \nu_b E_b^{char},\qquad k_{\nu} = \frac{k_b}{\nu_b},\qquad u_{\nu} = \nu_b u_b,\qquad \zeta_{\nu}^{char} = \nu_b \zeta_b^{char},\qquad \kappa_{\nu} = \nu_b \kappa_b, \\
    & \lambda_{\nu}^{crit} = \lambda_b^{crit},\qquad a_{\nu} = \frac{a_b}{\nu_b}, \qquad \alpha_{\nu} = \alpha_b,\qquad \overline{u}_{\nu} = \overline{u}_b,\qquad \tilde{u}_{\nu} = \tilde{u}_b,\qquad \xi_{\nu} = \nu_b \xi_b.
\end{align}
Considering all of this, the nondimensional Helmholtz free energy per bond, $\psi_{cb}$, is given as
\begin{align}
    & \psi_{cb}(\lambda_b, \lambda_c^{eq}) = s_{cb}(\lambda_b, \lambda_c^{eq}) + u_b(\lambda_b), \\
    & s_{cb}(\lambda_b, \lambda_c^{eq}) = \frac{1}{\nu_b}\left[[\lambda_c^{eq} - \lambda_b + 1]\mathcal{L}^{-1}(\lambda_c^{eq} - \lambda_b + 1) + \ln(\frac{\mathcal{L}^{-1}(\lambda_c^{eq} - \lambda_b + 1)}{\sinh(\mathcal{L}^{-1}(\lambda_c^{eq} - \lambda_b + 1))})\right], \\
    & \psi_{cb}(\lambda_b, \lambda_c^{eq}) = \frac{1}{\nu_b}\left[[\lambda_c^{eq} - \lambda_b + 1]\mathcal{L}^{-1}(\lambda_c^{eq} - \lambda_b + 1) + \ln(\frac{\mathcal{L}^{-1}(\lambda_c^{eq} - \lambda_b + 1)}{\sinh(\mathcal{L}^{-1}(\lambda_c^{eq} - \lambda_b + 1))})\right] + u_b(\lambda_b),
\end{align}
where $s_{cb}$ is the nondimensional chain-level entropic contributions per bond, $u_b$ is the nondimensional bond-level enthalpic contributions, and $\psi_{c\nu} = \nu_b \psi_{cb}$.

With $\lambda_{\nu} = \lambda_b$, then, by proxy, the functional form for $\lambda_b$ is provided by \cref{eq:segment-stretch-function}. As a result, the reference bond stretch $\Lambda_b^{ref} = \Lambda_{\nu}^{ref}$, where $\Lambda_b^{ref}$ is taken as the ratio of the reference bond length $l_b^{ref}$ with the equilibrium bond length $l_b^{eq}$, $\Lambda_b^{ref} = l_b^{ref}/l_b^{eq}$.

Finally, taking all of the bond-level formulation up to this point in consideration, a bond scission framework can be developed in a completely analogous manner to how the segment scission framework was developed in \cref{subsec:segment-scission}. Taking $\hat{e}_b^{sci}$ as the nondimensional bond scission activation energy barrier, then the rate-independent probability of bond scission $\hat{p}_b^{sci}$ and the rate-independent probability of bond survival $\hat{p}_b^{sur}$ are provided for a statistically significant number of bonds
\begin{equation}
    \hat{p}_b^{sci} = \exp{-\hat{e}_b^{sci}},\qquad\hat{p}_b^{sur} = 1 - \hat{p}_b^{sci}.
\end{equation}
The rate-dependent probability of bond survival $\rho_b$, and rate-dependent probability of bond scission $\gamma_b$ are analogously defined
\begin{equation}
    \frac{\dot{\rho}_b}{\rho_b} = -\omega_0\hat{p}_b^{sci},\qquad\gamma_b = 1-\rho_b \implies \frac{\dot{\gamma}_b}{1-\gamma_b} = \omega_0\hat{p}_b^{sci}.
\end{equation}
The nondimensional bond scission energy is given as
\begin{equation}
    \hat{\varepsilon}_b^{sci} \equiv s_{cb}\left(\hat{\lambda}_b, \hat{\lambda}_c^{eq}\right) +  u_b\left(\hat{\lambda}_b\right) + \zeta_b^{char} = \psi_{cb}\left(\hat{\lambda}_b, \hat{\lambda}_c^{eq}\right) + \zeta_b^{char}.
\end{equation}
The rate-dependent nondimensional dissipated bond scission energy $\hat{\varepsilon}_b^{diss}$ is defined via its time rate-of-change equation
\begin{equation}
    \dot{\hat{\varepsilon}}_b^{diss} \equiv \dot{\gamma}_b\hat{\varepsilon}_b^{sci}.
\end{equation}
In an analogous way, the rate-independent $\hat{\varepsilon}_b^{diss}$ is defined via its applied bond stretch-based rate-of-change equation
\begin{equation}
    \left(\hat{\varepsilon}_b^{diss}\right)^{\prime} \equiv \left(\hat{p}_b^{sci}\right)^{\prime}\hat{\varepsilon}_b^{sci},
\end{equation}
where derivatives are taken with respect to $\hat{\lambda}_b$. For $\hat{\lambda}_b \leq \lambda_b^{crit}$
\begin{align}
    & \hat{e}_b^{sci} = \frac{1}{2}\kappa_b\left[\hat{\lambda}_b - 1\right]^2 - \frac{3}{2}\sqrt[3]{\left[\zeta_b^{char}\right]^2\kappa_b\left[\hat{\lambda}_b - 1\right]^2} + \zeta_b^{char}, \\
    & \left(\hat{p}_b^{sci}\right)^{\prime} = \hat{p}_b^{sci}\left[\sqrt[3]{\frac{\left[\zeta_b^{char}\right]^2\kappa_b}{\hat{\lambda}_b - 1}} - \kappa_b[\hat{\lambda}_b - 1]\right], \\
    & \left(\hat{\varepsilon}_b^{diss}\right)^{\prime} = \hat{p}_b^{sci}\left[\sqrt[3]{\frac{\left[\zeta_b^{char}\right]^2\kappa_b}{\hat{\lambda}_b - 1}} - \kappa_b[\hat{\lambda}_b - 1]\right]\hat{\varepsilon}_b^{sci},
\end{align}
where $\hat{\varepsilon}_b^{diss} = 0$ at $\hat{\lambda}_b = 1$. If irreversible bond scission is assumed to take place in the network, then $\left(\hat{p}_b^{sci}\right)^{\prime} \geq 0$ and $\left(\hat{\varepsilon}_b^{diss}\right)^{\prime} \geq 0$ must hold during deformation. The rate-independent probability of segment scission $\hat{p}_{\nu}^{sci}$ and the rate-independent probability of segment survival $\hat{p}_{\nu}^{sur}$ are implied to be
\begin{equation}
    \hat{p}_{\nu}^{sci} = [\hat{p}_b^{sci}]^{\nu_b},\qquad\hat{p}_{\nu}^{sur} = 1 - \hat{p}_{\nu}^{sci}.
\end{equation}
Via probabilistic considerations from \citet{guo2021micromechanics}, the rate-independent probability of chain survival $\hat{p}_c^{sur}$ and the rate-independent probability of chain scission $\hat{p}_c^{sci}$, are given
\begin{align}
    \hat{p}_c^{sur} = [\hat{p}_b^{sur}]^{n_b},\qquad\hat{p}_c^{sci} = 1 - \hat{p}_c^{sur}.
\end{align}
The rate-dependent probability of chain survival $\rho_c$ is then related to $\rho_b$ via $\rho_c = [\rho_b]^{n_b}$. As a result, $\rho_c$ and its counterpart, the rate-dependent probability of chain scission $\gamma_c$, can be represented with respect to bond-level probabilistic quantities
\begin{equation}
    \frac{\dot{\rho}_c}{\rho_c} = -n_b\omega_0\hat{p}_b^{sci},\qquad \gamma_c = 1-\rho_c \implies \frac{\dot{\gamma}_c}{1-\gamma_c} = n_b\omega_0\hat{p}_b^{sci}.
\end{equation}
The time rate-of-change equation for the rate-dependent nondimensional dissipated chain scission energy per bond $\hat{\varepsilon}_{cb}^{diss}$ is defined as \citep{guo2021micromechanics}
\begin{equation}
    \dot{\hat{\varepsilon}}_{cb}^{diss} \equiv \dot{\gamma}_c\hat{\varepsilon}_{cb}^{sci}.
\end{equation}
The applied bond stretch-based rate-of-change for the rate-independent $\hat{\varepsilon}_{cb}^{diss}$ is defined as
\begin{align}
    & \left(\hat{\varepsilon}_{cb}^{diss}\right)^{\prime} \equiv \left(\hat{p}_c^{sci}\right)^{\prime}\hat{\varepsilon}_{cb}^{sci},\qquad  \left(\hat{p}_c^{sci}\right)^{\prime}= n_b[1 - \hat{p}_b^{sci}]^{n_b-1}\left(\hat{p}_b^{sci}\right)^{\prime}, \\
    &\left(\hat{\varepsilon}_{cb}^{diss}\right)^{\prime} =  n_b[1 - \hat{p}_b^{sci}]^{n_b-1}\hat{p}_b^{sci}\left[\sqrt[3]{\frac{\left[\zeta_b^{char}\right]^2\kappa_b}{\hat{\lambda}_b - 1}} - \kappa_b[\hat{\lambda}_b - 1]\right]\hat{\varepsilon}_b^{sci},
\end{align}
where $\hat{\varepsilon}_{cb}^{diss} = 0$ at $\hat{\lambda}_b = 1$.

\section*{CRediT authorship contribution statement}

\textbf{Jason Mulderrig:} Conceptualization, Methodology, Software, Formal analysis, Investigation, Writing - Original Draft, Writing - Review \& Editing. \textbf{Brandon Talamini:} Conceptualization, Writing - Review \& Editing. \textbf{Nikolaos Bouklas:} Conceptualization, Methodology, Resources, Writing - Review \& Editing, Supervision.

\section*{Declaration of Competing Interest}

The authors declare that they have no known competing financial interests or personal relationships that could have appeared to influence the work reported in this paper.

\section*{Acknowledgment}

This material is based upon work supported by the National Science Foundation Graduate Research Fellowship Program under Grant No. DGE-1650441. Any opinions, findings, and conclusions or recommendations expressed in this material are those of the author(s) and do not necessarily reflect the views of the National Science Foundation.

This work was performed in part under the auspices of the U.S. Department of Energy by Lawrence Livermore National Laboratory under Contract DE-AC52-07NA27344.

\bibliographystyle{elsarticle-harv}

\bibliography{bibdata}

\end{document}